\documentclass[aps,prd,twocolumn,showpacs,showkeys,preprintnumbers,superscriptaddress,bibnotes,floatfix,longbibliography,nofootinbib]{revtex4-1}

\usepackage{placeins}
\usepackage{soul}
\usepackage{chngcntr}
\usepackage{color}
\usepackage{epsfig} 
\usepackage{graphicx}
\usepackage{tabularx}
\usepackage{xspace}
\usepackage{float}
\usepackage{comment}
\usepackage[usenames, dvipsnames]{xcolor}

\usepackage{color}
\usepackage{halloweenmath}
\usepackage{verbatim}
\usepackage{amsmath}
\usepackage{amssymb}
\usepackage{mathtools}
\usepackage{url}
\usepackage{bbold}
\usepackage{slashed}
\usepackage{array}
\usepackage{comment}
\usepackage{multirow}
\usepackage{threeparttable}
\usepackage{paralist}
\usepackage{xspace}
\usepackage{upgreek}
\usepackage{lipsum}
\usepackage{mathrsfs}
\usepackage[compat=1.1.0]{tikz-feynman}
\usepackage[colorlinks=true,citecolor=blue,urlcolor=blue]{hyperref}
\usepackage{slashed}
\usetikzlibrary{shapes.geometric, arrows}

\newcommand{\apjs}{ApJS}
\DeclareRobustCommand{\Eq}[1]{Eq.~\eqref{#1}}

\newcommand{\real}{\mathrm{Re}}

\newcommand{\g}{\gamma}

\newcommand{\mAp}{m_{A^\prime}}

\newcommand{\eps}{\epsilon}

\newcommand{\resfreq}{\Omega_I}

\begin{document}

\title{Resonant axion and dark photon production in magnetic white dwarfs}
\author{Nirmalya Brahma}
\email{nirmalaya.brahma@mail.mcgill.ca}
\affiliation{Department of Physics \& Trottier Space Institute,
McGill University, Montr\'eal, Qu\'ebec H3A 2T8, Canada}
\author{Ella Iles}
\email{ella.iles@mail.mcgill.ca}
\affiliation{Department of Physics \& Trottier Space Institute,
McGill University, Montr\'eal, Qu\'ebec H3A 2T8, Canada}
\author{Hugo Sch\'erer}
\email{hugo.scherer@mail.mcgill.ca}
\affiliation{Department of Physics \& Trottier Space Institute,
McGill University, Montr\'eal, Qu\'ebec H3A 2T8, Canada}
\author{Katelin Schutz}
\email{katelin.schutz@mcgill.ca}
\affiliation{Department of Physics \& Trottier Space Institute,
McGill University, Montr\'eal, Qu\'ebec H3A 2T8, Canada}

\begin{abstract}\noindent
Using magnetic white dwarfs as a case study, we show that the emission of sub-MeV bosons from stellar plasmas can be substantially modified in the presence of a magnetic field. In particular, the magnetic field-induced anisotropy and cyclotron resonance both significantly affect the in-medium dispersion relations of the Standard Model photon. As a result, resonant level crossing between photons and other light bosons occurs under environmental conditions that differ from the resonance criteria in unmagnetized environments. We find that the magnetic field opens additional regions within magnetic white dwarfs where resonance can occur. These findings motivate revisiting astrophysical constraints on light bosons in systems where the cyclotron frequency is comparable to or larger than the plasma frequency.

\noindent
\end{abstract}
\maketitle
\section{Introduction}
Axions are one of the most well-motivated extensions of the Standard Model (SM) owing to the fact that their existence can address the strong-$CP$ problem~\cite{peccei1977cp, weinberg1978new, wilczek1978problem,abbott1983cosmological, preskill1983cosmology, dine1983not} and that they arise in the context of many string theories~\cite{witten1984some,svrcek2006axions, Arvanitaki:2009fg, acharya2010m, Cicoli:2012sz}. Axions can interact with SM photons via the interaction Lagrangian 
\begin{equation}
\label{photon-axion-eom}
\mathcal{L} \supset \frac{1}{2}\partial_{\mu}a\partial^{\mu}a-\frac{1}{2}m_{a}^{2}a^{2}-\frac{g_{a\gamma}}{4} aF_{\mu\nu}\tilde{F}^{\mu\nu}
\end{equation}
where $\tilde{F}^{\mu\nu} = \frac{1}{2}\eps^{\mu\nu\alpha\beta} F_{\alpha\beta}$ is the dual of the electromagnetic field strength $F^{\mu\nu}$, and where $g_{a\gamma}$ is a model-dependent coupling~\cite{Kim:1979if,Shifman:1978bx,Zhitnitsky:1980tq,Dine:1981rt}. Similarly, dark photons arise as the gauge bosons for the simplest extension of the SM gauge group with an additional $U(1)'$~\cite{Holdom:1985ag}, which is yet another feature of many string theoretic constructions~\cite{Abel:2008ai,Goodsell:2009xc}. Their dimension-four kinetic mixing with the SM photon, \begin{align}
 \mathcal{L} \supset 
 &-\frac{1}{4}\, F'_{\mu\nu} \, F'^{\mu\nu} +\frac{\kappa}{2} \, F'_{\mu\nu} \, F^{\mu\nu} +\frac{1}{2} \, \mAp^{2} \, A'_{\mu} \, {A}'^{\mu} ~, 
\end{align}
is one of the most well-studied portals to a dark sector~\cite{Pospelov:2008zw,Fabbrichesi:2020wbt}, characterized by the dimensionless kinetic mixing parameter $\kappa$ and St\"uckelberg mass $m_{A'}$~\cite{Stueckelberg:1938hvi}. 

The production of these light, beyond-SM (BSM) bosons can be efficient in hot, dense stellar plasmas~\cite{Brahma:2025wos}. Due to their long mean free path, BSM particle production can have a large impact on stellar energy transport and evolution, even if the emission rate is low. Various stellar observables are sensitive to this energy loss, enabling some of the tightest constraints on axions, dark photons, and other light degrees of freedom~\cite{davidson1991limits, Davidson:1993sj, blinnikov1994cooling,Davidson:2000hf, Vogel:2013raa, Hardy:2016kme, An:2013yfc, Gondolo:2008dd, Vinyoles:2015aba, Ayala:2014pea, raffelt1990core,Haft:1993jt,Viaux:2013hca,Viaux:2013lha, Arceo-Diaz:2015pva,MillerBertolami:2014rka, Isern:2008nt,Corsico:2012ki,Fung:2023euv,Dolan:2022kul,Dolan:2023cjs}. BSM particles emitted from stellar interiors can also convert back to photons in the stellar atmosphere~\cite{Dessert:2019sgw, Dessert:2021bkv,Dessert:2022yqq, Noordhuis:2022ljw,Noordhuis:2023wid,Prabhu:2023cgb}, providing more ways to constrain BSM physics using stars. 

In many stellar environments, BSM particles can be produced resonantly, leading to an emission rate that is enhanced by orders of magnitude compared to continuum emission. Such resonant production occurs when the kinematics of the SM photon and the BSM particle match, leading to a level crossing. In a vacuum, the photon dispersion relation $\omega = k$ can never cross the dispersion relation of a massive BSM particle. However, in a medium, photons (sometimes referred to as plasmons) have nontrivial dispersion relations that are described by the real, Hermitian part of the photon self-energy tensor, $\real\, \Pi^{\mu \nu}$, which can be computed using thermal field theory both for on-shell and off-shell excitations~\cite{Scherer:2024uui}. For instance, in a classical plasma, the on-shell transverse mode of the photon has the same dispersion relation as a massive particle with an effective mass given by the plasma frequency imparted by the free electrons, $\omega_p= e \sqrt{n_e/m_e}$. In general, $\Pi^{\mu \nu}$ is a function of the photon kinematics and the properties of the ambient environment such as the temperature and density. It is crucial to accurately determine $\Pi^{\mu \nu}$ in stellar environments in order to establish where resonant conversion between photons and BSM states can occur, as mismodeling of the criteria for resonant conversion can strongly impact stellar bounds on BSM physics~\cite{Hardy:2016kme}.

\begin{figure}[t!]
\centering
\includegraphics[width=0.35\textwidth]{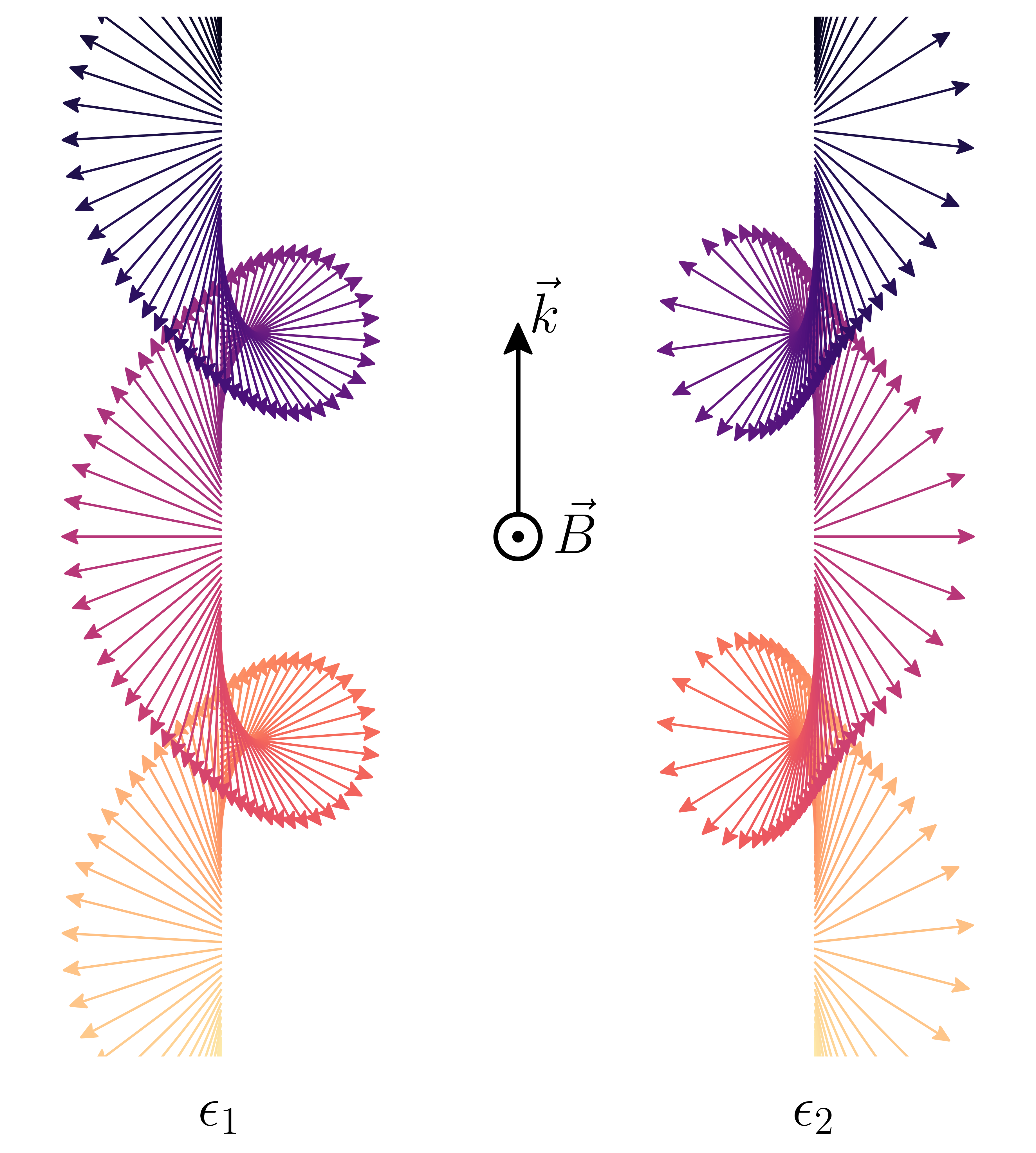}
\vspace{-0.3cm}
\caption{A depiction of the polarization vectors $\mathrm{Re}[\,\vec{\epsilon}_I e^{-i \vec{k} \cdot \vec{x}} \, ]$ for two propagating plasma normal modes in a configuration where the magnetic field (coming out of the page) is perpendicular to the direction of propagation. These two normal modes, which are orthogonal to the magnetic field (purely in the plane of the page), have components that are both transverse and longitudinal to the direction of propagation. The third normal mode is parallel to the magnetic field and is purely transverse to the direction of propagation. All three normal modes have distinct dispersion relations, providing new opportunities for resonant conversion to BSM states.}
\label{fig:cartoon_pol_perp}
\vspace{-0.3cm}
\end{figure}

The photon self-energy in a \emph{magnetized} plasma is qualitatively very different than in plasmas without magnetic fields, primarily due to the local anisotropy imparted by a background vector field and the Landau quantization of the electrons~\cite{Ganguly:1999ts, Ganguly:2000vt, DOlivo:2002omk, Hattori:2012je, Hattori:2012ny, Visinelli:2018zif, Hattori:2022uzp, Wang:2021ebh, Wang:2021eud, Brahma:2024vxb}. In an isotropic plasma, the photon equations of motion (EoM) are diagonalized by decomposing the photon field into two degenerate transverse modes and one longitudinal mode. In contrast, the transverse and longitudinal EoM are highly coupled in magnetized plasmas, making it necessary to find the new independently propagating ``normal modes'' of the photon by diagonalizing the EoM. These propagating modes, which are generally all distinct from each other, are rotated away from the transverse and longitudinal directions to a degree that depends on both the magnitude and direction of the magnetic field. The magnetic field does not have to be strong in an absolute sense in order to affect the propagation of photons, i.e. it can be well below the Schwinger field value for QED, $B_\text{crit} = m_e^2/e \sim 4.4\times 10^{13}$~G. Rather, the relevant comparison is between the electron cyclotron frequency, $\omega_B = eB/m_e$, and the plasma frequency $\omega_p$. Environments with $\omega_B \gtrsim \omega_p$ will have propagating electromagnetic excitations that are substantially different from their transverse and longitudinal counterparts in an isotropic plasma~\cite{Brahma:2024vxb}, as depicted in Fig.~\ref{fig:cartoon_pol_perp}.

Previous treatments of stellar energy loss from axions and dark photons have typically not accounted for the strong magnetic fields in stars, in part because they can be challenging to model astrophysically. In this work, we therefore treat magnetic white dwarfs (MWDs) as a case study to highlight the impact of magnetic fields in determining the resonant emission rate of axions and dark photons. Despite their Fermi-degenerate electron densities, MWDs can host strong enough magnetic fields such that $\omega_B \gtrsim \omega_p$. Their surface magnetic fields have been measured to be as large as $\sim10^9$~G~\cite{Liebert_2003,Kepler_2013,Ferrario_2015,Ferrario_2020,Amorim_2023}, with models generally predicting internal field strengths that are orders of magnitude stronger. Furthermore, MWDs have relatively simple equations of state compared to other strongly magnetized compact objects such as neutron stars, circumventing the need to account for more exotic states of matter. We find that in large volumes of MWDs, resonant production of BSM particles can be enabled purely by the presence of a magnetic field. This underscores the need for more detailed studies on how magnetic fields affect stellar constraints on BSM physics.

The rest of this article is organized as follows. In Section~\ref{sec:theory}, we compute the emissivity that arises from the resonant mixing of SM photons with axions and dark photons in an ambient magnetic field. In Section~\ref{sec: profiles} we discuss the properties of MWD interiors and relevant observables. In Section~\ref{sec:MWDresults}, we compute the emissivity of axions and dark photons from MWDs, finding that it can potentially exceed the SM emissivity in unconstrained regions of parameter space. Concluding remarks follow in Section~\ref{sec:conclusions}.

\section{Particle production via resonant conversion in magnetized plasmas}
\label{sec:theory}
\subsection{Dispersion relations in a magnetized plasma}
The anisotropy generated by magnetic fields has a strong impact on the dispersive properties of photons in a plasma. Specifically, modifications to the plasma dispersion relations can be captured by the \textit{plasma mixing matrix} $\boldsymbol{\pi}$ whose elements are given by, 
\begin{equation}
 \pi^{IJ}\equiv -(\eps_\mu^{I})^* \Pi^{\mu\nu}_R \epsilon^{J}_\nu
\end{equation} where $\epsilon_\mu^I$ represents the polarization vector of the $I$th polarization and where $\Pi^{\mu\nu}_R $ is the retarded photon self-energy tensor. We take the self-energy from Ref.~\cite{Brahma:2024vxb}, where it was computed in a magnetized background using the real-time formalism of thermal field theory in the long-wavelength limit, $\omega \gg v k$, where $v$ is the typical velocity of a charged particle in the plasma. In this limit, over the course of one oscillation, the wavelength of plasma excitations is longer than the distance traveled by thermal particles. This allows us to neglect the effects of backreactions and electron diffusion. Even in this limit, we treat the plasma as locally homogeneous because the plasma frequency and magnetic field vary on scales much longer than the wavelength of plasma excitations.

A common way of decomposing the self-energy tensor is to project onto the transverse and longitudinal directions with respect to the photon propagation $K^\mu$, \begin{align}\label{TL_polvec}
&\epsilon_{T1}^{\mu}=(0,1,0,0),\quad\epsilon_{T2}^{\mu}=(0,0,\cos{\theta_B}, -\sin{\theta_B}), \nonumber\\
&\epsilon_{L}^{\mu}=\frac{1}{\sqrt{\omega^{2}-k^{2}}}(k,0,\omega \sin{\theta_B}, \omega \cos{\theta_B} ),
\end{align} 
where we have chosen a coordinate system such that the external magnetic field $\vec{B}_0$ points along the $z$-axis at an angle $\theta_B$ from the direction of photon propagation, i.e. $K^\mu = (\omega, 0, k \sin{\theta_B}, k \cos{\theta_B})$. In the basis of transverse and longitudinal modes, $\boldsymbol{\pi}$ is diagonal in an \textit{isotropic} plasma. The diagonal entries of $\boldsymbol{\pi}$ have real parts that are given by the transverse and longitudinal form factors $\pi_T$ and $\pi_L$ appearing in the dispersion relations, $\omega^2_{T,L} = k^2_{T,L} + \pi_{T,L}$. More generally, in the presence of a background magnetic field, in the basis of transverse and longitudinal modes, the \emph{Hermitian} part of the plasma mixing matrix takes the form
\begin{widetext}
\begin{equation}
\label{pimixB}
\mathcal{H} \left[\pi^{IJ}\right]=\mathcal{H}\left[-(\eps^I_\mu)^*\hat{\Pi}^{\mu \nu}_\text{vac}\eps^J_\nu\right]
 +\left[\begin{array}{ccc}
\pi_{\perp} & -i\pi_{\times}c_\theta & -i\frac{\sqrt{K^{2}}}{\omega}\pi_{\times}s_\theta\\
i\pi_{\times}c_\theta & \pi_{\perp}c^{2}_\theta+\pi_{\parallel}s^{2}_\theta & \frac{\sqrt{K^{2}}}{\omega}\left(\pi_{\parallel}-\pi_{\perp}\right)c_\theta s_\theta\\
i\frac{\sqrt{K^{2}}}{\omega}\pi_{\times}s_\theta & \frac{\sqrt{K^{2}}}{\omega}\left(\pi_{\parallel}-\pi_{\perp}\right)c_\theta s_\theta & \frac{K^{2}}{\omega^{2}}\left(\pi_{\parallel}c^{2}_\theta+\pi_{\perp}s^{2}_\theta\right)
\end{array}\right] 
\end{equation}
where $\mathcal{H}$ denotes the Hermitian part,\footnote{Note that in the literature, it is common to denote the Hermitian part of the self-energy tensor and the corresponding mixing matrix as $\real[\Pi^{\mu \nu}]$ and $\real[\pi^{IJ}]$. This is a moot point for isotropic plasmas, where the self-energy tensor is purely symmetric. However, this distinction is important in the context of this work where we deal with anisotropic plasmas where these tensors have anti-symmetric components.} $c_\theta \equiv \cos \theta_B$, and $s_\theta \equiv \sin \theta_B $. The first term corresponds to contributions from the magnetized vacuum (i.e. at zero electron density) due to non-linear QED effects that are relevant for $B\gtrsim B_\text{crit}$. Since we restrict ourselves to magnetic fields below the Schwinger value, we neglect this term in the remainder of this work. The second term, which is the focus of this paper, arises in a magnetized plasma. The form-factors $\pi_\parallel$, $\pi_\perp$, and $\pi_\times$ were computed under different environmental conditions using thermal field theory as detailed in Ref.~\cite{Brahma:2024vxb}. In MWDs where the electrons are degenerate, the form-factors can be approximated in the long-wavelength limit ($\vec{k}\to 0$) as 
\begin{subequations}\label{J_a}
\begin{align}
&\pi_{\perp}=\frac{\omega_p^2 }{1-\frac{(\omega_B/\mu_m)^2}{ \omega^2}},\\
&\pi_{\times}=\omega_p^2 \frac{\omega_B}{\omega}\frac{3\mu_m}{(\mu_m^2 -1)^{3/2}}\left[\sqrt{\frac{\mu_m^2 -1}{12 \mu_m^2}}\frac{4(\mu_m^2 -1) - 3\omega_m^2}{1-\frac{(\omega_B/\mu_m)^2}{ \omega^2}}-\frac{\omega_{m}^{2}}{4\sqrt{\left(\frac{\omega_B^2}{\omega^2}\right)^{2}-1}}\tanh^{-1}\left(\frac{\sqrt{\mu_m^2 -1}}{\sqrt{\left(\frac{\omega_B}{\omega}\right)^2 -1}}\right)\right],\\
&\pi_{\parallel}=\omega_p^2\frac{\mu_m}{(\mu_m^2 -1)^{3/2}}\left[\mu_m\sqrt{\mu_m^2 -1} +\frac{\omega_{m}^{2}}{2}\tanh^{-1}\left(\frac{\sqrt{\mu_m^2 -1}}{\mu_m}\right)-\left(1+\frac{\omega_{m}^{2}}{2}\right)\frac{\sqrt{4-\omega_{m}^{2}}}{\omega_{m}}\tan^{-1}\left(\frac{\omega_{m}\sqrt{\mu_m^2-1}}{\mu_m\sqrt{4-\omega_{m}^{2}}}\right)\right],
\end{align}
\end{subequations}
\end{widetext}
where $\omega_m \equiv \omega/m_e$ and where we note that in a degenerate plasma the plasma frequency is given by $\omega_p^2 = 4 \pi \alpha n_e/m_e\mu_m$ with the non-dimensionalized chemical potential, $\mu_m =\mu/m_e = \sqrt{1 + (3\pi^2 n_e)^{2/3}/m_e^2}$. In this work, we primarily consider the electron contribution to the response tensor. Specifically, we are interested in scenarios where $\omega_{B} \gtrsim \omega_{p}$, which is much more easily satisfied for electrons since $\omega_B/\omega_p \sim 1/\sqrt{m}$. 

In the low-frequency limit, ($\omega^2 \sim \omega_p^2 < \omega_B m_e$), the form factors can be well approximated (at the sub-percent level) as $\pi_{\parallel}\approx\omega_{p}^{2}$ and 
\begin{align}
\label{pi-deg}
\pi_{\perp}\approx\frac{\omega_{p}^{2}\omega^{2}}{\omega^{2}-(\omega_B/\mu_m)^2}, \quad 
\pi_{\times}\approx\frac{\omega_{p}^{2}\omega (\omega_B/\mu_m)}{\omega^{2}-(\omega_B/\mu_m)^2}.
\end{align}
In this limit, the form factors share a similar functional form as the one derived using classical kinetic theory in a cold, magnetized plasma~\cite{stix1962theory,melrose1986instabilities}, with the main difference being the form of the plasma and cyclotron resonance frequencies. For instance, note that the cyclotron resonance happens at the frequency $\omega_B/\mu_m$ in a degenerate plasma.

The non-diagonal form of $\boldsymbol{\pi}$ couples the equations of motion for the transverse and longitudinal modes in the plasma. To obtain the independently propagating normal modes, this mixing matrix has to be diagonalized by solving for the eigenvalues $\{\pi_I\}$ and eigenvectors $\{\epsilon_I^\mu\}$ with $I=1,2,3$ for the three normal modes. Note that the eigenvalues are guaranteed to be real since we are considering the Hermitian part of the plasma mixing matrix. Each normal mode then has its own distinct dispersion relation, $\omega^2_I = k^2_I + \pi_I$. As an example, in Fig.~\ref{fig:cartoon_pol_perp} we show two of the propagating normal modes in a degenerate plasma propagating in a direction perpendicular to the magnetic field. 

\subsection{Production rate}\label{sec: theoryB}
The resonant production of a BSM state occurs when the propagating plasma normal modes have a kinematic level crossing with that BSM state. In such a situation, the net BSM production rate $\Gamma^P$ in some part of phase space can be computed from detailed balance in terms of the absorption rate $\Gamma^A$, giving $\Gamma^P = e^{-\omega/T}\Gamma^A$. The resonant production rate of a particle $X$ per unit volume from the $I$th plasma normal mode then has the form 
\begin{equation}
 \frac{dn_{A^{I}\rightarrow X}}{dt} = \int \frac{d^3 p}{(2 \pi)^3} \ \Gamma^P_{A^{I}\rightarrow X} \ ,
\end{equation}
where $P^\mu = (\omega, \vec{p})$ is the four-momentum of the BSM state $X$ with $P^2 = m_X^2$. This rate determines the particle flux, which is relevant for direct detection experiments like axion helioscopes. In this paper, we instead focus on the energy-loss rate per unit volume (emissivity),
\begin{equation}
 Q_{A^{I}\rightarrow X} = \int \frac{d^3 p}{(2 \pi)^3} \ \omega \ \Gamma^P_{A^{I}\rightarrow X} \ .
\end{equation}
Integrating $Q$ over the whole volume of the star yields a ``dark luminosity'' which is the relevant quantity for anomalous energy loss signatures. For the explicit cases at hand \cite{Brahma:2024vxb}, 
\begin{widetext}
\begin{equation}\label{GammaI}
Q_{A^{I}\rightarrow X}
= \int \frac{d^3 p}{(2 \pi)^3} 
\frac{m_X^2}{\left(e^{\omega/T}-1\right)} 
\frac{\omega \Gamma_\gamma^I}{\left(m_X^{2}-\pi_I\right)^2 + \left(\omega \Gamma_\gamma^I\right)^2} 
\times\begin{cases}
g_{a\gamma}^{2}
\left|\epsilon_{I}\cdot \mathcal{B}_0\right|^2 & X =\text{axions}\vspace{0.2cm}\\ 
\kappa^2 \mAp^2 & X =\text{dark photons}
\end{cases}
\end{equation}
where $\mathcal{B}^\mu_0\equiv P^{-1}\tilde{F}_{0}^{\mu\nu}P_\nu$ is the magnetic field four-vector in the rest frame of the BSM particle. The production rate in \Eq{GammaI} has a Lorentzian resonance width $\Gamma^I_\gamma$, related to the in-medium damping associated with the eigenvalues of the anti-Hermitian part of $\boldsymbol{\pi}$.

For the purposes of this article, we mainly focus on the resonant production of $X$ from on-shell photons in the plasma which happens when there is a kinematic level crossing, $P^2 \approx K^2$ (i.e. $m_X^2 \approx \pi_I^2$). In astrophysical environments, the eigenvalues of the Hermitian part of the plasma mixing matrix are generally much larger than the eigenvalues of the anti-Hermitian part, $\Gamma^I_\g \ll \pi^I/\omega$ ~\cite{Redondo:2013lna, Hardy:2016kme}. We have explicitly checked that this weak-damping limit is applicable to the plasmas of interest in this work, even including the effects of strong magnetic fields. The production rate in Eq.~\eqref{GammaI} can then be expressed in terms of a delta function 
\begin{equation}\label{prodrate_I}
Q_{A^{I}\rightarrow X}
= \int \frac{d^3 p}{(2 \pi)^3}
\frac{m_X^2 \pi \delta \left(m_X^{2}-\pi_I\right)}{\left(e^{\omega/T}-1\right)}
\times\begin{cases}
g_{a\gamma}^{2}\left|\epsilon_{I}\cdot \mathcal{B}_0\right|^2 & X =\text{axions}\vspace{0.2cm}\\ 
\kappa^2 \mAp^2 & X =\text{dark photons.}
\end{cases}
\end{equation}
In general, $\pi_I$ is a function of both $\omega$ and $\vec{k}$. However, having imposed a resonant level crossing with a BSM state, $\pi_I$ is now formally a function of the energy and the orientation angle with respect to the magnetic field direction, $\pi_I \equiv \pi_I(\omega,k=\sqrt{\omega^2 - m_X^2},\theta_B)$. The delta function can thus be expressed as
\begin{equation}\label{eq:res_Z}
\delta\left(m_X^{2}-\pi_{I}\right)=\frac{Z_{I}}{2\omega}\delta\left(\omega-\resfreq\right), \qquad \qquad \text{with} \qquad \qquad
Z_{I}\equiv\left|\frac{d \pi_{I}}{d \omega^{2}}\right|^{-1}_{\omega = \resfreq},
\end{equation}
and where the resonant frequency $\resfreq = \resfreq(\theta_B)$ is obtained by solving for the resonance condition $\left.{\pi_I}\right|_{\omega = \resfreq} = m_X^2$. We emphasize that we ultimately enforce the resonance criterion with a delta function in the energy (or frequency) domain. With the exception of situations where $\pi_I$ is independent of $\omega$ (e.g. for the transverse modes in an isotropic plasma), resonance occurs at one particular value of $\omega$ given a fixed magnetic field and plasma frequency. This determines a resonant ``shell'' in frequency space that depends on the ambient conditions in the plasma. Performing the $\omega$ and $\phi$ integrals, we then obtain
\begin{equation}\label{eq:Q_I}
Q = \int d \cos{\theta_B} \sum_I \frac{Z_{I} m_X^2 \sqrt{\resfreq^{2}-m_{X}^{2}}}{8\pi\left(e^{\resfreq/T}-1\right)}\times\begin{cases}
g_{a\gamma}^{2}\left|\epsilon_{I}\cdot \mathcal{B}_0\right|^2 & \text{X = axions}\vspace{0.2cm}\\
\kappa^{2}m_{A^\prime}^{2} & \text{X = dark photons}
\end{cases}
\end{equation}
\end{widetext}
where we note that $Z_I$, $\resfreq$, and $\epsilon_I$ are all functions of $\theta_B$. A subtle point to emphasize here is that different values of $\theta_B$ will generate a unique set of eigenmodes with distinct eigenvalues and eigenvectors. As such, the sum over $I$ has to be done before performing the integral over $\theta_B$. The two cases of parallel ($\theta_B = 0$) and perpendicular ($\theta_B = \pi/2$) propagation are analytically tractable, and hence in the following Subsections we explore these limits before discussing arbitrary values of $\theta_B$. 

\subsection{Propagation parallel to the magnetic field}
\begin{figure}[t]
\includegraphics[width=0.47\textwidth]{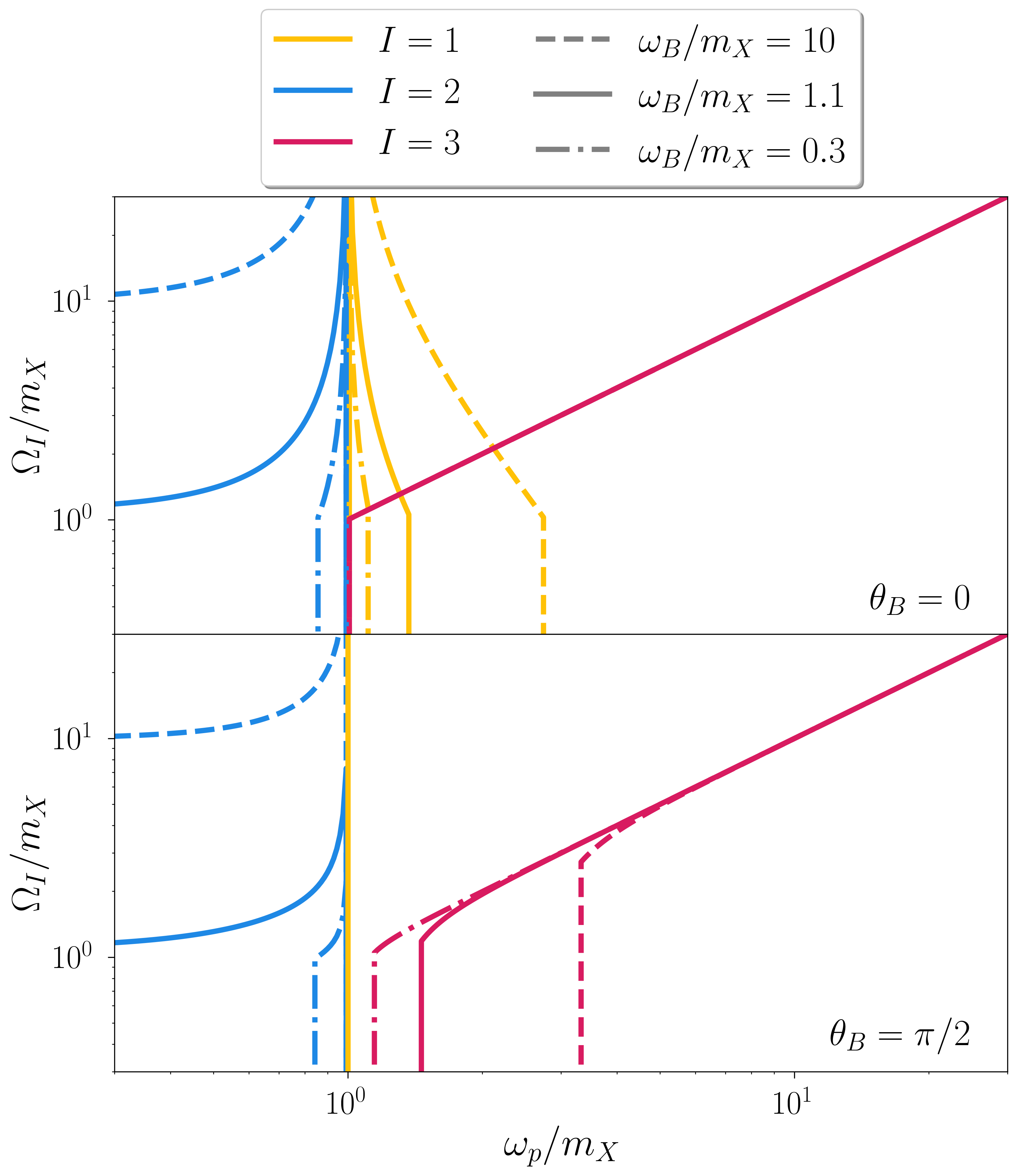}
 \vspace{-0.4cm}
 \caption{Resonant frequencies $\omega = \Omega_I$ from Eq.~\eqref{eq:res_Z} as a function of the plasma frequency (a proxy for stellar radius) assuming different cyclotron frequencies for modes propagating parallel (top) and perpendicular (bottom) to the magnetic field. In an isotropic plasma, resonant conversion occurs when $\omega_p = m_X$ (transverse) or when $\omega = \omega_p$ (longitudinal). The former condition is satisfied over a broad range of energies in a thin shell inside the star, while the latter is satisfied over a broad range of stellar radii at one radius-dependent energy. Magnetic fields broaden the resonance at $\omega_p = m_X$ to a wider range of plasma frequencies (radii), while preventing the resonance at $\omega = \omega_p$ for some plasma frequencies. In the regime $\omega_p \gtrsim \omega_B$, the resonance condition reduces to that of an isotropic plasma. The temperature enters as a relevant scale in the problem by determining the degree to which different resonance frequencies $\Omega$ are Boltzmann suppressed. Overall, strong magnetic fields enable resonant conversion in more regions of position-energy space as compared with an isotropic plasma of the same temperature and density.} 
 \vspace{-0.4cm}
 \label{fig: resonantFrequency}
\end{figure}

The dispersion relations for propagation parallel to the magnetic field are $\omega^2 = k^2 + \pi_I$ for the magnetized plasma normal modes $I=1,2,3$ with effective masses 
\begin{equation}\label{eigval_theta0}
\pi_{1,2}=\frac{\omega_{p}^{2}}{1\pm(\omega_{B}/\mu_m \omega )},\quad \pi_{3}=\frac{K^2}{\omega^2}\omega_{p}^{2}
\end{equation}
and polarization vectors
\begin{equation}\label{polvec_theta0}
\eps_{1,2}^\mu=\frac{1}{\sqrt{2}}(0,\pm i,1,0), \quad
\eps_{3}^\mu=\frac{1}{\sqrt{\omega^{2}-k^{2}}}(k,0,0, \omega ).
\end{equation}
The resonant frequencies for level crossing between the BSM state and the photon in Eq.~\eqref{eq:res_Z} are given by 
\begin{equation}
 \Omega_{1,2}(\theta_B=0) = \pm \frac{\omega_B/\mu_m}{\left(\frac{\omega_p}{m_X}\right)^2-1}, \quad \Omega_3(\theta_B=0) = \omega_p, 
 \label{eq:resFreqParallel}
\end{equation}
with corresponding wave renormalization $Z$-factors
\begin{equation}
Z_{1,2}=\frac{2(\omega_{B}\omega_{p})^{2}}{m_{X}^{4} \mu_m^2}\left|\frac{\omega_{p}^{2}}{m_{X}^{2}}-1\right|^{-3},\quad Z_3 = \frac{\omega_p^2}{m_X^2}.
\end{equation}

Evidently, the longitudinal mode of propagation remains unaffected by the magnetic field, which is expected due to the inherent symmetry when $\theta_B=0$ (i.e. the motion of charged species in the plasma is unaffected when moving parallel to the magnetic field). In {the top panel of} Fig.~\ref{fig: resonantFrequency}, we show the resonant frequencies $\resfreq(\theta_B=0)$ as a function of $\omega_p/m_X$ for several values of $\omega_B/m_X$. One key effect of strong magnetic fields is to broaden and shift the resonance near $\omega_p \sim m_X$ so that resonance can occur over a broader range of plasma frequencies. This can allow for resonant conversion in large volumes of a star where resonance would not occur in the absence of a magnetic field. Since $\Omega_{1,2} \geq m_X > 0 $ is required for resonance to occur, the resonance criteria for these two modes can be expressed as
\begin{align}
I=1:\quad &1<\frac{\omega_{p}}{m_{X}}\leq \sqrt{1+\frac{\omega_{B}}{m_{X}\mu_m}}\\
I=2:\quad & \sqrt{\max\left[0,1-\frac{\omega_{B}}{m_{X} \mu_m}\right]}\leq\frac{\omega_{p}}{m_{X}}<1.
\end{align}
We thus take $dQ_I(\theta_B=0)/d\cos{\theta_B}$ to be zero in any regions of a WD where these criteria are not met. For $\omega_B/(m_X\mu_m) \ll 1$, resonant production from transverse modes only occurs in a thin spherical shell where $\omega_p \sim m_X$, matching the isotropic case studied in earlier works.

Since the propagation of longitudinal modes is exactly the same as it would be in an isotropic plasma, the resonant production rate is~\cite{Redondo:2013lna, Caputo:2020quz}
\begin{widetext}
\begin{equation}
\frac{dQ_{3}(\theta_B=0)}{d\cos{\theta_B}}=\frac{\omega_{p}^{2}\sqrt{\omega_{p}^{2}-m_{X}^{2}}}{8\pi\left(e^{\omega_{p}/T}-1\right)}\times\begin{cases}
g_{a\gamma}^{2}\left|\vec{B}_{0}\right|^{2} & X=\text{axions}\vspace{0.2cm}\\
\kappa^{2}m_{A^\prime}^{2} & X=\text{dark photons}
\end{cases}
\end{equation}
where $\vec{B}_0$ is the external magnetic field in the rest frame of the medium. However, the production rate from the transverse modes $I=1,2$ is modified by the anisotropy caused by the magnetic field, 
\begin{equation}
\frac{dQ_{I}(\theta_{B}=0)}{d\cos{\theta_B}}
=\frac{\omega_{B}^{2}\omega_{p}^{2}\sqrt{\resfreq^{2}-m_{X}^{2}}}{4\pi \mu_m^2 \left(e^{\resfreq/T}-1\right)}
\left|\frac{\omega_{p}^{2}}{m_{X}^{2}}-1\right|^{-3}
\times\begin{cases}
0 & X= \mathrm{axions} \vspace{0.2cm}\\
\kappa^{2} & X= \mathrm{dark\, photons.} 
\end{cases}
\end{equation} 
\end{widetext}
Note that the axion production rate from transverse modes vanishes when $\theta_B=0$, since the transverse polarization vectors are by definition orthogonal to the background magnetic field. 

\subsection{Propagation perpendicular to the magnetic field}
A similar analysis can be carried out for $\theta_B = \pi/2$, which also allows for simple analytical diagonalization of the plasma mixing matrix. The effective masses in this case are found to be 
\begin{align}\label{eigval_theta0}
\pi_{1,2}&=\frac{\omega_{p}^{2} \mu_m^2}{\omega^2 \mu_m^2- \omega_B^2} \left(\omega^2 - \frac{k^2}{2} 
 \pm \frac{1}{2}\sqrt{k^4 + \frac{ 4 \omega_B^2 (\omega^2 - k^2)}{\mu_m^2}}
 ~\right) \nonumber 
 \\
\pi_{3}&=\omega_{p}^{2},
\end{align}
with the $I=1,2$ polarization vectors lying perpendicular to the magnetic field and the $I=3$ polarization vector lying parallel to the magnetic field. 
The corresponding resonant frequencies of Eq.~\eqref{eq:res_Z} 
are given by 
\begin{align}
\Omega_{1,2}(\theta_{B}= \frac{\pi}{2})&= \Theta\left[\mp (\omega_p - m_X)\right]\sqrt{\omega_{p}^{2}-\frac{(\omega_{B}/\mu_m)^{2}}{\left(\frac{\omega_{p}}{m_{X}}\right)^{2}-1}} \\
\Omega_{3}(\theta_{B}= \frac{\pi}{2}) &= \sqrt{k^2 + \omega_p^2}
\end{align}
where $\Theta$ is the Heaviside step function. 

In Fig.~\ref{fig: resonantFrequency}, we show these resonant frequencies $\Omega_I(\theta_B = \pi/2)$ to contrast against the $\Omega_I(\theta_B = 0)$ case. We see that the resonant frequencies differ significantly between these two special orientations, implying anisotropic production of dark photons and axions as discussed further below.

The resonance criteria for $\theta_B=\pi/2$ are 
\begin{align}\label{res_region_perp}
I=1:\quad & \frac{\omega_{p}}{m_{X}} \geq \sqrt{1+\frac{\omega_{B}}{m_{X}\mu_m}}\\
I=2:\quad & \sqrt{\max\left[0,1-\frac{\omega_{B}}{m_{X}\mu_m}\right]}\leq\frac{\omega_{p}}{m_{X}}<1,
\end{align}
with $dQ_I(\theta_B = \pi/2)/d\cos{\theta_B}=0$ 
in regions of a WD where these criteria are not satisfied. 
As in the $\theta_B=0$ case, in the limit $\omega_B/m_X \rightarrow 0$ the resonance criteria and emissivities reduce to the values obtained assuming an isotropic plasma. Similarly, the $I=3$ mode in this case is identical to the transverse mode in an isotropic plasma, with resonant conversion occurring in a thin spherical shell of the WD where $\omega_p \sim m_X$. For axion production, the conversion rate also depends on the projection of the B-field along the respective polarization vectors,
\begin{equation}\label{eq:ax_coupling_piby2}
 \left|\epsilon_{1,2}\cdot\mathcal{B}_{0}\right|^{2}=0, \quad \left|\epsilon_{3}\cdot\mathcal{B}_{0}\right|^{2}= \omega^2 B^2/(\omega^2 - k^2)
\end{equation}
whereby it is evident that the $1,2$ modes do not resonantly produce axions if $\theta_B =\pi/2$. 

\begin{figure}[t!]
\centering
\includegraphics[width = 0.48\textwidth]{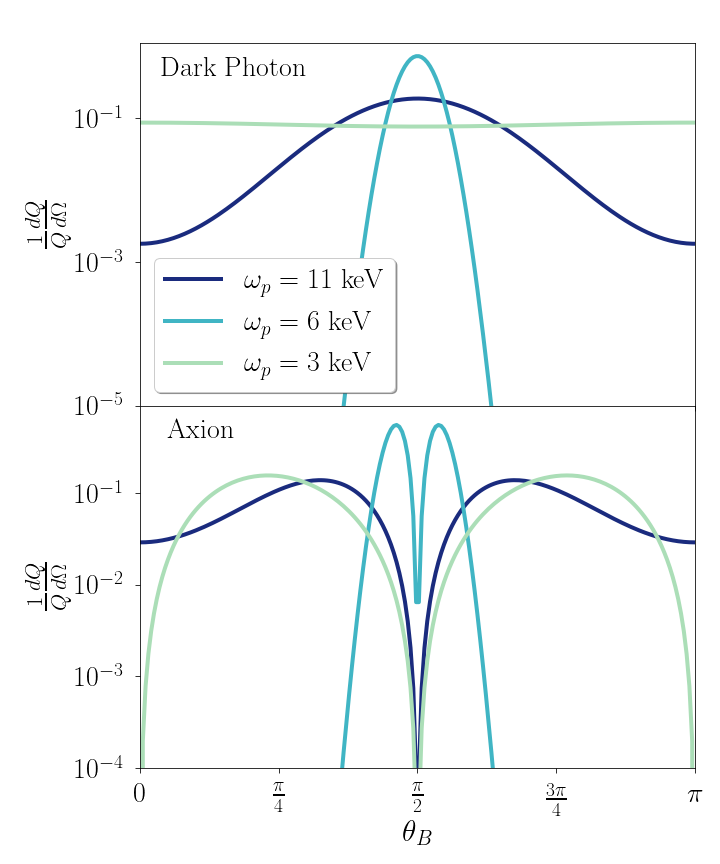}
\vspace{-0.8cm}
\caption{Angular profile of the differential emissivity of 7~keV dark photons and axions, summed over all eigenmodes at three representative plasma frequencies. Here we assume a temperature $T=0.2$ keV relevant to MWDs and a cyclotron frequency $\omega_B = 6$~keV, corresponding to a magnetic field strength of $\sim 5.17\times 10^{11}$ G inside the MWD. See Section~\ref{sec: profiles} for further details on the range of possible internal magnetic field strengths in MWDs.}
\label{fig: angprof}
\vspace{-0.6cm}
\end{figure}

\subsection{Propagation in any direction}
While the $\theta_B = 0$ and $\theta_B = \pi/2$ cases are helpful to build intuition, in realistic astrophysical environments it is necessary to integrate over arbitrary magnetic field orientations. For arbitrary values of $\theta_B$, the primary challenge lies in solving the characteristic equation of $\boldsymbol{\pi}$ (in the $T,L$ basis), which is a cubic polynomial. While this equation simplifies considerably in the two special cases discussed above, it becomes more intricate for general $\theta_B$. We employ a version of Cardano’s method to obtain the roots~\cite{fernandez2022rapidly}. Once the eigenvalues are determined, the corresponding eigenvectors -- which are particularly crucial for computing axion couplings due to the dot product with $\mathcal{B}_{0}$ -- are extracted using standard techniques. The new polarization vectors $\tilde{\boldsymbol{\epsilon}}=\{\eps_1^\mu, \eps_2^\mu, \eps_3^\mu\}$ are obtained from $\boldsymbol{\eps}=\{\eps_{T1}^\mu, \eps_{T2}^\mu, \eps_{L}^\mu\}$ by using $\tilde{\boldsymbol{\eps}} = \boldsymbol{\eps} \cdot \boldsymbol{\Lambda}$ where the diagonalizing matrix $\boldsymbol{\Lambda}$ is obtained by adjoining all the eigenvectors of $\boldsymbol{\pi}$.

By solving for the eigenvalues and eigenvectors at each point inside a WD and across all plasma propagation directions, we can compute the emissivity via \Eq{eq:Q_I}. Fig.~\ref{fig: angprof} shows the differential emissivity, $dQ/d\Omega$, as a function of $\theta_B$ at three representative plasma frequencies. The relative contribution of different $\theta_B$ to the total emissivity varies substantially, particularly for axion production due to the sensitivity to $(\eps_I \cdot \mathcal{B}_0)$. For instance, the differential axion emissivity drops sharply at $\theta_B=\pi/2$ because of the lack of transverse production. These results emphasize that in magnetized environments, the emissivity is highly directional, highlighting a significant qualitative difference compared to emission from an isotropic plasma.

Despite the fact that the differential emissivity $dQ/d\Omega$ is strongly dependent on $\theta_B$, we emphasize that \emph{the total emissivity $Q$ does not depend on the orientation of the magnetic field}. The integral over $\theta_B$ and the sum over $I$ in Eq.~\eqref{eq:Q_I} account for all possible sources of energy loss from every configuration. To see this, note that we can always choose local coordinates where the magnetic field points along the $z$-axis to perform the sum and integral before moving to another region of the star and constructing similar local coordinates there. Thus, the only thing that affects the integrated dark luminosity is the magnitude of the magnetic field as a function of position, even if the direction of the magnetic field varies within the star.

\section{Magnetic White Dwarfs}
\label{sec: profiles}
\begin{figure}[t!]
 \centering
\includegraphics[width = 0.38\textwidth]{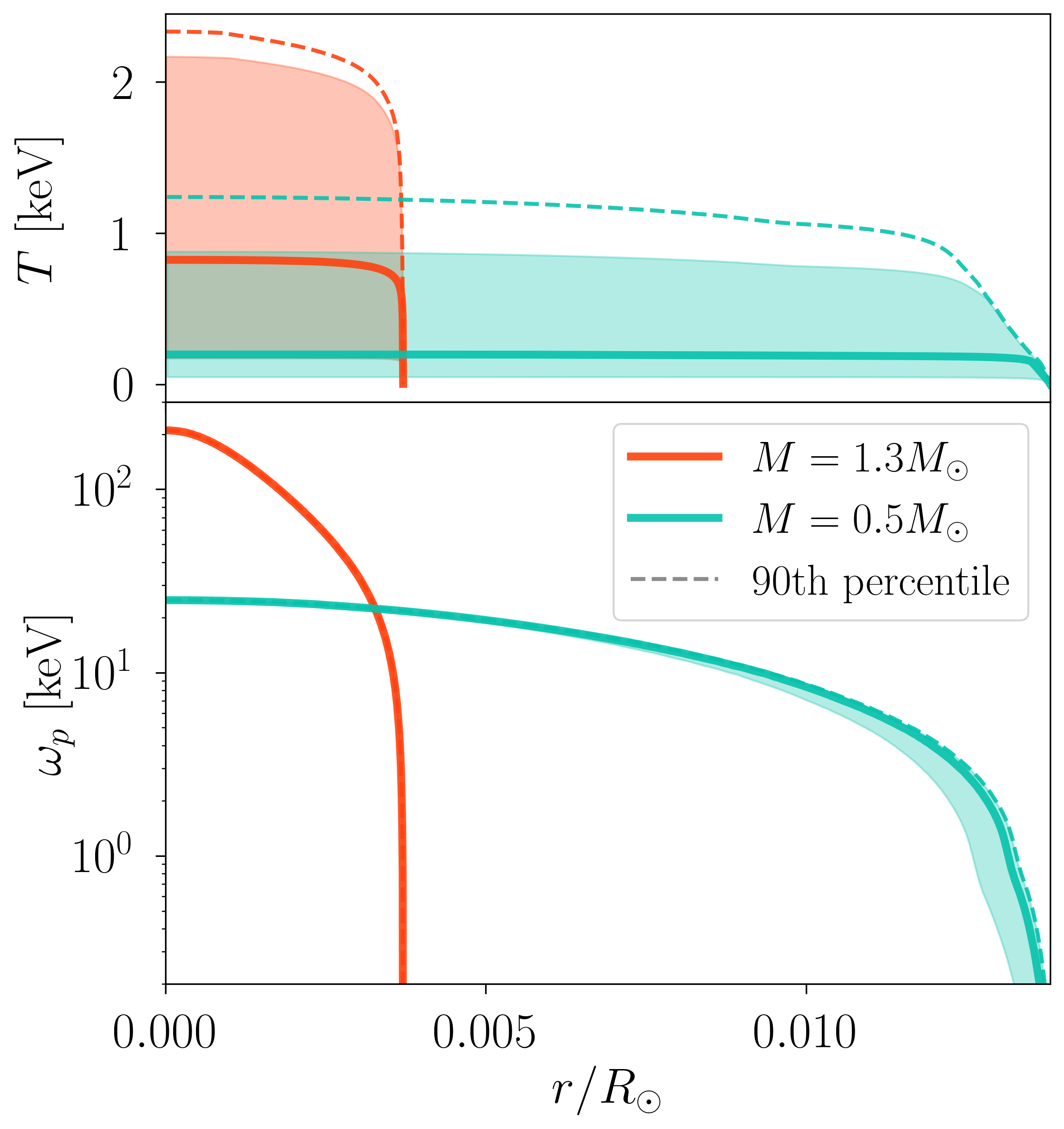}
\vspace{-0.3cm}
\caption{Temperature and plasma frequency profiles for realizations of MWDs with different masses simulated using \texttt{MESA}~\cite{Castro-Tapia_2024, 2025arXiv250616529C}. The median profiles are shown as solid lines within the 68\% containment bands.}
\vspace{-0.4cm}
\label{WDprofiles}
\end{figure}

MWDs are particularly compelling astrophysical targets to search for new physics. One appealing feature of WDs is the absence of nuclear reactions to generate energy inside the star. This implies that the primary SM cooling mechanisms are plasmon production of neutrinos from inside the WD and photon emission from the surface~\cite{raffelt1996stars}. Adding novel BSM cooling channels can therefore have a relatively clear impact on the observed properties of WDs. For instance, their luminosity function appears to favor additional energy loss mechanisms beyond these standard predictions~\cite{Isern:2008fs,Isern:2008nt}, although these findings could be the result of survey selection effects~\cite{hansen2015constraining}. Some WDs have pulsating brightness variations, and the rate at which this period decreases depends on the cooling rate of the star. The pulsation period change rate is thus another observable from which we can constrain extra BSM cooling mechanisms~\cite{corsico2012rate}. The initial-final mass relation, which relates the final mass of a WD to the initial stellar mass ($\lesssim 8 M_\odot$), is also affected by extra BSM energy-loss channels throughout the lifetime of the star~\cite{Dolan:2021rya}. All of these observable WD properties highlight the importance of the \textit{dark luminosity}, defined as the total energy loss rate through BSM particle production channels. 

Aside from the BSM constraints they provide, WDs are appealing targets because of their relative simplicity. The nearly isothermal temperature profiles and density profiles of WDs are relatively well known and can be calculated for different WD masses and ages. To capture the variation in density and temperature between different WDs, we consider an ensemble of WDs at two representative masses simulated using the Modules for Experiments in Stellar Astrophysics (\texttt{MESA}) framework~\cite{Paxton2011, Paxton2013, Paxton2015, Paxton2018, Paxton2019, Jermyn2023}. In particular, we adopt the $0.5\,M_\odot$ model from Ref.~\cite{Castro-Tapia_2024} and the $1.3\,M_\odot$ model from Ref.~\cite{2025arXiv250616529C}, which make use of a grid of cooling models from Ref.~\cite{Bauer_2023}. The temperature and plasma frequency as a function of radius for these models are shown in Fig.~\ref{WDprofiles}. The plasma frequency profile is relatively robust to different realizations at the same WD mass, but varies substantially between WDs with different masses. The heaviest WDs we consider, which have smaller radii and correspondingly larger densities, have plasma frequencies that are about an order of magnitude larger than those of the lightest WDs. This indicates that the magnetic fields in heavy WDs have to be correspondingly larger to have the same impact as they would in lighter WDs. Meanwhile, the temperature profiles exhibit a somewhat broader variation between WDs of a given mass due to the impacts of different cooling models and ages, but do not vary quite as much between WDs of different masses. Since the temperatures in a given realization are relatively uniform throughout the WD, we take the median temperature profile as a suitable representative for WDs of a given mass. Because the rates are exponentially sensitive to the temperature (due to the phase space factor in the emissivity integral), we also consider WDs with temperatures in the 90th percentile, corresponding to younger WDs whose SM cooling is dominated by plasmon decay to neutrinos.

Several hundred MWDs have been observed, with measured surface magnetic fields ranging from $10^3$–$10^9$~G~\cite{Liebert_2003,Kepler_2013,Ferrario_2015,Ferrario_2020,Amorim_2023}. Several models have been put forward to describe the magnetic field configuration of the WD interior~\cite{Ferrario_2020, Euchner_2002, Braithwaite_2004, Braithwaite_2009,Fujisawa_2012, Bera_2014, Peterson_2021, Drewes:2021fjx}. However, from an observational point of view, the internal field structure remains poorly constrained and may differ substantially from the measured values at the surface. Various models predict that the core magnetic field strength can be more than $100\times$ stronger than the surface magnetic field~\cite{Fujisawa_2012,Castro-Tapia_2024}, while other models establish theoretical upper bounds on the internal field strength extending up to $10^{15}- 10^{16}$~G~\cite{Franzon:2015gda,PhysRevD.90.043002,PhysRevD.86.042001}. Given this uncertainty, we adopt a simplified model assuming a uniform magnetic field inside the MWD, which we vary in order to quantify the effects of different magnetic field strengths. This simplified model is purely for illustrative purposes, and the use of more realistic magnetic field profiles will be the topic of future work. As argued in the previous Section, our calculation of the dark luminosity depends neither on the orientation of the magnetic field, nor on the magnetic field pointing coherently in some direction over any appreciable volume of the star. 

\section{Emission from Magnetic White Dwarfs} 
\label{sec:MWDresults}
To facilitate the comparison of energy loss rates between different models, we adopt fiducial values of the dark photon kinetic mixing $\kappa_{10} \equiv \kappa / 10^{-10}$ and axion-photon coupling $g_{10} \equiv g_{a\g }/10^{-10}\,$GeV$^{-1}$. We also present BSM luminosities in terms of two benchmark luminosities that meaningfully contribute to WD energy loss: (1) photon cooling from the surface of the WD (i.e. Mestel's cooling law), which is dominant for the majority of WDs, and (2) neutrino emission via the plasmon decay process, which is the dominant cooling channel for young, hot WDs. For the purposes of comparison with resonant conversion to BSM states, which leads to emission throughout a large volume of the WD, we adopt an effective photon energy loss rate per unit mass of $3.3\times 10^{-3}\, \text{ergs} ~\text{g}^{-1} ~ \text{s}^{-1} \times (T/10^7~\text{K})^{3.5}$ throughout the entire WD~\cite{raffelt1996stars}. To obtain a volumetric emissivity (i.e. $Q_\gamma$) analogous to Eq.~\eqref{eq:Q_I}, we multiply the effective photon energy loss rate per unit mass by the MWD densities as appropriate for the 0.5~$M_\odot$ and 1.3~$M_\odot$ MWD density profiles considered in this work. Meanwhile, we adopt the energy loss rate per unit volume, $Q_\nu$, from Ref.~\cite{Haft:1993jt} for the neutrino plasmon process. Since this process involves in-medium photons in the initial state, the phase space factors in the emissivity are the same as those appearing for resonant conversion, making this an especially relevant comparison. 
 
\begin{figure}[t!]
\centering
\includegraphics[width = 0.48\textwidth]{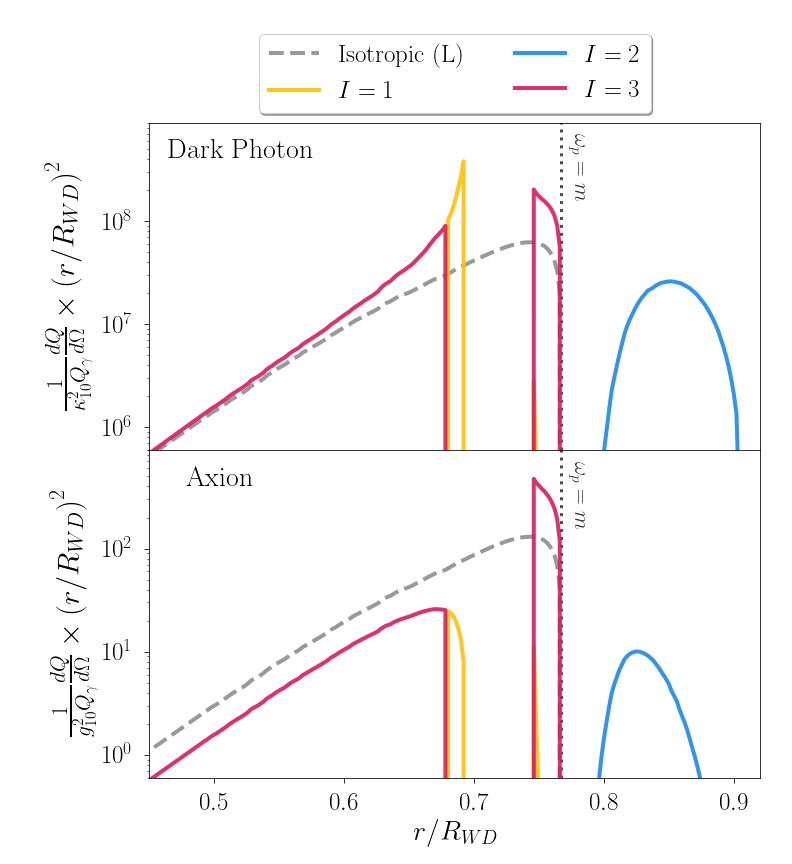}
\vspace{-0.3cm}
\caption{Differential emissivities for 7~keV dark photons and axions at $\theta_B = \pi/4$ assuming the $0.5~M_\odot$ MWD plasma frequency profile and assuming that the temperature is in the 90th percentile of MWDs at that mass. We assume a cyclotron frequency of $\omega_B = 6$ keV ($5.17 \times 10^{11}$ G) inside the MWD. The dashed line denotes the emissivity from the resonant conversion of longitudinal modes, computed assuming dispersion relations for an isotropic plasma. Similarly, the vertical dotted line denotes the radius $r_0^*$ where resonant conversion of transverse modes is possible.} 
\vspace{-0.3cm}
\label{fig:emissivity}
\end{figure}

In Fig.~\ref{fig:emissivity}, we show an example of the differential emissivity profile, $dQ/d\Omega$, at $\theta_B = \pi/4$. For resonant conversion to axions and dark photons in an isotropic plasma, the longitudinal mode dominates the emissivity over a broad range of stellar radii, while the transverse modes appear as a sharply peaked resonance at one specific stellar radius $r = r_0^*$ where the plasma frequency matches the mass of the BSM state. In contrast, in the presence of a magnetic field, resonant conversion to axions and dark photons can occur over broad ranges of radii for all three normal modes of propagation. Consequently, magnetic fields allow multiple plasma modes to simultaneously contribute to the resonant production of BSM states across different radii where such production was previously thought to be kinematically forbidden. A common trend among the three normal modes is that resonant conversion tends to have the largest local emissivity at the largest kinematically accessible radius. This is primarily because of the relatively steep density profile and flat temperature profile, which causes the thermal number density of photons to be most highly suppressed at small radii where $\omega_p > \omega_B > T$.

\begin{figure}[t!]
 \centering
 \includegraphics[width = 0.48\textwidth]{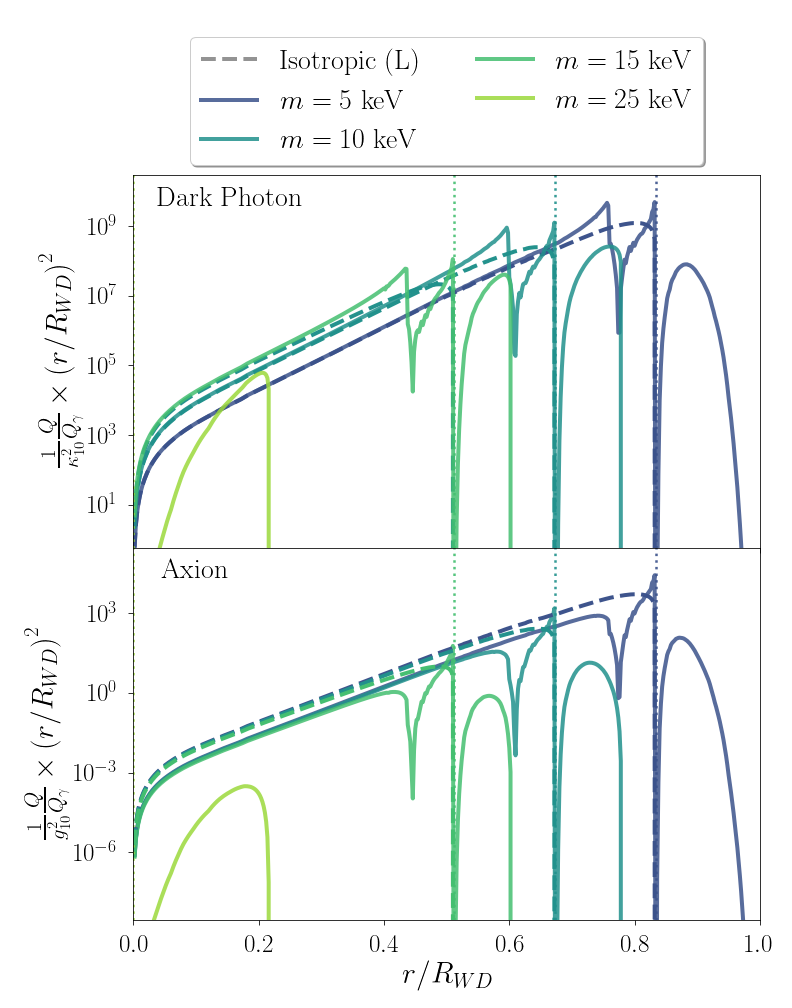}
 \vspace{-0.3cm}
 \caption{Total emissivity for several dark photon and axion masses assuming the $0.5~M_\odot$ MWD plasma frequency profile and assuming that the temperature is in the 90th percentile of MWDs at that mass. We assume a cyclotron frequency of $\omega_B = 6$~keV ($5.17 \times 10^{11}$~G) inside the MWD. The dashed lines denote the emissivity from the resonant conversion of longitudinal modes, computed assuming dispersion relations for an isotropic plasma. Similarly, the vertical dotted lines denote the radii where resonant conversion of transverse modes is possible. Note that even when using the dispersion relations of an isotropic plasma, we still assume $\omega_B = 6$~keV since a magnetic field is necessary for axion production.}
 \vspace{-0.3cm}
 \label{fig: mass emissivity}
\end{figure}

In Fig.~\ref{fig: mass emissivity}, we show the total emissivity from resonant conversion for several dark photon and axion masses summed over all three propagating normal modes. Isotropic plasmas have no emissivity from resonant conversion in regions where $r>r_0^*$ because it is not kinematically possible. In the magnetized case, because of the altered plasma dispersion relations, the outer layers of the star ($r>r_0^*$) can support resonant conversion into axions and dark photons, even when $\omega_p < m_X$. 
This opens the possibility of resonant BSM particle production for values of $m_X$ that are larger than the maximum plasma frequency, which would otherwise be impossible in the absence of a magnetic field. 

\begin{figure}[t!]
\centering
\includegraphics[width = 0.48\textwidth]{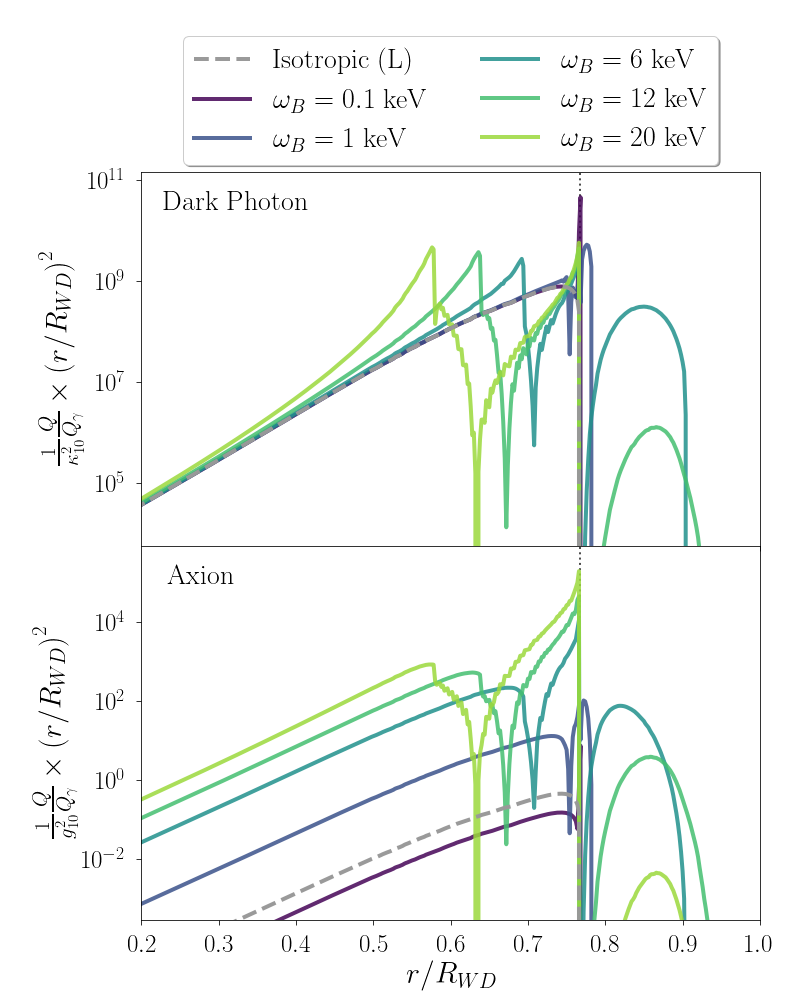}
\vspace{-0.5cm}
\caption{Same as Fig.~\ref{fig: mass emissivity}, but fixing the BSM particle mass at 7~keV and varying the strength of the magnetic field.}
\label{fig:B emissivity}
\end{figure}

In Fig.~\ref{fig:B emissivity}, we show how varying the internal magnetic field impacts the local emissivity. In the limit where $\omega_B < \omega_p$ everywhere in the star, the emissivity converges to the isotropic limit, as expected. As the magnetic field strength increases, the differences become more pronounced, for instance with emission in regions where $r > r_0^*$. For large enough values of the cyclotron frequency, the resonant frequencies in the outermost regions of the star become highly Boltzmann-suppressed, quenching resonant conversion rates.

\begin{figure}[t!]
\centering
\includegraphics[width = 0.45\textwidth]{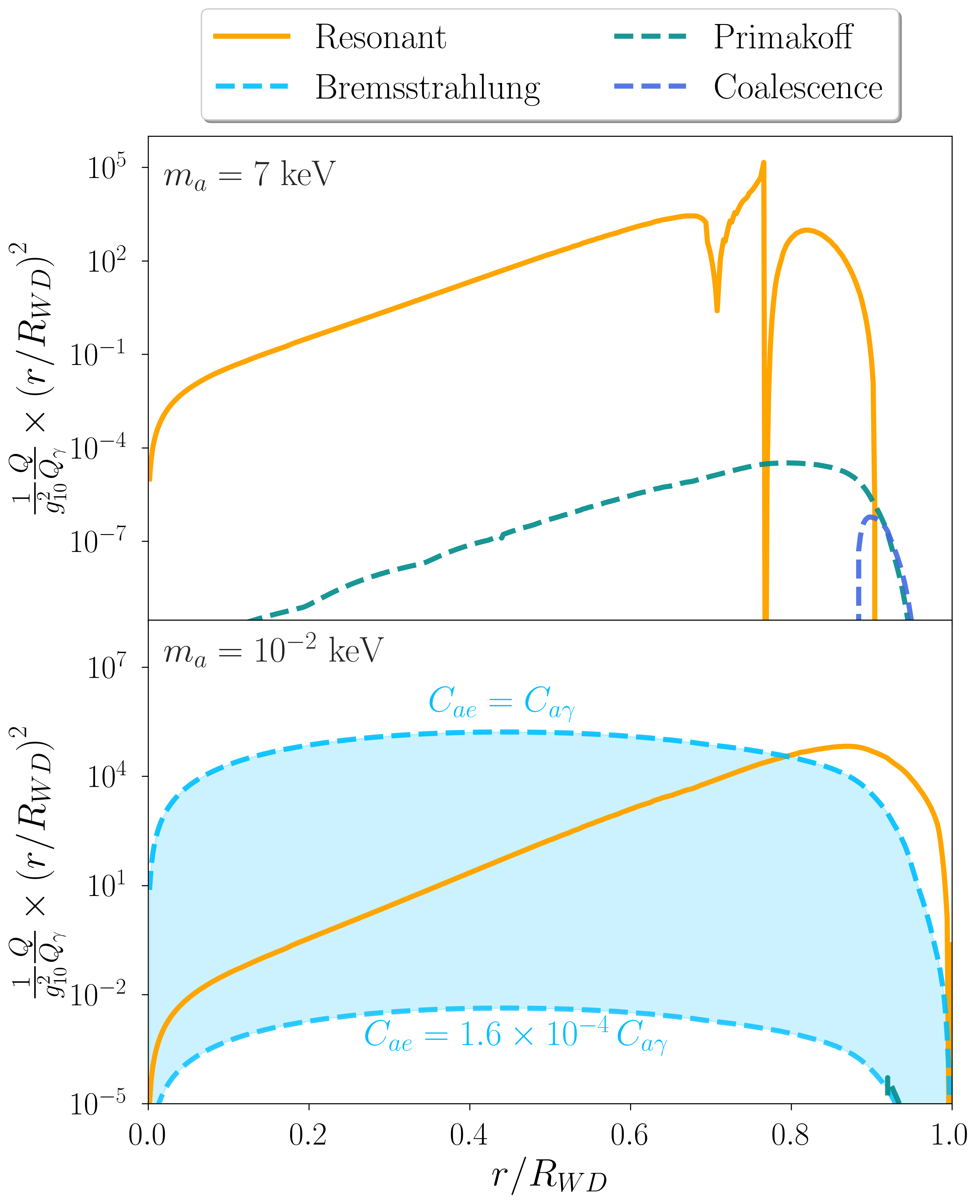}
\vspace{-0.3cm}
\caption{Axion emissivity via resonant conversion for $m_a\gtrsim T$ (top) and $m_a \lesssim T$ (bottom) compared to emissivity from other axion production channels. For heavier axions, the relevant channels are Primakoff off of nuclei and $2\to1$ coalescence. For lighter axions, bremsstrahlung is the relevant channel. In computing these rates, we assume the same MWD properties as in Fig.~\ref{fig: mass emissivity} and further assume that the MWD is comprised of carbon.}
\label{fig:emissivity_comparison}
\end{figure}

\begin{figure*}[t!]
\centering
\includegraphics[width =\textwidth]{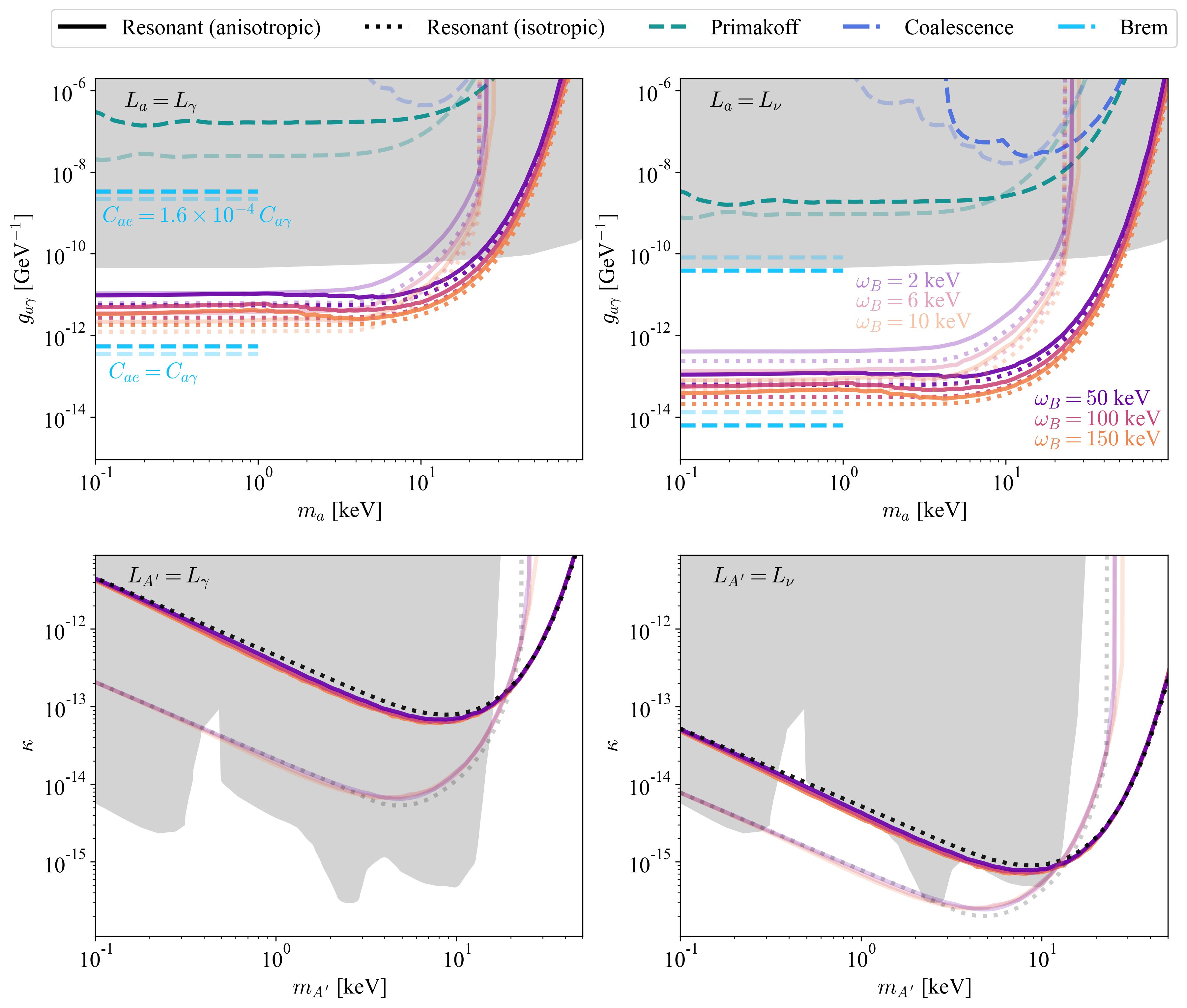}
\vspace{-0.3cm}
\caption{Contours of constant dark luminosity where $L_X = L_\gamma$ (left) and $L_X = L_\nu$ (right) for axions (top) and dark photons (bottom). Note that these contours do not constitute a constraint. We assume plasma frequency profiles corresponding to 0.5~$M_\odot$ (transparent lines) and 1.3~$M_\odot$ (opaque lines) MWDs and assume temperatures corresponding to the 90th percentile values for each WD mass. Solid lines of different colors correspond to resonant conversion rates determined using different cyclotron frequencies. We also show the luminosity contours arising from resonant conversion computed assuming the dispersion relations for an isotropic plasma (dotted). For axion emission, the Primakoff and coalescence luminosities are subdominant to resonant conversion. Bremsstrahlung emission at low masses is also subdominant to resonant conversion unless $C_{ae} \sim C_{a\gamma}$. Existing limits from stellar energy loss~\cite{Dolan:2022kul, Dolan:2021rya, Dolan:2023cjs,Li:2023vpv} are shown as grey shaded regions.}
\label{fig:lumplot}
\end{figure*}

For axion production, emission via resonant conversion has been relatively overlooked compared to other production channels. However, as discussed in Ref.~\cite{Caputo:2020quz} and shown in Fig.~\ref{fig:emissivity_comparison}, the production rate from resonant conversion can be comparable to or even much larger than other production channels. For keV-scale axion masses, the typical production mechanisms considered in previous studies are Primakoff production (via photons scattering on nuclei) and $2\to 1$ coalescence~\cite{DiLella:2000dn,Carenza:2020zil, Dolan:2021rya, Dolan:2022kul}. Comparing these rates (calculated using the expressions from Refs.~\cite{DiLella:2000dn, Raffelt:1985nk}) to resonant conversion in Fig.~\ref{fig:emissivity_comparison}, we find that in the presence of strong magnetic fields, resonant conversion is by far the dominant mode of production. This is primarily due to the $\sim B^2$ scaling of the resonant conversion rate, in contrast to the other rates which are independent of the magnetic field. 

Meanwhile, sub-keV axions with $m_a \lesssim T$ are typically assumed to be produced via electron-nucleus bremsstrahlung, a process which is Boltzmann suppressed at high axion masses~\cite{Bottaro:2023gep,Raffelt:1985nj, Nakagawa:1987pga, Nakagawa:1988rhp}. There is some model dependence in relating processes purely involving factors of $g_{a\g}$ with ones involving the coupling to electrons, described by the Lagrangian 
\begin{equation}
 \mathcal{L}\supset \frac{g_{ae}}{2 m_e}(\partial_\mu a)\bar{e}\g^\mu\g^5 e.
\end{equation}
A useful parametrization for comparing the axion-electron and axion-photon couplings is~\cite{ParticleDataGroup:2024cfk}
\begin{equation}
g_{ae} = \frac{C_{ae} m_e}{f_a}, \qquad 
g_{a\g} = \frac{C_{a\g} \alpha}{2\pi f_a},
\end{equation}
where $C_{ae}$ and $C_{a\g}$ are dimensionless coefficients and where $f_a$ is the axion decay constant. One representative scenario is the DFSZ-type axion, which has a tree-level coupling to electrons such that $C_{ae} = C_{a\gamma}$~\cite{Dine:1981rt, Dine:1982ah, Zhitnitsky:1980tq}. Meanwhile, for KSVZ-type~\cite{Kim:1979if, Shifman:1979if} axions (also known as $W$-phobic axions), the coupling to electrons does not appear fundamentally but rather as an effective coupling through loop-suppressed diagrams, yielding $C_{ae} = 1.6 \times 10^{-4} \, C_{a\gamma}$~\cite{Srednicki:1985xd, Dessert:2021bkv}. We emphasize that, while we refer to QCD axion scenarios to motivate representative values of $C_{ae}/C_{a\gamma}$, the axions considered in this work belong to the broader class of axion-like particles and are not limited to the QCD axion paradigm. Despite the fact that resonant conversion can only occur near the surface of the WD, we find that it can be comparable to or dominate over bremsstrahlung production of sub-keV axions for models with 
$C_{ae}/C_{a\g}\in[1.6\times 10^{-4}, 1]$, as shown in Fig.~\ref{fig:emissivity_comparison}. For completeness, we also show the rate for electro-Primakoff (EP) production, where an axion is emitted from the photon exchanged between an electron and a nucleus undergoing Coulomb scattering. The ratio of the emissivities between EP and bremsstrahlung is given by~\cite{Raffelt:1985nk,Raffelt:1985nj,Dessert:2021bkv}
\begin{equation}
 \frac{Q_\text{EP}}{Q_\text{brem}} \sim \frac{\alpha^2}{4 \pi^2}\left(\frac{C_{a\g}}{C_{ae}}\right)^2 \left(\frac{T}{m_e}\right)^2.
\end{equation}
For WD temperatures ($T \lesssim \mathrm{keV}$), the EP emissivity is always highly sub-dominant compared to bremsstrahlung. 

The cooling of MWDs can be significantly enhanced by the emission of BSM particles, which carry energy away as they freely escape the stellar interior. This energy loss is quantified by the {dark luminosity}, $L_X = \int dV\, Q_X$, which is shown in Fig.~\ref{fig:lumplot} for both dark photons and axions. Specifically, we show contours in parameter space where $L_X = L_\gamma$ and $L_X = L_\nu$, the fiducial photon luminosity via Mestel's law~\cite{raffelt1996stars} and the neutrino luminosity via the plasmon process~\cite{Haft:1993jt}. Note that \emph{these contours do not constitute bounds on the parameter space}, but rather indicate interesting regions of parameter space where the BSM and relevant SM luminosities are comparable. 

When including the effects of magnetic fields on the plasma dispersion relations, there are competing effects that contribute to the overall luminosity: resonant conversion is possible in distinct regions of the star, and the local emissivity can either be suppressed or enhanced depending on the exact configuration and local environment. Therefore, the question of whether the emission is enhanced or suppressed by the magnetic field is likely sensitive to the assumed magnetic field profile. When adopting a uniform magnetic field in this work, we find that the emission can be enhanced or suppressed by a factor of a few depending on the parameter space, but the effect could in principle be much larger for a spatially varying magnetic field. Regardless of whether we use the dispersion relations for an isotropic or magnetized plasma, we find that resonant conversion to axions and dark photons can potentially be a very efficient source of dark luminosity in MWDs. This strongly motivates a more detailed study involving self-consistent simulations of MWDs with realistic magnetic field profiles, which we leave to future work.

\section{Discussion and Conclusions}\label{sec:conclusions}
Magnetic fields are ubiquitous in astrophysical plasmas where tests of BSM physics often rely on resonant level crossing between photons and BSM states. However, there has been relatively little work exploring the effect of magnetic fields on the photon dispersion relations that are responsible for these level crossings. Therefore, in this proof-of-concept work, we have used MWDs as a case study to demonstrate how the resonant emission of axions and dark photons is strongly affected by the altered dispersion relations inside a magnetized plasma. We have focused purely on the kinematic aspects of this effect, which alter dark photon production despite the fact that the coupling of dark photons to photons does not require the presence of an ambient magnetic field (in contrast to axions). The same kinematic arguments would also affect the emission of other light bosons like $B-L$~vectors and scalars coupling to leptons and nuclei. We leave the exploration of such models to future work.

We find that the local BSM emissivity inside MWDs is qualitatively completely different compared to what one would obtain using the dispersion relations of an unmagnetized, isotropic plasma. Different normal modes of propagation can resonantly convert to dark photons and axions in distinct regions of the stellar plasma, and the emissivities in those regions can be strongly enhanced or suppressed (by many orders of magnitude) relative to the isotropic case. Therefore, if they exist, BSM particles could in principle probe the densities and magnetic fields in different layers of the stellar interior. This raises the tantalizing possibility of \textit{axion} or \textit{dark photon astronomy}, where the internal structure of dense stars could be inferred from their BSM emission profiles. We also found that resonant conversion, which has long been known to be the dominant channel for dark photon production in stars, can also dominate over bremsstrahlung and Primakoff emission for axions, in agreement with Ref.~\cite{Caputo:2020quz}. 

The present work, which serves to emphasize the impact of internal magnetic fields, can be extended to treat BSM particle emission assuming more realistic MWD magnetic field profiles in a self-consistent way. Notably, we assumed a static \texttt{MESA}-inspired model without accounting for any backreactions from BSM energy loss. These backreactions have been shown to be substantial in other stellar constraints on BSM physics~\cite{Dolan:2022kul, Dolan:2023cjs, Fung:2023euv}, and could also be important for MWDs. For simplicity, we also assumed a uniform internal magnetic field; a more realistic configuration could potentially increase the impact of anisotropy on the emissivities and dark luminosity. With the simplifying assumptions we made in this work, we found that the emissivity from BSM states in a magnetized plasma could be comparable to the relevant SM energy-loss channels in unconstrained regions of parameter space. This indicates that with self-consistent simulations, it may be possible to set very strong new constraints on axions and dark photons. We leave an exploration of this to future work. 

While we focused on the dark emissivity and luminosity, other properties of MWDs can be used to constrain BSM physics. For instance, axions produced in the core of MWDs can convert to photons before they escape the magnetosphere, giving rise to X-ray emission from the star. The null observation of this signal yields strong constraints on axions~\cite{Dessert:2019sgw, Dessert:2021bkv,Dessert:2022yqq}. It would be interesting to explore whether including the resonant conversion channel in these analyses could provide a strong constraint on $g_{a \g}$ alone (since the dominant production channel is typically assumed to be from bremsstrahlung emission, these searches have thus far been sensitive to the product $g_{a e} g_{a \g}$).

Our work raises several other possible extensions of these types of searches. For example, BSM particles emitted from a magnetized plasma are produced from one of the three normal modes, whose properties (dispersion and polarization) are highly dependent on the ambient density, temperature, and magnetic field. If the BSM particles reconvert in a different ambient environment (whether in the stellar magnetosphere or in a terrestrial detector~\cite{Berlin:2024pzi}), the resulting photon will be in some nontrivial superposition of the local normal modes, possibly leading to distinct polarimetric or absorption signatures~\cite{Berlin:2023ppd, Alonso-Alvarez:2024ypq}. Another interesting distinction is that, in contrast to the integrated dark luminosity, searches for BSM particles converting back to photons can be highly directionally dependent. Since we find a very strong, model-dependent directional dependence in the emissivity (i.e. depending on the relative orientation of the magnetic field and outgoing particle), there could potentially be a large impact on the expected signal strength. This further motivates future work incorporating more realistic models of internal magnetic fields, including the directional profile rather than just its magnitude. Finally, we have focused primarily on MWDs as a model system due to their relative simplicity, but any environment where $\omega_B \gtrsim \omega_p$ will be subject to the same considerations that are discussed here. Neutron stars are a noteworthy example. We leave exploration of all of these extensions to future work.

\section*{Acknowledgements}
It is a pleasure to thank Matias Castro-Tapia and Andrew Cumming for useful discussions and for providing us with their MWD density and temperature profiles computed using \texttt{MESA}. We also thank Saniya Heeba and Yitian Sun for useful discussions and Asher Berlin, Jamie McDonald, and Sam Witte for useful comments on the manuscript. NB was supported in part by a Doctoral Research Scholarship from the Fonds de Recherche du Qu\'ebec -- Nature et Technologies and by the Canada First Research Excellence Fund through the Arthur B. McDonald Canadian Astroparticle Physics Research Institute. HS acknowledges support from the Natural Sciences and Engineering Research Council of Canada as a Vanier Scholar. NB, EI, HS, and KS acknowledge support from a Natural Sciences and Engineering Research Council of Canada Subatomic Physics Discovery Grant, from the Canada Research Chairs program, and from the CIFAR Global Scholars program. 
\bibliography{main}

\begin{thebibliography}{109}%
\makeatletter
\providecommand \@ifxundefined [1]{%
 \@ifx{#1\undefined}
}%
\providecommand \@ifnum [1]{%
 \ifnum #1\expandafter \@firstoftwo
 \else \expandafter \@secondoftwo
 \fi
}%
\providecommand \@ifx [1]{%
 \ifx #1\expandafter \@firstoftwo
 \else \expandafter \@secondoftwo
 \fi
}%
\providecommand \natexlab [1]{#1}%
\providecommand \enquote  [1]{``#1''}%
\providecommand \bibnamefont  [1]{#1}%
\providecommand \bibfnamefont [1]{#1}%
\providecommand \citenamefont [1]{#1}%
\providecommand \href@noop [0]{\@secondoftwo}%
\providecommand \href [0]{\begingroup \@sanitize@url \@href}%
\providecommand \@href[1]{\@@startlink{#1}\@@href}%
\providecommand \@@href[1]{\endgroup#1\@@endlink}%
\providecommand \@sanitize@url [0]{\catcode `\\12\catcode `\$12\catcode `\&12\catcode `\#12\catcode `\^12\catcode `\_12\catcode `\%12\relax}%
\providecommand \@@startlink[1]{}%
\providecommand \@@endlink[0]{}%
\providecommand \url  [0]{\begingroup\@sanitize@url \@url }%
\providecommand \@url [1]{\endgroup\@href {#1}{\urlprefix }}%
\providecommand \urlprefix  [0]{URL }%
\providecommand \Eprint [0]{\href }%
\providecommand \doibase [0]{http://dx.doi.org/}%
\providecommand \selectlanguage [0]{\@gobble}%
\providecommand \bibinfo  [0]{\@secondoftwo}%
\providecommand \bibfield  [0]{\@secondoftwo}%
\providecommand \translation [1]{[#1]}%
\providecommand \BibitemOpen [0]{}%
\providecommand \bibitemStop [0]{}%
\providecommand \bibitemNoStop [0]{.\EOS\space}%
\providecommand \EOS [0]{\spacefactor3000\relax}%
\providecommand \BibitemShut  [1]{\csname bibitem#1\endcsname}%
\let\auto@bib@innerbib\@empty
\bibitem [{\citenamefont {Peccei}\ and\ \citenamefont {Quinn}(1977)}]{peccei1977cp}%
  \BibitemOpen
  \bibfield  {author} {\bibinfo {author} {\bibfnamefont {Roberto~D}\ \bibnamefont {Peccei}}\ and\ \bibinfo {author} {\bibfnamefont {Helen~R}\ \bibnamefont {Quinn}},\ }\bibfield  {title} {\enquote {\bibinfo {title} {$cp$ conservation in the presence of pseudoparticles},}\ }\href@noop {} {\bibfield  {journal} {\bibinfo  {journal} {Physical Review Letters}\ }\textbf {\bibinfo {volume} {38}},\ \bibinfo {pages} {1440} (\bibinfo {year} {1977})}\BibitemShut {NoStop}%
\bibitem [{\citenamefont {Weinberg}(1978)}]{weinberg1978new}%
  \BibitemOpen
  \bibfield  {author} {\bibinfo {author} {\bibfnamefont {Steven}\ \bibnamefont {Weinberg}},\ }\bibfield  {title} {\enquote {\bibinfo {title} {A new light boson?}}\ }\href@noop {} {\bibfield  {journal} {\bibinfo  {journal} {Physical Review Letters}\ }\textbf {\bibinfo {volume} {40}},\ \bibinfo {pages} {223} (\bibinfo {year} {1978})}\BibitemShut {NoStop}%
\bibitem [{\citenamefont {Wilczek}(1978)}]{wilczek1978problem}%
  \BibitemOpen
  \bibfield  {author} {\bibinfo {author} {\bibfnamefont {Frank}\ \bibnamefont {Wilczek}},\ }\bibfield  {title} {\enquote {\bibinfo {title} {Problem of strong $p$ and $t$ invariance in the presence of instantons},}\ }\href@noop {} {\bibfield  {journal} {\bibinfo  {journal} {Physical Review Letters}\ }\textbf {\bibinfo {volume} {40}},\ \bibinfo {pages} {279} (\bibinfo {year} {1978})}\BibitemShut {NoStop}%
\bibitem [{\citenamefont {Abbott}\ and\ \citenamefont {Sikivie}(1983)}]{abbott1983cosmological}%
  \BibitemOpen
  \bibfield  {author} {\bibinfo {author} {\bibfnamefont {Laurence~F}\ \bibnamefont {Abbott}}\ and\ \bibinfo {author} {\bibfnamefont {P}~\bibnamefont {Sikivie}},\ }\bibfield  {title} {\enquote {\bibinfo {title} {A cosmological bound on the invisible axion},}\ }\href@noop {} {\bibfield  {journal} {\bibinfo  {journal} {Physics Letters B}\ }\textbf {\bibinfo {volume} {120}},\ \bibinfo {pages} {133--136} (\bibinfo {year} {1983})}\BibitemShut {NoStop}%
\bibitem [{\citenamefont {Preskill}\ \emph {et~al.}(1983)\citenamefont {Preskill}, \citenamefont {Wise},\ and\ \citenamefont {Wilczek}}]{preskill1983cosmology}%
  \BibitemOpen
  \bibfield  {author} {\bibinfo {author} {\bibfnamefont {John}\ \bibnamefont {Preskill}}, \bibinfo {author} {\bibfnamefont {Mark~B}\ \bibnamefont {Wise}}, \ and\ \bibinfo {author} {\bibfnamefont {Frank}\ \bibnamefont {Wilczek}},\ }\bibfield  {title} {\enquote {\bibinfo {title} {Cosmology of the invisible axion},}\ }\href@noop {} {\bibfield  {journal} {\bibinfo  {journal} {Physics Letters B}\ }\textbf {\bibinfo {volume} {120}},\ \bibinfo {pages} {127--132} (\bibinfo {year} {1983})}\BibitemShut {NoStop}%
\bibitem [{\citenamefont {Dine}\ and\ \citenamefont {Fischler}(1983{\natexlab{a}})}]{dine1983not}%
  \BibitemOpen
  \bibfield  {author} {\bibinfo {author} {\bibfnamefont {Michael}\ \bibnamefont {Dine}}\ and\ \bibinfo {author} {\bibfnamefont {Willy}\ \bibnamefont {Fischler}},\ }\bibfield  {title} {\enquote {\bibinfo {title} {The not-so-harmless axion},}\ }\href@noop {} {\bibfield  {journal} {\bibinfo  {journal} {Physics Letters B}\ }\textbf {\bibinfo {volume} {120}},\ \bibinfo {pages} {137--141} (\bibinfo {year} {1983}{\natexlab{a}})}\BibitemShut {NoStop}%
\bibitem [{\citenamefont {Witten}(1984)}]{witten1984some}%
  \BibitemOpen
  \bibfield  {author} {\bibinfo {author} {\bibfnamefont {Edward}\ \bibnamefont {Witten}},\ }\bibfield  {title} {\enquote {\bibinfo {title} {Some properties of $o(32)$ superstrings},}\ }\href@noop {} {\bibfield  {journal} {\bibinfo  {journal} {Physics Letters B}\ }\textbf {\bibinfo {volume} {149}},\ \bibinfo {pages} {351--356} (\bibinfo {year} {1984})}\BibitemShut {NoStop}%
\bibitem [{\citenamefont {Svrcek}\ and\ \citenamefont {Witten}(2006)}]{svrcek2006axions}%
  \BibitemOpen
  \bibfield  {author} {\bibinfo {author} {\bibfnamefont {Peter}\ \bibnamefont {Svrcek}}\ and\ \bibinfo {author} {\bibfnamefont {Edward}\ \bibnamefont {Witten}},\ }\bibfield  {title} {\enquote {\bibinfo {title} {Axions in string theory},}\ }\href@noop {} {\bibfield  {journal} {\bibinfo  {journal} {Journal of High Energy Physics}\ }\textbf {\bibinfo {volume} {2006}},\ \bibinfo {pages} {051} (\bibinfo {year} {2006})}\BibitemShut {NoStop}%
\bibitem [{\citenamefont {Arvanitaki}\ \emph {et~al.}(2010)\citenamefont {Arvanitaki}, \citenamefont {Dimopoulos}, \citenamefont {Dubovsky}, \citenamefont {Kaloper},\ and\ \citenamefont {March-Russell}}]{Arvanitaki:2009fg}%
  \BibitemOpen
  \bibfield  {author} {\bibinfo {author} {\bibfnamefont {Asimina}\ \bibnamefont {Arvanitaki}}, \bibinfo {author} {\bibfnamefont {Savas}\ \bibnamefont {Dimopoulos}}, \bibinfo {author} {\bibfnamefont {Sergei}\ \bibnamefont {Dubovsky}}, \bibinfo {author} {\bibfnamefont {Nemanja}\ \bibnamefont {Kaloper}}, \ and\ \bibinfo {author} {\bibfnamefont {John}\ \bibnamefont {March-Russell}},\ }\bibfield  {title} {\enquote {\bibinfo {title} {{String Axiverse}},}\ }\href {\doibase 10.1103/PhysRevD.81.123530} {\bibfield  {journal} {\bibinfo  {journal} {Phys. Rev. D}\ }\textbf {\bibinfo {volume} {81}},\ \bibinfo {pages} {123530} (\bibinfo {year} {2010})},\ \Eprint {http://arxiv.org/abs/0905.4720} {arXiv:0905.4720 [hep-th]} \BibitemShut {NoStop}%
\bibitem [{\citenamefont {Acharya}\ \emph {et~al.}(2010)\citenamefont {Acharya}, \citenamefont {Bobkov},\ and\ \citenamefont {Kumar}}]{acharya2010m}%
  \BibitemOpen
  \bibfield  {author} {\bibinfo {author} {\bibfnamefont {Bobby~Samir}\ \bibnamefont {Acharya}}, \bibinfo {author} {\bibfnamefont {Konstantin}\ \bibnamefont {Bobkov}}, \ and\ \bibinfo {author} {\bibfnamefont {Piyush}\ \bibnamefont {Kumar}},\ }\bibfield  {title} {\enquote {\bibinfo {title} {An m theory solution to the strong cp-problem, and constraints on the axiverse},}\ }\href@noop {} {\bibfield  {journal} {\bibinfo  {journal} {Journal of High Energy Physics}\ }\textbf {\bibinfo {volume} {2010}},\ \bibinfo {pages} {105} (\bibinfo {year} {2010})}\BibitemShut {NoStop}%
\bibitem [{\citenamefont {Cicoli}\ \emph {et~al.}(2012)\citenamefont {Cicoli}, \citenamefont {Goodsell},\ and\ \citenamefont {Ringwald}}]{Cicoli:2012sz}%
  \BibitemOpen
  \bibfield  {author} {\bibinfo {author} {\bibfnamefont {Michele}\ \bibnamefont {Cicoli}}, \bibinfo {author} {\bibfnamefont {Mark}\ \bibnamefont {Goodsell}}, \ and\ \bibinfo {author} {\bibfnamefont {Andreas}\ \bibnamefont {Ringwald}},\ }\bibfield  {title} {\enquote {\bibinfo {title} {{The type IIB string axiverse and its low-energy phenomenology}},}\ }\href {\doibase 10.1007/JHEP10(2012)146} {\bibfield  {journal} {\bibinfo  {journal} {JHEP}\ }\textbf {\bibinfo {volume} {10}},\ \bibinfo {pages} {146} (\bibinfo {year} {2012})},\ \Eprint {http://arxiv.org/abs/1206.0819} {arXiv:1206.0819 [hep-th]} \BibitemShut {NoStop}%
\bibitem [{\citenamefont {Kim}(1979)}]{Kim:1979if}%
  \BibitemOpen
  \bibfield  {author} {\bibinfo {author} {\bibfnamefont {Jihn~E.}\ \bibnamefont {Kim}},\ }\bibfield  {title} {\enquote {\bibinfo {title} {{Weak Interaction Singlet and Strong CP Invariance}},}\ }\href {\doibase 10.1103/PhysRevLett.43.103} {\bibfield  {journal} {\bibinfo  {journal} {Phys. Rev. Lett.}\ }\textbf {\bibinfo {volume} {43}},\ \bibinfo {pages} {103} (\bibinfo {year} {1979})}\BibitemShut {NoStop}%
\bibitem [{\citenamefont {Shifman}\ \emph {et~al.}(1979)\citenamefont {Shifman}, \citenamefont {Vainshtein},\ and\ \citenamefont {Zakharov}}]{Shifman:1978bx}%
  \BibitemOpen
  \bibfield  {author} {\bibinfo {author} {\bibfnamefont {Mikhail~A.}\ \bibnamefont {Shifman}}, \bibinfo {author} {\bibfnamefont {A.~I.}\ \bibnamefont {Vainshtein}}, \ and\ \bibinfo {author} {\bibfnamefont {Valentin~I.}\ \bibnamefont {Zakharov}},\ }\bibfield  {title} {\enquote {\bibinfo {title} {{QCD and Resonance Physics. Theoretical Foundations}},}\ }\href {\doibase 10.1016/0550-3213(79)90022-1} {\bibfield  {journal} {\bibinfo  {journal} {Nucl. Phys. B}\ }\textbf {\bibinfo {volume} {147}},\ \bibinfo {pages} {385--447} (\bibinfo {year} {1979})}\BibitemShut {NoStop}%
\bibitem [{\citenamefont {Zhitnitsky}(1980)}]{Zhitnitsky:1980tq}%
  \BibitemOpen
  \bibfield  {author} {\bibinfo {author} {\bibfnamefont {A.~R.}\ \bibnamefont {Zhitnitsky}},\ }\bibfield  {title} {\enquote {\bibinfo {title} {{On Possible Suppression of the Axion Hadron Interactions. (In Russian)}},}\ }\href@noop {} {\bibfield  {journal} {\bibinfo  {journal} {Sov. J. Nucl. Phys.}\ }\textbf {\bibinfo {volume} {31}},\ \bibinfo {pages} {260} (\bibinfo {year} {1980})}\BibitemShut {NoStop}%
\bibitem [{\citenamefont {Dine}\ \emph {et~al.}(1981)\citenamefont {Dine}, \citenamefont {Fischler},\ and\ \citenamefont {Srednicki}}]{Dine:1981rt}%
  \BibitemOpen
  \bibfield  {author} {\bibinfo {author} {\bibfnamefont {Michael}\ \bibnamefont {Dine}}, \bibinfo {author} {\bibfnamefont {Willy}\ \bibnamefont {Fischler}}, \ and\ \bibinfo {author} {\bibfnamefont {Mark}\ \bibnamefont {Srednicki}},\ }\bibfield  {title} {\enquote {\bibinfo {title} {{A Simple Solution to the Strong CP Problem with a Harmless Axion}},}\ }\href {\doibase 10.1016/0370-2693(81)90590-6} {\bibfield  {journal} {\bibinfo  {journal} {Phys. Lett. B}\ }\textbf {\bibinfo {volume} {104}},\ \bibinfo {pages} {199--202} (\bibinfo {year} {1981})}\BibitemShut {NoStop}%
\bibitem [{\citenamefont {Holdom}(1986)}]{Holdom:1985ag}%
  \BibitemOpen
  \bibfield  {author} {\bibinfo {author} {\bibfnamefont {Bob}\ \bibnamefont {Holdom}},\ }\bibfield  {title} {\enquote {\bibinfo {title} {{Two U(1)'s and Epsilon Charge Shifts}},}\ }\href {\doibase 10.1016/0370-2693(86)91377-8} {\bibfield  {journal} {\bibinfo  {journal} {Phys. Lett. B}\ }\textbf {\bibinfo {volume} {166}},\ \bibinfo {pages} {196--198} (\bibinfo {year} {1986})}\BibitemShut {NoStop}%
\bibitem [{\citenamefont {Abel}\ \emph {et~al.}(2008)\citenamefont {Abel}, \citenamefont {Goodsell}, \citenamefont {Jaeckel}, \citenamefont {Khoze},\ and\ \citenamefont {Ringwald}}]{Abel:2008ai}%
  \BibitemOpen
  \bibfield  {author} {\bibinfo {author} {\bibfnamefont {S.~A.}\ \bibnamefont {Abel}}, \bibinfo {author} {\bibfnamefont {M.~D.}\ \bibnamefont {Goodsell}}, \bibinfo {author} {\bibfnamefont {J.}~\bibnamefont {Jaeckel}}, \bibinfo {author} {\bibfnamefont {V.~V.}\ \bibnamefont {Khoze}}, \ and\ \bibinfo {author} {\bibfnamefont {A.}~\bibnamefont {Ringwald}},\ }\bibfield  {title} {\enquote {\bibinfo {title} {{Kinetic Mixing of the Photon with Hidden U(1)s in String Phenomenology}},}\ }\href {\doibase 10.1088/1126-6708/2008/07/124} {\bibfield  {journal} {\bibinfo  {journal} {JHEP}\ }\textbf {\bibinfo {volume} {07}},\ \bibinfo {pages} {124} (\bibinfo {year} {2008})},\ \Eprint {http://arxiv.org/abs/0803.1449} {arXiv:0803.1449 [hep-ph]} \BibitemShut {NoStop}%
\bibitem [{\citenamefont {Goodsell}\ \emph {et~al.}(2009)\citenamefont {Goodsell}, \citenamefont {Jaeckel}, \citenamefont {Redondo},\ and\ \citenamefont {Ringwald}}]{Goodsell:2009xc}%
  \BibitemOpen
  \bibfield  {author} {\bibinfo {author} {\bibfnamefont {Mark}\ \bibnamefont {Goodsell}}, \bibinfo {author} {\bibfnamefont {Joerg}\ \bibnamefont {Jaeckel}}, \bibinfo {author} {\bibfnamefont {Javier}\ \bibnamefont {Redondo}}, \ and\ \bibinfo {author} {\bibfnamefont {Andreas}\ \bibnamefont {Ringwald}},\ }\bibfield  {title} {\enquote {\bibinfo {title} {{Naturally Light Hidden Photons in LARGE Volume String Compactifications}},}\ }\href {\doibase 10.1088/1126-6708/2009/11/027} {\bibfield  {journal} {\bibinfo  {journal} {JHEP}\ }\textbf {\bibinfo {volume} {11}},\ \bibinfo {pages} {027} (\bibinfo {year} {2009})},\ \Eprint {http://arxiv.org/abs/0909.0515} {arXiv:0909.0515 [hep-ph]} \BibitemShut {NoStop}%
\bibitem [{\citenamefont {Pospelov}(2009)}]{Pospelov:2008zw}%
  \BibitemOpen
  \bibfield  {author} {\bibinfo {author} {\bibfnamefont {Maxim}\ \bibnamefont {Pospelov}},\ }\bibfield  {title} {\enquote {\bibinfo {title} {{Secluded U(1) below the weak scale}},}\ }\href {\doibase 10.1103/PhysRevD.80.095002} {\bibfield  {journal} {\bibinfo  {journal} {Phys. Rev. D}\ }\textbf {\bibinfo {volume} {80}},\ \bibinfo {pages} {095002} (\bibinfo {year} {2009})},\ \Eprint {http://arxiv.org/abs/0811.1030} {arXiv:0811.1030 [hep-ph]} \BibitemShut {NoStop}%
\bibitem [{\citenamefont {Fabbrichesi}\ \emph {et~al.}(2020)\citenamefont {Fabbrichesi}, \citenamefont {Gabrielli},\ and\ \citenamefont {Lanfranchi}}]{Fabbrichesi:2020wbt}%
  \BibitemOpen
  \bibfield  {author} {\bibinfo {author} {\bibfnamefont {Marco}\ \bibnamefont {Fabbrichesi}}, \bibinfo {author} {\bibfnamefont {Emidio}\ \bibnamefont {Gabrielli}}, \ and\ \bibinfo {author} {\bibfnamefont {Gaia}\ \bibnamefont {Lanfranchi}},\ }\bibfield  {title} {\enquote {\bibinfo {title} {{The Dark Photon}},}\ }\href {\doibase 10.1007/978-3-030-62519-1} {\  (\bibinfo {year} {2020}),\ 10.1007/978-3-030-62519-1},\ \Eprint {http://arxiv.org/abs/2005.01515} {arXiv:2005.01515 [hep-ph]} \BibitemShut {NoStop}%
\bibitem [{\citenamefont {Stueckelberg}(1938)}]{Stueckelberg:1938hvi}%
  \BibitemOpen
  \bibfield  {author} {\bibinfo {author} {\bibfnamefont {E.~C.~G.}\ \bibnamefont {Stueckelberg}},\ }\bibfield  {title} {\enquote {\bibinfo {title} {{Interaction energy in electrodynamics and in the field theory of nuclear forces}},}\ }\href {\doibase 10.5169/seals-110852} {\bibfield  {journal} {\bibinfo  {journal} {Helv. Phys. Acta}\ }\textbf {\bibinfo {volume} {11}},\ \bibinfo {pages} {225--244} (\bibinfo {year} {1938})}\BibitemShut {NoStop}%
\bibitem [{\citenamefont {Brahma}\ \emph {et~al.}(2025)\citenamefont {Brahma}, \citenamefont {Heeba}, \citenamefont {Sch{\'e}rer},\ and\ \citenamefont {Schutz}}]{Brahma:2025wos}%
  \BibitemOpen
  \bibfield  {author} {\bibinfo {author} {\bibfnamefont {Nirmalya}\ \bibnamefont {Brahma}}, \bibinfo {author} {\bibfnamefont {Saniya}\ \bibnamefont {Heeba}}, \bibinfo {author} {\bibfnamefont {Hugo}\ \bibnamefont {Sch{\'e}rer}}, \ and\ \bibinfo {author} {\bibfnamefont {Katelin}\ \bibnamefont {Schutz}},\ }\bibfield  {title} {\enquote {\bibinfo {title} {{Spectral Surgery in a Heat Bath: a finite-temperature guide to particle production for phenomenologists}},}\ }\href@noop {} {\  (\bibinfo {year} {2025})},\ \Eprint {http://arxiv.org/abs/2507.14277} {arXiv:2507.14277 [hep-ph]} \BibitemShut {NoStop}%
\bibitem [{\citenamefont {Davidson}\ \emph {et~al.}(1991)\citenamefont {Davidson}, \citenamefont {Campbell},\ and\ \citenamefont {Bailey}}]{davidson1991limits}%
  \BibitemOpen
  \bibfield  {author} {\bibinfo {author} {\bibfnamefont {S}~\bibnamefont {Davidson}}, \bibinfo {author} {\bibfnamefont {B}~\bibnamefont {Campbell}}, \ and\ \bibinfo {author} {\bibfnamefont {D}~\bibnamefont {Bailey}},\ }\bibfield  {title} {\enquote {\bibinfo {title} {Limits on particles of small electric charge},}\ }\href@noop {} {\bibfield  {journal} {\bibinfo  {journal} {Physical Review D}\ }\textbf {\bibinfo {volume} {43}},\ \bibinfo {pages} {2314} (\bibinfo {year} {1991})}\BibitemShut {NoStop}%
\bibitem [{\citenamefont {Davidson}\ and\ \citenamefont {Peskin}(1994)}]{Davidson:1993sj}%
  \BibitemOpen
  \bibfield  {author} {\bibinfo {author} {\bibfnamefont {Sacha}\ \bibnamefont {Davidson}}\ and\ \bibinfo {author} {\bibfnamefont {Michael~E.}\ \bibnamefont {Peskin}},\ }\bibfield  {title} {\enquote {\bibinfo {title} {{Astrophysical bounds on millicharged particles in models with a paraphoton}},}\ }\href {\doibase 10.1103/PhysRevD.49.2114} {\bibfield  {journal} {\bibinfo  {journal} {Phys. Rev. D}\ }\textbf {\bibinfo {volume} {49}},\ \bibinfo {pages} {2114--2117} (\bibinfo {year} {1994})},\ \Eprint {http://arxiv.org/abs/hep-ph/9310288} {arXiv:hep-ph/9310288} \BibitemShut {NoStop}%
\bibitem [{\citenamefont {Blinnikov}\ and\ \citenamefont {Dunina-Barkovskaya}(1994)}]{blinnikov1994cooling}%
  \BibitemOpen
  \bibfield  {author} {\bibinfo {author} {\bibfnamefont {Sergej~I}\ \bibnamefont {Blinnikov}}\ and\ \bibinfo {author} {\bibfnamefont {NV}~\bibnamefont {Dunina-Barkovskaya}},\ }\bibfield  {title} {\enquote {\bibinfo {title} {The cooling of hot white dwarfs: a theory with non-standard weak interactions, and a comparison with observations},}\ }\href@noop {} {\bibfield  {journal} {\bibinfo  {journal} {Monthly Notices of the Royal Astronomical Society}\ }\textbf {\bibinfo {volume} {266}},\ \bibinfo {pages} {289--304} (\bibinfo {year} {1994})}\BibitemShut {NoStop}%
\bibitem [{\citenamefont {Davidson}\ \emph {et~al.}(2000)\citenamefont {Davidson}, \citenamefont {Hannestad},\ and\ \citenamefont {Raffelt}}]{Davidson:2000hf}%
  \BibitemOpen
  \bibfield  {author} {\bibinfo {author} {\bibfnamefont {Sacha}\ \bibnamefont {Davidson}}, \bibinfo {author} {\bibfnamefont {Steen}\ \bibnamefont {Hannestad}}, \ and\ \bibinfo {author} {\bibfnamefont {Georg}\ \bibnamefont {Raffelt}},\ }\bibfield  {title} {\enquote {\bibinfo {title} {{Updated bounds on millicharged particles}},}\ }\href {\doibase 10.1088/1126-6708/2000/05/003} {\bibfield  {journal} {\bibinfo  {journal} {JHEP}\ }\textbf {\bibinfo {volume} {05}},\ \bibinfo {pages} {003} (\bibinfo {year} {2000})},\ \Eprint {http://arxiv.org/abs/hep-ph/0001179} {arXiv:hep-ph/0001179} \BibitemShut {NoStop}%
\bibitem [{\citenamefont {Vogel}\ and\ \citenamefont {Redondo}(2014)}]{Vogel:2013raa}%
  \BibitemOpen
  \bibfield  {author} {\bibinfo {author} {\bibfnamefont {Hendrik}\ \bibnamefont {Vogel}}\ and\ \bibinfo {author} {\bibfnamefont {Javier}\ \bibnamefont {Redondo}},\ }\bibfield  {title} {\enquote {\bibinfo {title} {{Dark Radiation constraints on minicharged particles in models with a hidden photon}},}\ }\href {\doibase 10.1088/1475-7516/2014/02/029} {\bibfield  {journal} {\bibinfo  {journal} {JCAP}\ }\textbf {\bibinfo {volume} {1402}},\ \bibinfo {pages} {029} (\bibinfo {year} {2014})},\ \Eprint {http://arxiv.org/abs/1311.2600} {arXiv:1311.2600 [hep-ph]} \BibitemShut {NoStop}%
\bibitem [{\citenamefont {Hardy}\ and\ \citenamefont {Lasenby}(2017)}]{Hardy:2016kme}%
  \BibitemOpen
  \bibfield  {author} {\bibinfo {author} {\bibfnamefont {Edward}\ \bibnamefont {Hardy}}\ and\ \bibinfo {author} {\bibfnamefont {Robert}\ \bibnamefont {Lasenby}},\ }\bibfield  {title} {\enquote {\bibinfo {title} {{Stellar cooling bounds on new light particles: plasma mixing effects}},}\ }\href {\doibase 10.1007/JHEP02(2017)033} {\bibfield  {journal} {\bibinfo  {journal} {JHEP}\ }\textbf {\bibinfo {volume} {02}},\ \bibinfo {pages} {033} (\bibinfo {year} {2017})},\ \Eprint {http://arxiv.org/abs/1611.05852} {arXiv:1611.05852 [hep-ph]} \BibitemShut {NoStop}%
\bibitem [{\citenamefont {An}\ \emph {et~al.}(2013)\citenamefont {An}, \citenamefont {Pospelov},\ and\ \citenamefont {Pradler}}]{An:2013yfc}%
  \BibitemOpen
  \bibfield  {author} {\bibinfo {author} {\bibfnamefont {Haipeng}\ \bibnamefont {An}}, \bibinfo {author} {\bibfnamefont {Maxim}\ \bibnamefont {Pospelov}}, \ and\ \bibinfo {author} {\bibfnamefont {Josef}\ \bibnamefont {Pradler}},\ }\bibfield  {title} {\enquote {\bibinfo {title} {{New stellar constraints on dark photons}},}\ }\href {\doibase 10.1016/j.physletb.2013.07.008} {\bibfield  {journal} {\bibinfo  {journal} {Phys. Lett. B}\ }\textbf {\bibinfo {volume} {725}},\ \bibinfo {pages} {190--195} (\bibinfo {year} {2013})},\ \Eprint {http://arxiv.org/abs/1302.3884} {arXiv:1302.3884 [hep-ph]} \BibitemShut {NoStop}%
\bibitem [{\citenamefont {Gondolo}\ and\ \citenamefont {Raffelt}(2009)}]{Gondolo:2008dd}%
  \BibitemOpen
  \bibfield  {author} {\bibinfo {author} {\bibfnamefont {Paolo}\ \bibnamefont {Gondolo}}\ and\ \bibinfo {author} {\bibfnamefont {Georg~G.}\ \bibnamefont {Raffelt}},\ }\bibfield  {title} {\enquote {\bibinfo {title} {{Solar neutrino limit on axions and keV-mass bosons}},}\ }\href {\doibase 10.1103/PhysRevD.79.107301} {\bibfield  {journal} {\bibinfo  {journal} {Phys. Rev. D}\ }\textbf {\bibinfo {volume} {79}},\ \bibinfo {pages} {107301} (\bibinfo {year} {2009})},\ \Eprint {http://arxiv.org/abs/0807.2926} {arXiv:0807.2926 [astro-ph]} \BibitemShut {NoStop}%
\bibitem [{\citenamefont {Vinyoles}\ \emph {et~al.}(2015)\citenamefont {Vinyoles}, \citenamefont {Serenelli}, \citenamefont {Villante}, \citenamefont {Basu}, \citenamefont {Redondo},\ and\ \citenamefont {Isern}}]{Vinyoles:2015aba}%
  \BibitemOpen
  \bibfield  {author} {\bibinfo {author} {\bibfnamefont {N\'uria}\ \bibnamefont {Vinyoles}}, \bibinfo {author} {\bibfnamefont {Aldo}\ \bibnamefont {Serenelli}}, \bibinfo {author} {\bibfnamefont {Francesco~L.}\ \bibnamefont {Villante}}, \bibinfo {author} {\bibfnamefont {Sarbani}\ \bibnamefont {Basu}}, \bibinfo {author} {\bibfnamefont {Javier}\ \bibnamefont {Redondo}}, \ and\ \bibinfo {author} {\bibfnamefont {Jordi}\ \bibnamefont {Isern}},\ }\bibfield  {title} {\enquote {\bibinfo {title} {{New axion and hidden photon constraints from a solar data global fit}},}\ }\href {\doibase 10.1088/1475-7516/2015/10/015} {\bibfield  {journal} {\bibinfo  {journal} {JCAP}\ }\textbf {\bibinfo {volume} {10}},\ \bibinfo {pages} {015} (\bibinfo {year} {2015})},\ \Eprint {http://arxiv.org/abs/1501.01639} {arXiv:1501.01639 [astro-ph.SR]} \BibitemShut {NoStop}%
\bibitem [{\citenamefont {Ayala}\ \emph {et~al.}(2014)\citenamefont {Ayala}, \citenamefont {Dom\'\i{}nguez}, \citenamefont {Giannotti}, \citenamefont {Mirizzi},\ and\ \citenamefont {Straniero}}]{Ayala:2014pea}%
  \BibitemOpen
  \bibfield  {author} {\bibinfo {author} {\bibfnamefont {Adrian}\ \bibnamefont {Ayala}}, \bibinfo {author} {\bibfnamefont {Inma}\ \bibnamefont {Dom\'\i{}nguez}}, \bibinfo {author} {\bibfnamefont {Maurizio}\ \bibnamefont {Giannotti}}, \bibinfo {author} {\bibfnamefont {Alessandro}\ \bibnamefont {Mirizzi}}, \ and\ \bibinfo {author} {\bibfnamefont {Oscar}\ \bibnamefont {Straniero}},\ }\bibfield  {title} {\enquote {\bibinfo {title} {{Revisiting the bound on axion-photon coupling from Globular Clusters}},}\ }\href {\doibase 10.1103/PhysRevLett.113.191302} {\bibfield  {journal} {\bibinfo  {journal} {Phys. Rev. Lett.}\ }\textbf {\bibinfo {volume} {113}},\ \bibinfo {pages} {191302} (\bibinfo {year} {2014})},\ \Eprint {http://arxiv.org/abs/1406.6053} {arXiv:1406.6053 [astro-ph.SR]} \BibitemShut {NoStop}%
\bibitem [{\citenamefont {Raffelt}(1990)}]{raffelt1990core}%
  \BibitemOpen
  \bibfield  {author} {\bibinfo {author} {\bibfnamefont {Georg~G}\ \bibnamefont {Raffelt}},\ }\bibfield  {title} {\enquote {\bibinfo {title} {Core mass at the helium flash from observations and a new bound on neutrino electromagnetic properties},}\ }\href@noop {} {\bibfield  {journal} {\bibinfo  {journal} {The Astrophysical Journal}\ }\textbf {\bibinfo {volume} {365}},\ \bibinfo {pages} {559--568} (\bibinfo {year} {1990})}\BibitemShut {NoStop}%
\bibitem [{\citenamefont {Haft}\ \emph {et~al.}(1994)\citenamefont {Haft}, \citenamefont {Raffelt},\ and\ \citenamefont {Weiss}}]{Haft:1993jt}%
  \BibitemOpen
  \bibfield  {author} {\bibinfo {author} {\bibfnamefont {Martin}\ \bibnamefont {Haft}}, \bibinfo {author} {\bibfnamefont {Georg}\ \bibnamefont {Raffelt}}, \ and\ \bibinfo {author} {\bibfnamefont {Achim}\ \bibnamefont {Weiss}},\ }\bibfield  {title} {\enquote {\bibinfo {title} {{Standard and nonstandard plasma neutrino emission revisited}},}\ }\href {\doibase 10.1086/173978} {\bibfield  {journal} {\bibinfo  {journal} {Astrophys. J.}\ }\textbf {\bibinfo {volume} {425}},\ \bibinfo {pages} {222--230} (\bibinfo {year} {1994})},\ \bibinfo {note} {[Erratum: Astrophys.J. 438, 1017 (1995)]},\ \Eprint {http://arxiv.org/abs/astro-ph/9309014} {arXiv:astro-ph/9309014} \BibitemShut {NoStop}%
\bibitem [{\citenamefont {Viaux}\ \emph {et~al.}(2013{\natexlab{a}})\citenamefont {Viaux}, \citenamefont {Catelan}, \citenamefont {Stetson}, \citenamefont {Raffelt}, \citenamefont {Redondo}, \citenamefont {Valcarce},\ and\ \citenamefont {Weiss}}]{Viaux:2013hca}%
  \BibitemOpen
  \bibfield  {author} {\bibinfo {author} {\bibfnamefont {Nicol\'as}\ \bibnamefont {Viaux}}, \bibinfo {author} {\bibfnamefont {M\'arcio}\ \bibnamefont {Catelan}}, \bibinfo {author} {\bibfnamefont {Peter~B.}\ \bibnamefont {Stetson}}, \bibinfo {author} {\bibfnamefont {Georg}\ \bibnamefont {Raffelt}}, \bibinfo {author} {\bibfnamefont {Javier}\ \bibnamefont {Redondo}}, \bibinfo {author} {\bibfnamefont {Aldo A.~R.}\ \bibnamefont {Valcarce}}, \ and\ \bibinfo {author} {\bibfnamefont {Achim}\ \bibnamefont {Weiss}},\ }\bibfield  {title} {\enquote {\bibinfo {title} {{Particle-physics constraints from the globular cluster M5: Neutrino Dipole Moments}},}\ }\href {\doibase 10.1051/0004-6361/201322004} {\bibfield  {journal} {\bibinfo  {journal} {Astron. Astrophys.}\ }\textbf {\bibinfo {volume} {558}},\ \bibinfo {pages} {A12} (\bibinfo {year} {2013}{\natexlab{a}})},\ \Eprint {http://arxiv.org/abs/1308.4627} {arXiv:1308.4627 [astro-ph.SR]} \BibitemShut {NoStop}%
\bibitem [{\citenamefont {Viaux}\ \emph {et~al.}(2013{\natexlab{b}})\citenamefont {Viaux}, \citenamefont {Catelan}, \citenamefont {Stetson}, \citenamefont {Raffelt}, \citenamefont {Redondo}, \citenamefont {Valcarce},\ and\ \citenamefont {Weiss}}]{Viaux:2013lha}%
  \BibitemOpen
  \bibfield  {author} {\bibinfo {author} {\bibfnamefont {Nicolás}\ \bibnamefont {Viaux}}, \bibinfo {author} {\bibfnamefont {Márcio}\ \bibnamefont {Catelan}}, \bibinfo {author} {\bibfnamefont {Peter~B.}\ \bibnamefont {Stetson}}, \bibinfo {author} {\bibfnamefont {Georg}\ \bibnamefont {Raffelt}}, \bibinfo {author} {\bibfnamefont {Javier}\ \bibnamefont {Redondo}}, \bibinfo {author} {\bibfnamefont {Aldo A.~R.}\ \bibnamefont {Valcarce}}, \ and\ \bibinfo {author} {\bibfnamefont {Achim}\ \bibnamefont {Weiss}},\ }\bibfield  {title} {\enquote {\bibinfo {title} {{Neutrino and axion bounds from the globular cluster M5 (NGC 5904)}},}\ }\href {\doibase 10.1103/PhysRevLett.111.231301} {\bibfield  {journal} {\bibinfo  {journal} {Phys. Rev. Lett.}\ }\textbf {\bibinfo {volume} {111}},\ \bibinfo {pages} {231301} (\bibinfo {year} {2013}{\natexlab{b}})},\ \Eprint {http://arxiv.org/abs/1311.1669} {arXiv:1311.1669 [astro-ph.SR]} \BibitemShut {NoStop}%
\bibitem [{\citenamefont {Arceo-D\'\i{}az}\ \emph {et~al.}(2015)\citenamefont {Arceo-D\'\i{}az}, \citenamefont {Schr\"oder}, \citenamefont {Zuber},\ and\ \citenamefont {Jack}}]{Arceo-Diaz:2015pva}%
  \BibitemOpen
  \bibfield  {author} {\bibinfo {author} {\bibfnamefont {S.}~\bibnamefont {Arceo-D\'\i{}az}}, \bibinfo {author} {\bibfnamefont {K.~P.}\ \bibnamefont {Schr\"oder}}, \bibinfo {author} {\bibfnamefont {K.}~\bibnamefont {Zuber}}, \ and\ \bibinfo {author} {\bibfnamefont {D.}~\bibnamefont {Jack}},\ }\bibfield  {title} {\enquote {\bibinfo {title} {{Constraint on the magnetic dipole moment of neutrinos by the tip-RGB luminosity in \ensuremath{\omega} -Centauri}},}\ }\href {\doibase 10.1016/j.astropartphys.2015.03.006} {\bibfield  {journal} {\bibinfo  {journal} {Astropart. Phys.}\ }\textbf {\bibinfo {volume} {70}},\ \bibinfo {pages} {1--11} (\bibinfo {year} {2015})}\BibitemShut {NoStop}%
\bibitem [{\citenamefont {Miller~Bertolami}\ \emph {et~al.}(2014)\citenamefont {Miller~Bertolami}, \citenamefont {Melendez}, \citenamefont {Althaus},\ and\ \citenamefont {Isern}}]{MillerBertolami:2014rka}%
  \BibitemOpen
  \bibfield  {author} {\bibinfo {author} {\bibfnamefont {Marcelo~M.}\ \bibnamefont {Miller~Bertolami}}, \bibinfo {author} {\bibfnamefont {Brenda~E.}\ \bibnamefont {Melendez}}, \bibinfo {author} {\bibfnamefont {Leandro~G.}\ \bibnamefont {Althaus}}, \ and\ \bibinfo {author} {\bibfnamefont {Jordi}\ \bibnamefont {Isern}},\ }\bibfield  {title} {\enquote {\bibinfo {title} {{Revisiting the axion bounds from the Galactic white dwarf luminosity function}},}\ }\href {\doibase 10.1088/1475-7516/2014/10/069} {\bibfield  {journal} {\bibinfo  {journal} {JCAP}\ }\textbf {\bibinfo {volume} {10}},\ \bibinfo {pages} {069} (\bibinfo {year} {2014})},\ \Eprint {http://arxiv.org/abs/1406.7712} {arXiv:1406.7712 [hep-ph]} \BibitemShut {NoStop}%
\bibitem [{\citenamefont {Isern}\ \emph {et~al.}(2008)\citenamefont {Isern}, \citenamefont {Garcia-Berro}, \citenamefont {Torres},\ and\ \citenamefont {Catalan}}]{Isern:2008nt}%
  \BibitemOpen
  \bibfield  {author} {\bibinfo {author} {\bibfnamefont {J.}~\bibnamefont {Isern}}, \bibinfo {author} {\bibfnamefont {E.}~\bibnamefont {Garcia-Berro}}, \bibinfo {author} {\bibfnamefont {S.}~\bibnamefont {Torres}}, \ and\ \bibinfo {author} {\bibfnamefont {S.}~\bibnamefont {Catalan}},\ }\bibfield  {title} {\enquote {\bibinfo {title} {{Axions and the cooling of white dwarf stars}},}\ }\href {\doibase 10.1086/591042} {\bibfield  {journal} {\bibinfo  {journal} {Astrophys. J. Lett.}\ }\textbf {\bibinfo {volume} {682}},\ \bibinfo {pages} {L109} (\bibinfo {year} {2008})},\ \Eprint {http://arxiv.org/abs/0806.2807} {arXiv:0806.2807 [astro-ph]} \BibitemShut {NoStop}%
\bibitem [{\citenamefont {Corsico}\ \emph {et~al.}(2012{\natexlab{a}})\citenamefont {Corsico}, \citenamefont {Althaus}, \citenamefont {Bertolami}, \citenamefont {Romero}, \citenamefont {Garcia-Berro}, \citenamefont {Isern},\ and\ \citenamefont {Kepler}}]{Corsico:2012ki}%
  \BibitemOpen
  \bibfield  {author} {\bibinfo {author} {\bibfnamefont {Alejandro~H.}\ \bibnamefont {Corsico}}, \bibinfo {author} {\bibfnamefont {Leandro~G.}\ \bibnamefont {Althaus}}, \bibinfo {author} {\bibfnamefont {Marcelo M.~Miller}\ \bibnamefont {Bertolami}}, \bibinfo {author} {\bibfnamefont {Alejandra~D.}\ \bibnamefont {Romero}}, \bibinfo {author} {\bibfnamefont {Enrique}\ \bibnamefont {Garcia-Berro}}, \bibinfo {author} {\bibfnamefont {Jordi}\ \bibnamefont {Isern}}, \ and\ \bibinfo {author} {\bibfnamefont {S.~O.}\ \bibnamefont {Kepler}},\ }\bibfield  {title} {\enquote {\bibinfo {title} {{The rate of cooling of the pulsating white dwarf star G117$-$B15A: a new asteroseismological inference of the axion mass}},}\ }\href {\doibase 10.1111/j.1365-2966.2012.21401.x} {\bibfield  {journal} {\bibinfo  {journal} {Mon. Not. Roy. Astron. Soc.}\ }\textbf {\bibinfo {volume} {424}},\ \bibinfo {pages} {2792} (\bibinfo {year} {2012}{\natexlab{a}})},\ \Eprint {http://arxiv.org/abs/1205.6180} {arXiv:1205.6180 [astro-ph.SR]}
  \BibitemShut {NoStop}%
\bibitem [{\citenamefont {Fung}\ \emph {et~al.}(2024)\citenamefont {Fung}, \citenamefont {Heeba}, \citenamefont {Liu}, \citenamefont {Muralidharan}, \citenamefont {Schutz},\ and\ \citenamefont {Vincent}}]{Fung:2023euv}%
  \BibitemOpen
  \bibfield  {author} {\bibinfo {author} {\bibfnamefont {Audrey}\ \bibnamefont {Fung}}, \bibinfo {author} {\bibfnamefont {Saniya}\ \bibnamefont {Heeba}}, \bibinfo {author} {\bibfnamefont {Qinrui}\ \bibnamefont {Liu}}, \bibinfo {author} {\bibfnamefont {Varun}\ \bibnamefont {Muralidharan}}, \bibinfo {author} {\bibfnamefont {Katelin}\ \bibnamefont {Schutz}}, \ and\ \bibinfo {author} {\bibfnamefont {Aaron~C.}\ \bibnamefont {Vincent}},\ }\bibfield  {title} {\enquote {\bibinfo {title} {{New bounds on light millicharged particles from the tip of the red-giant branch}},}\ }\href {\doibase 10.1103/PhysRevD.109.083011} {\bibfield  {journal} {\bibinfo  {journal} {Phys. Rev. D}\ }\textbf {\bibinfo {volume} {109}},\ \bibinfo {pages} {083011} (\bibinfo {year} {2024})},\ \Eprint {http://arxiv.org/abs/2309.06465} {arXiv:2309.06465 [hep-ph]} \BibitemShut {NoStop}%
\bibitem [{\citenamefont {Dolan}\ \emph {et~al.}(2022)\citenamefont {Dolan}, \citenamefont {Hiskens},\ and\ \citenamefont {Volkas}}]{Dolan:2022kul}%
  \BibitemOpen
  \bibfield  {author} {\bibinfo {author} {\bibfnamefont {Matthew~J.}\ \bibnamefont {Dolan}}, \bibinfo {author} {\bibfnamefont {Frederick~J.}\ \bibnamefont {Hiskens}}, \ and\ \bibinfo {author} {\bibfnamefont {Raymond~R.}\ \bibnamefont {Volkas}},\ }\bibfield  {title} {\enquote {\bibinfo {title} {{Advancing globular cluster constraints on the axion-photon coupling}},}\ }\href {\doibase 10.1088/1475-7516/2022/10/096} {\bibfield  {journal} {\bibinfo  {journal} {JCAP}\ }\textbf {\bibinfo {volume} {10}},\ \bibinfo {pages} {096} (\bibinfo {year} {2022})},\ \Eprint {http://arxiv.org/abs/2207.03102} {arXiv:2207.03102 [hep-ph]} \BibitemShut {NoStop}%
\bibitem [{\citenamefont {Dolan}\ \emph {et~al.}(2023)\citenamefont {Dolan}, \citenamefont {Hiskens},\ and\ \citenamefont {Volkas}}]{Dolan:2023cjs}%
  \BibitemOpen
  \bibfield  {author} {\bibinfo {author} {\bibfnamefont {Matthew~J.}\ \bibnamefont {Dolan}}, \bibinfo {author} {\bibfnamefont {Frederick~J.}\ \bibnamefont {Hiskens}}, \ and\ \bibinfo {author} {\bibfnamefont {Raymond~R.}\ \bibnamefont {Volkas}},\ }\bibfield  {title} {\enquote {\bibinfo {title} {{Constraining Dark Photons with Self-consistent Simulations of Globular Cluster Stars}},}\ }\href@noop {} {\  (\bibinfo {year} {2023})},\ \Eprint {http://arxiv.org/abs/2306.13335} {arXiv:2306.13335 [hep-ph]} \BibitemShut {NoStop}%
\bibitem [{\citenamefont {Dessert}\ \emph {et~al.}(2019)\citenamefont {Dessert}, \citenamefont {Long},\ and\ \citenamefont {Safdi}}]{Dessert:2019sgw}%
  \BibitemOpen
  \bibfield  {author} {\bibinfo {author} {\bibfnamefont {Christopher}\ \bibnamefont {Dessert}}, \bibinfo {author} {\bibfnamefont {Andrew~J.}\ \bibnamefont {Long}}, \ and\ \bibinfo {author} {\bibfnamefont {Benjamin~R.}\ \bibnamefont {Safdi}},\ }\bibfield  {title} {\enquote {\bibinfo {title} {{X-ray Signatures of Axion Conversion in Magnetic White Dwarf Stars}},}\ }\href {\doibase 10.1103/PhysRevLett.123.061104} {\bibfield  {journal} {\bibinfo  {journal} {Phys. Rev. Lett.}\ }\textbf {\bibinfo {volume} {123}},\ \bibinfo {pages} {061104} (\bibinfo {year} {2019})},\ \Eprint {http://arxiv.org/abs/1903.05088} {arXiv:1903.05088 [hep-ph]} \BibitemShut {NoStop}%
\bibitem [{\citenamefont {Dessert}\ \emph {et~al.}(2022{\natexlab{a}})\citenamefont {Dessert}, \citenamefont {Long},\ and\ \citenamefont {Safdi}}]{Dessert:2021bkv}%
  \BibitemOpen
  \bibfield  {author} {\bibinfo {author} {\bibfnamefont {Christopher}\ \bibnamefont {Dessert}}, \bibinfo {author} {\bibfnamefont {Andrew~J.}\ \bibnamefont {Long}}, \ and\ \bibinfo {author} {\bibfnamefont {Benjamin~R.}\ \bibnamefont {Safdi}},\ }\bibfield  {title} {\enquote {\bibinfo {title} {{No Evidence for Axions from Chandra Observation of the Magnetic White Dwarf RE J0317-853}},}\ }\href {\doibase 10.1103/PhysRevLett.128.071102} {\bibfield  {journal} {\bibinfo  {journal} {Phys. Rev. Lett.}\ }\textbf {\bibinfo {volume} {128}},\ \bibinfo {pages} {071102} (\bibinfo {year} {2022}{\natexlab{a}})},\ \Eprint {http://arxiv.org/abs/2104.12772} {arXiv:2104.12772 [hep-ph]} \BibitemShut {NoStop}%
\bibitem [{\citenamefont {Dessert}\ \emph {et~al.}(2022{\natexlab{b}})\citenamefont {Dessert}, \citenamefont {Dunsky},\ and\ \citenamefont {Safdi}}]{Dessert:2022yqq}%
  \BibitemOpen
  \bibfield  {author} {\bibinfo {author} {\bibfnamefont {Christopher}\ \bibnamefont {Dessert}}, \bibinfo {author} {\bibfnamefont {David}\ \bibnamefont {Dunsky}}, \ and\ \bibinfo {author} {\bibfnamefont {Benjamin~R.}\ \bibnamefont {Safdi}},\ }\bibfield  {title} {\enquote {\bibinfo {title} {{Upper limit on the axion-photon coupling from magnetic white dwarf polarization}},}\ }\href {\doibase 10.1103/PhysRevD.105.103034} {\bibfield  {journal} {\bibinfo  {journal} {Phys. Rev. D}\ }\textbf {\bibinfo {volume} {105}},\ \bibinfo {pages} {103034} (\bibinfo {year} {2022}{\natexlab{b}})},\ \Eprint {http://arxiv.org/abs/2203.04319} {arXiv:2203.04319 [hep-ph]} \BibitemShut {NoStop}%
\bibitem [{\citenamefont {Noordhuis}\ \emph {et~al.}(2023)\citenamefont {Noordhuis}, \citenamefont {Prabhu}, \citenamefont {Witte}, \citenamefont {Chen}, \citenamefont {Cruz},\ and\ \citenamefont {Weniger}}]{Noordhuis:2022ljw}%
  \BibitemOpen
  \bibfield  {author} {\bibinfo {author} {\bibfnamefont {Dion}\ \bibnamefont {Noordhuis}}, \bibinfo {author} {\bibfnamefont {Anirudh}\ \bibnamefont {Prabhu}}, \bibinfo {author} {\bibfnamefont {Samuel~J.}\ \bibnamefont {Witte}}, \bibinfo {author} {\bibfnamefont {Alexander~Y.}\ \bibnamefont {Chen}}, \bibinfo {author} {\bibfnamefont {F{\'a}bio}\ \bibnamefont {Cruz}}, \ and\ \bibinfo {author} {\bibfnamefont {Christoph}\ \bibnamefont {Weniger}},\ }\bibfield  {title} {\enquote {\bibinfo {title} {{Novel Constraints on Axions Produced in Pulsar Polar-Cap Cascades}},}\ }\href {\doibase 10.1103/PhysRevLett.131.111004} {\bibfield  {journal} {\bibinfo  {journal} {Phys. Rev. Lett.}\ }\textbf {\bibinfo {volume} {131}},\ \bibinfo {pages} {111004} (\bibinfo {year} {2023})},\ \Eprint {http://arxiv.org/abs/2209.09917} {arXiv:2209.09917 [hep-ph]} \BibitemShut {NoStop}%
\bibitem [{\citenamefont {Noordhuis}\ \emph {et~al.}(2024)\citenamefont {Noordhuis}, \citenamefont {Prabhu}, \citenamefont {Weniger},\ and\ \citenamefont {Witte}}]{Noordhuis:2023wid}%
  \BibitemOpen
  \bibfield  {author} {\bibinfo {author} {\bibfnamefont {Dion}\ \bibnamefont {Noordhuis}}, \bibinfo {author} {\bibfnamefont {Anirudh}\ \bibnamefont {Prabhu}}, \bibinfo {author} {\bibfnamefont {Christoph}\ \bibnamefont {Weniger}}, \ and\ \bibinfo {author} {\bibfnamefont {Samuel~J.}\ \bibnamefont {Witte}},\ }\bibfield  {title} {\enquote {\bibinfo {title} {{Axion Clouds around Neutron Stars}},}\ }\href {\doibase 10.1103/PhysRevX.14.041015} {\bibfield  {journal} {\bibinfo  {journal} {Phys. Rev. X}\ }\textbf {\bibinfo {volume} {14}},\ \bibinfo {pages} {041015} (\bibinfo {year} {2024})},\ \Eprint {http://arxiv.org/abs/2307.11811} {arXiv:2307.11811 [hep-ph]} \BibitemShut {NoStop}%
\bibitem [{\citenamefont {Prabhu}(2023)}]{Prabhu:2023cgb}%
  \BibitemOpen
  \bibfield  {author} {\bibinfo {author} {\bibfnamefont {Anirudh}\ \bibnamefont {Prabhu}},\ }\bibfield  {title} {\enquote {\bibinfo {title} {{Axion-mediated Transport of Fast Radio Bursts Originating in Inner Magnetospheres of Magnetars}},}\ }\href {\doibase 10.3847/2041-8213/acc7a7} {\bibfield  {journal} {\bibinfo  {journal} {Astrophys. J. Lett.}\ }\textbf {\bibinfo {volume} {946}},\ \bibinfo {pages} {L52} (\bibinfo {year} {2023})},\ \Eprint {http://arxiv.org/abs/2302.11645} {arXiv:2302.11645 [astro-ph.HE]} \BibitemShut {NoStop}%
\bibitem [{\citenamefont {Sch{\'e}rer}\ and\ \citenamefont {Schutz}(2024)}]{Scherer:2024uui}%
  \BibitemOpen
  \bibfield  {author} {\bibinfo {author} {\bibfnamefont {Hugo}\ \bibnamefont {Sch{\'e}rer}}\ and\ \bibinfo {author} {\bibfnamefont {Katelin}\ \bibnamefont {Schutz}},\ }\bibfield  {title} {\enquote {\bibinfo {title} {{Photon self-energy at all temperatures and densities in all of phase space}},}\ }\href {\doibase 10.1007/JHEP11(2024)139} {\bibfield  {journal} {\bibinfo  {journal} {JHEP}\ }\textbf {\bibinfo {volume} {11}},\ \bibinfo {pages} {139} (\bibinfo {year} {2024})},\ \Eprint {http://arxiv.org/abs/2405.18466} {arXiv:2405.18466 [hep-ph]} \BibitemShut {NoStop}%
\bibitem [{\citenamefont {Ganguly}\ \emph {et~al.}(1999)\citenamefont {Ganguly}, \citenamefont {Konar},\ and\ \citenamefont {Pal}}]{Ganguly:1999ts}%
  \BibitemOpen
  \bibfield  {author} {\bibinfo {author} {\bibfnamefont {Avijit~K.}\ \bibnamefont {Ganguly}}, \bibinfo {author} {\bibfnamefont {Sushan}\ \bibnamefont {Konar}}, \ and\ \bibinfo {author} {\bibfnamefont {Palash~B.}\ \bibnamefont {Pal}},\ }\bibfield  {title} {\enquote {\bibinfo {title} {{Faraday effect: A Field theoretical point of view}},}\ }\href {\doibase 10.1103/PhysRevD.60.105014} {\bibfield  {journal} {\bibinfo  {journal} {Phys. Rev. D}\ }\textbf {\bibinfo {volume} {60}},\ \bibinfo {pages} {105014} (\bibinfo {year} {1999})},\ \Eprint {http://arxiv.org/abs/hep-ph/9905206} {arXiv:hep-ph/9905206} \BibitemShut {NoStop}%
\bibitem [{\citenamefont {Ganguly}\ and\ \citenamefont {Konar}(2001)}]{Ganguly:2000vt}%
  \BibitemOpen
  \bibfield  {author} {\bibinfo {author} {\bibfnamefont {Avijit~K.}\ \bibnamefont {Ganguly}}\ and\ \bibinfo {author} {\bibfnamefont {Sushan}\ \bibnamefont {Konar}},\ }\bibfield  {title} {\enquote {\bibinfo {title} {{Absorption of electromagnetic waves in a magnetized medium}},}\ }\href {\doibase 10.1103/PhysRevD.63.065001} {\bibfield  {journal} {\bibinfo  {journal} {Phys. Rev. D}\ }\textbf {\bibinfo {volume} {63}},\ \bibinfo {pages} {065001} (\bibinfo {year} {2001})},\ \Eprint {http://arxiv.org/abs/hep-ph/0007027} {arXiv:hep-ph/0007027} \BibitemShut {NoStop}%
\bibitem [{\citenamefont {D'Olivo}\ \emph {et~al.}(2003)\citenamefont {D'Olivo}, \citenamefont {Nieves},\ and\ \citenamefont {Sahu}}]{DOlivo:2002omk}%
  \BibitemOpen
  \bibfield  {author} {\bibinfo {author} {\bibfnamefont {Juan~Carlos}\ \bibnamefont {D'Olivo}}, \bibinfo {author} {\bibfnamefont {Jose~F.}\ \bibnamefont {Nieves}}, \ and\ \bibinfo {author} {\bibfnamefont {Sarira}\ \bibnamefont {Sahu}},\ }\bibfield  {title} {\enquote {\bibinfo {title} {{Field theory of the photon selfenergy in a medium with a magnetic field and the Faraday effect}},}\ }\href {\doibase 10.1103/PhysRevD.67.025018} {\bibfield  {journal} {\bibinfo  {journal} {Phys. Rev. D}\ }\textbf {\bibinfo {volume} {67}},\ \bibinfo {pages} {025018} (\bibinfo {year} {2003})},\ \Eprint {http://arxiv.org/abs/hep-ph/0208146} {arXiv:hep-ph/0208146} \BibitemShut {NoStop}%
\bibitem [{\citenamefont {Hattori}\ and\ \citenamefont {Itakura}(2013{\natexlab{a}})}]{Hattori:2012je}%
  \BibitemOpen
  \bibfield  {author} {\bibinfo {author} {\bibfnamefont {Koichi}\ \bibnamefont {Hattori}}\ and\ \bibinfo {author} {\bibfnamefont {Kazunori}\ \bibnamefont {Itakura}},\ }\bibfield  {title} {\enquote {\bibinfo {title} {{Vacuum birefringence in strong magnetic fields: (I) Photon polarization tensor with all the Landau levels}},}\ }\href {\doibase 10.1016/j.aop.2012.11.010} {\bibfield  {journal} {\bibinfo  {journal} {Annals Phys.}\ }\textbf {\bibinfo {volume} {330}},\ \bibinfo {pages} {23--54} (\bibinfo {year} {2013}{\natexlab{a}})},\ \Eprint {http://arxiv.org/abs/1209.2663} {arXiv:1209.2663 [hep-ph]} \BibitemShut {NoStop}%
\bibitem [{\citenamefont {Hattori}\ and\ \citenamefont {Itakura}(2013{\natexlab{b}})}]{Hattori:2012ny}%
  \BibitemOpen
  \bibfield  {author} {\bibinfo {author} {\bibfnamefont {Koichi}\ \bibnamefont {Hattori}}\ and\ \bibinfo {author} {\bibfnamefont {Kazunori}\ \bibnamefont {Itakura}},\ }\bibfield  {title} {\enquote {\bibinfo {title} {{Vacuum birefringence in strong magnetic fields: (II) Complex refractive index from the lowest Landau level}},}\ }\href {\doibase 10.1016/j.aop.2013.03.016} {\bibfield  {journal} {\bibinfo  {journal} {Annals Phys.}\ }\textbf {\bibinfo {volume} {334}},\ \bibinfo {pages} {58--82} (\bibinfo {year} {2013}{\natexlab{b}})},\ \Eprint {http://arxiv.org/abs/1212.1897} {arXiv:1212.1897 [hep-ph]} \BibitemShut {NoStop}%
\bibitem [{\citenamefont {Visinelli}\ and\ \citenamefont {Ter\c{c}as}(2022)}]{Visinelli:2018zif}%
  \BibitemOpen
  \bibfield  {author} {\bibinfo {author} {\bibfnamefont {Luca}\ \bibnamefont {Visinelli}}\ and\ \bibinfo {author} {\bibfnamefont {Hugo}\ \bibnamefont {Ter\c{c}as}},\ }\bibfield  {title} {\enquote {\bibinfo {title} {{B-field induced mixing between Langmuir waves and axions}},}\ }\href {\doibase 10.1103/PhysRevD.105.096024} {\bibfield  {journal} {\bibinfo  {journal} {Phys. Rev. D}\ }\textbf {\bibinfo {volume} {105}},\ \bibinfo {pages} {096024} (\bibinfo {year} {2022})},\ \Eprint {http://arxiv.org/abs/1807.06828} {arXiv:1807.06828 [hep-ph]} \BibitemShut {NoStop}%
\bibitem [{\citenamefont {Hattori}\ and\ \citenamefont {Itakura}(2022)}]{Hattori:2022uzp}%
  \BibitemOpen
  \bibfield  {author} {\bibinfo {author} {\bibfnamefont {Koichi}\ \bibnamefont {Hattori}}\ and\ \bibinfo {author} {\bibfnamefont {Kazunori}\ \bibnamefont {Itakura}},\ }\bibfield  {title} {\enquote {\bibinfo {title} {{In-medium polarization tensor in strong magnetic fields (I): Magneto-birefringence at finite temperature and density}},}\ }\href {\doibase 10.1016/j.aop.2022.169114} {\bibfield  {journal} {\bibinfo  {journal} {Annals Phys.}\ }\textbf {\bibinfo {volume} {446}},\ \bibinfo {pages} {169114} (\bibinfo {year} {2022})},\ \Eprint {http://arxiv.org/abs/2205.04312} {arXiv:2205.04312 [hep-ph]} \BibitemShut {NoStop}%
\bibitem [{\citenamefont {Wang}\ and\ \citenamefont {Shovkovy}(2021{\natexlab{a}})}]{Wang:2021ebh}%
  \BibitemOpen
  \bibfield  {author} {\bibinfo {author} {\bibfnamefont {Xinyang}\ \bibnamefont {Wang}}\ and\ \bibinfo {author} {\bibfnamefont {Igor}\ \bibnamefont {Shovkovy}},\ }\bibfield  {title} {\enquote {\bibinfo {title} {{Photon polarization tensor in a magnetized plasma: Absorptive part}},}\ }\href {\doibase 10.1103/PhysRevD.104.056017} {\bibfield  {journal} {\bibinfo  {journal} {Phys. Rev. D}\ }\textbf {\bibinfo {volume} {104}},\ \bibinfo {pages} {056017} (\bibinfo {year} {2021}{\natexlab{a}})},\ \Eprint {http://arxiv.org/abs/2103.01967} {arXiv:2103.01967 [nucl-th]} \BibitemShut {NoStop}%
\bibitem [{\citenamefont {Wang}\ and\ \citenamefont {Shovkovy}(2021{\natexlab{b}})}]{Wang:2021eud}%
  \BibitemOpen
  \bibfield  {author} {\bibinfo {author} {\bibfnamefont {Xinyang}\ \bibnamefont {Wang}}\ and\ \bibinfo {author} {\bibfnamefont {Igor}\ \bibnamefont {Shovkovy}},\ }\bibfield  {title} {\enquote {\bibinfo {title} {{Polarization tensor of magnetized quark-gluon plasma at nonzero baryon density}},}\ }\href {\doibase 10.1140/epjc/s10052-021-09650-3} {\bibfield  {journal} {\bibinfo  {journal} {Eur. Phys. J. C}\ }\textbf {\bibinfo {volume} {81}},\ \bibinfo {pages} {901} (\bibinfo {year} {2021}{\natexlab{b}})},\ \Eprint {http://arxiv.org/abs/2106.09029} {arXiv:2106.09029 [nucl-th]} \BibitemShut {NoStop}%
\bibitem [{\citenamefont {Brahma}\ and\ \citenamefont {Schutz}(2024)}]{Brahma:2024vxb}%
  \BibitemOpen
  \bibfield  {author} {\bibinfo {author} {\bibfnamefont {Nirmalya}\ \bibnamefont {Brahma}}\ and\ \bibinfo {author} {\bibfnamefont {Katelin}\ \bibnamefont {Schutz}},\ }\bibfield  {title} {\enquote {\bibinfo {title} {{Photon conversion to axions and dark photons in magnetized plasmas: a finite-temperature field theory approach}},}\ }\href {\doibase 10.1007/JHEP12(2024)191} {\bibfield  {journal} {\bibinfo  {journal} {JHEP}\ }\textbf {\bibinfo {volume} {12}},\ \bibinfo {pages} {191} (\bibinfo {year} {2024})},\ \Eprint {http://arxiv.org/abs/2410.14771} {arXiv:2410.14771 [hep-ph]} \BibitemShut {NoStop}%
\bibitem [{\citenamefont {Liebert}\ \emph {et~al.}(2003)\citenamefont {Liebert}, \citenamefont {Bergeron},\ and\ \citenamefont {Holberg}}]{Liebert_2003}%
  \BibitemOpen
  \bibfield  {author} {\bibinfo {author} {\bibfnamefont {James}\ \bibnamefont {Liebert}}, \bibinfo {author} {\bibfnamefont {P.}~\bibnamefont {Bergeron}}, \ and\ \bibinfo {author} {\bibfnamefont {J.~B.}\ \bibnamefont {Holberg}},\ }\bibfield  {title} {\enquote {\bibinfo {title} {The true incidence of magnetism among field white dwarfs},}\ }\href {\doibase 10.1086/345573} {\bibfield  {journal} {\bibinfo  {journal} {The Astronomical Journal}\ }\textbf {\bibinfo {volume} {125}},\ \bibinfo {pages} {348–353} (\bibinfo {year} {2003})}\BibitemShut {NoStop}%
\bibitem [{\citenamefont {Kepler}\ \emph {et~al.}(2013)\citenamefont {Kepler}, \citenamefont {Pelisoli}, \citenamefont {Jordan}, \citenamefont {Kleinman}, \citenamefont {Koester}, \citenamefont {Külebi}, \citenamefont {Peçanha}, \citenamefont {Castanheira}, \citenamefont {Nitta}, \citenamefont {Costa}, \citenamefont {Winget}, \citenamefont {Kanaan},\ and\ \citenamefont {Fraga}}]{Kepler_2013}%
  \BibitemOpen
  \bibfield  {author} {\bibinfo {author} {\bibfnamefont {S.~O.}\ \bibnamefont {Kepler}}, \bibinfo {author} {\bibfnamefont {I.}~\bibnamefont {Pelisoli}}, \bibinfo {author} {\bibfnamefont {S.}~\bibnamefont {Jordan}}, \bibinfo {author} {\bibfnamefont {S.~J.}\ \bibnamefont {Kleinman}}, \bibinfo {author} {\bibfnamefont {D.}~\bibnamefont {Koester}}, \bibinfo {author} {\bibfnamefont {B.}~\bibnamefont {Külebi}}, \bibinfo {author} {\bibfnamefont {V.}~\bibnamefont {Peçanha}}, \bibinfo {author} {\bibfnamefont {B.~G.}\ \bibnamefont {Castanheira}}, \bibinfo {author} {\bibfnamefont {A.}~\bibnamefont {Nitta}}, \bibinfo {author} {\bibfnamefont {J.~E.~S.}\ \bibnamefont {Costa}}, \bibinfo {author} {\bibfnamefont {D.~E.}\ \bibnamefont {Winget}}, \bibinfo {author} {\bibfnamefont {A.}~\bibnamefont {Kanaan}}, \ and\ \bibinfo {author} {\bibfnamefont {L.}~\bibnamefont {Fraga}},\ }\bibfield  {title} {\enquote {\bibinfo {title} {Magnetic white dwarf stars in the sloan digital sky survey},}\ }\href {\doibase 10.1093/mnras/sts522}
  {\bibfield  {journal} {\bibinfo  {journal} {Monthly Notices of the Royal Astronomical Society}\ }\textbf {\bibinfo {volume} {429}},\ \bibinfo {pages} {2934–2944} (\bibinfo {year} {2013})}\BibitemShut {NoStop}%
\bibitem [{\citenamefont {Ferrario}\ \emph {et~al.}(2015)\citenamefont {Ferrario}, \citenamefont {de~Martino},\ and\ \citenamefont {Gänsicke}}]{Ferrario_2015}%
  \BibitemOpen
  \bibfield  {author} {\bibinfo {author} {\bibfnamefont {Lilia}\ \bibnamefont {Ferrario}}, \bibinfo {author} {\bibfnamefont {Domitilla}\ \bibnamefont {de~Martino}}, \ and\ \bibinfo {author} {\bibfnamefont {Boris~T.}\ \bibnamefont {Gänsicke}},\ }\bibfield  {title} {\enquote {\bibinfo {title} {Magnetic white dwarfs},}\ }\href {\doibase 10.1007/s11214-015-0152-0} {\bibfield  {journal} {\bibinfo  {journal} {Space Science Reviews}\ }\textbf {\bibinfo {volume} {191}},\ \bibinfo {pages} {111–169} (\bibinfo {year} {2015})}\BibitemShut {NoStop}%
\bibitem [{\citenamefont {Ferrario}\ \emph {et~al.}(2020)\citenamefont {Ferrario}, \citenamefont {Wickramasinghe},\ and\ \citenamefont {Kawka}}]{Ferrario_2020}%
  \BibitemOpen
  \bibfield  {author} {\bibinfo {author} {\bibfnamefont {Lilia}\ \bibnamefont {Ferrario}}, \bibinfo {author} {\bibfnamefont {Dayal}\ \bibnamefont {Wickramasinghe}}, \ and\ \bibinfo {author} {\bibfnamefont {Adela}\ \bibnamefont {Kawka}},\ }\bibfield  {title} {\enquote {\bibinfo {title} {Magnetic fields in isolated and interacting white dwarfs},}\ }\href {\doibase 10.1016/j.asr.2019.11.012} {\bibfield  {journal} {\bibinfo  {journal} {Advances in Space Research}\ }\textbf {\bibinfo {volume} {66}},\ \bibinfo {pages} {1025–1056} (\bibinfo {year} {2020})}\BibitemShut {NoStop}%
\bibitem [{\citenamefont {Amorim}\ \emph {et~al.}(2023)\citenamefont {Amorim}, \citenamefont {Kepler}, \citenamefont {Külebi}, \citenamefont {Jordan},\ and\ \citenamefont {Romero}}]{Amorim_2023}%
  \BibitemOpen
  \bibfield  {author} {\bibinfo {author} {\bibfnamefont {L.~L.}\ \bibnamefont {Amorim}}, \bibinfo {author} {\bibfnamefont {S.~O.}\ \bibnamefont {Kepler}}, \bibinfo {author} {\bibfnamefont {Baybars}\ \bibnamefont {Külebi}}, \bibinfo {author} {\bibfnamefont {S.}~\bibnamefont {Jordan}}, \ and\ \bibinfo {author} {\bibfnamefont {A.~D.}\ \bibnamefont {Romero}},\ }\bibfield  {title} {\enquote {\bibinfo {title} {Catalog of magnetic white dwarfs with hydrogen dominated atmospheres},}\ }\href {\doibase 10.3847/1538-4357/acaf6e} {\bibfield  {journal} {\bibinfo  {journal} {The Astrophysical Journal}\ }\textbf {\bibinfo {volume} {944}},\ \bibinfo {pages} {56} (\bibinfo {year} {2023})}\BibitemShut {NoStop}%
\bibitem [{\citenamefont {Stix}(1962)}]{stix1962theory}%
  \BibitemOpen
  \bibfield  {author} {\bibinfo {author} {\bibfnamefont {Thomas~Howard}\ \bibnamefont {Stix}},\ }\bibfield  {title} {\enquote {\bibinfo {title} {The theory of plasma waves},}\ }\href@noop {} {\bibfield  {journal} {\bibinfo  {journal} {The theory of plasma waves}\ } (\bibinfo {year} {1962})}\BibitemShut {NoStop}%
\bibitem [{\citenamefont {Melrose}(1986)}]{melrose1986instabilities}%
  \BibitemOpen
  \bibfield  {author} {\bibinfo {author} {\bibfnamefont {Donald~B}\ \bibnamefont {Melrose}},\ }\href@noop {} {\emph {\bibinfo {title} {Instabilities in space and laboratory plasmas}}}\ (\bibinfo {year} {1986})\BibitemShut {NoStop}%
\bibitem [{\citenamefont {Redondo}\ and\ \citenamefont {Raffelt}(2013)}]{Redondo:2013lna}%
  \BibitemOpen
  \bibfield  {author} {\bibinfo {author} {\bibfnamefont {Javier}\ \bibnamefont {Redondo}}\ and\ \bibinfo {author} {\bibfnamefont {Georg}\ \bibnamefont {Raffelt}},\ }\bibfield  {title} {\enquote {\bibinfo {title} {{Solar constraints on hidden photons re-visited}},}\ }\href {\doibase 10.1088/1475-7516/2013/08/034} {\bibfield  {journal} {\bibinfo  {journal} {JCAP}\ }\textbf {\bibinfo {volume} {08}},\ \bibinfo {pages} {034} (\bibinfo {year} {2013})},\ \Eprint {http://arxiv.org/abs/1305.2920} {arXiv:1305.2920 [hep-ph]} \BibitemShut {NoStop}%
\bibitem [{\citenamefont {Caputo}\ \emph {et~al.}(2020)\citenamefont {Caputo}, \citenamefont {Millar},\ and\ \citenamefont {Vitagliano}}]{Caputo:2020quz}%
  \BibitemOpen
  \bibfield  {author} {\bibinfo {author} {\bibfnamefont {Andrea}\ \bibnamefont {Caputo}}, \bibinfo {author} {\bibfnamefont {Alexander~J.}\ \bibnamefont {Millar}}, \ and\ \bibinfo {author} {\bibfnamefont {Edoardo}\ \bibnamefont {Vitagliano}},\ }\bibfield  {title} {\enquote {\bibinfo {title} {{Revisiting longitudinal plasmon-axion conversion in external magnetic fields}},}\ }\href {\doibase 10.1103/PhysRevD.101.123004} {\bibfield  {journal} {\bibinfo  {journal} {Phys. Rev. D}\ }\textbf {\bibinfo {volume} {101}},\ \bibinfo {pages} {123004} (\bibinfo {year} {2020})},\ \Eprint {http://arxiv.org/abs/2005.00078} {arXiv:2005.00078 [hep-ph]} \BibitemShut {NoStop}%
\bibitem [{\citenamefont {Fern{\'a}ndez~Molina}\ \emph {et~al.}(2022)\citenamefont {Fern{\'a}ndez~Molina}, \citenamefont {Sigalotti}, \citenamefont {Rend{\'o}n},\ and\ \citenamefont {Mejias}}]{fernandez2022rapidly}%
  \BibitemOpen
  \bibfield  {author} {\bibinfo {author} {\bibfnamefont {Ram{\'o}n~A}\ \bibnamefont {Fern{\'a}ndez~Molina}}, \bibinfo {author} {\bibfnamefont {Leonardo Di~G}\ \bibnamefont {Sigalotti}}, \bibinfo {author} {\bibfnamefont {Otto}\ \bibnamefont {Rend{\'o}n}}, \ and\ \bibinfo {author} {\bibfnamefont {Antonio~J}\ \bibnamefont {Mejias}},\ }\bibfield  {title} {\enquote {\bibinfo {title} {A rapidly convergent method for solving third-order polynomials},}\ }\href@noop {} {\bibfield  {journal} {\bibinfo  {journal} {AIP Advances}\ }\textbf {\bibinfo {volume} {12}} (\bibinfo {year} {2022})}\BibitemShut {NoStop}%
\bibitem [{\citenamefont {Castro-Tapia}\ \emph {et~al.}(2024)\citenamefont {Castro-Tapia}, \citenamefont {Zhang},\ and\ \citenamefont {Cumming}}]{Castro-Tapia_2024}%
  \BibitemOpen
  \bibfield  {author} {\bibinfo {author} {\bibfnamefont {Matias}\ \bibnamefont {Castro-Tapia}}, \bibinfo {author} {\bibfnamefont {Shu}\ \bibnamefont {Zhang}}, \ and\ \bibinfo {author} {\bibfnamefont {Andrew}\ \bibnamefont {Cumming}},\ }\bibfield  {title} {\enquote {\bibinfo {title} {Magnetic field evolution for crystallization-driven dynamos in c/o white dwarfs},}\ }\href {\doibase 10.3847/1538-4357/ad7a6a} {\bibfield  {journal} {\bibinfo  {journal} {The Astrophysical Journal}\ }\textbf {\bibinfo {volume} {975}},\ \bibinfo {pages} {63} (\bibinfo {year} {2024})}\BibitemShut {NoStop}%
\bibitem [{\citenamefont {{Castro-Tapia}}\ and\ \citenamefont {{Cumming}}(2025)}]{2025arXiv250616529C}%
  \BibitemOpen
  \bibfield  {author} {\bibinfo {author} {\bibfnamefont {Matias}\ \bibnamefont {{Castro-Tapia}}}\ and\ \bibinfo {author} {\bibfnamefont {Andrew}\ \bibnamefont {{Cumming}}},\ }\bibfield  {title} {\enquote {\bibinfo {title} {{Three-component Phase Separation for Ultramassive White Dwarf Models}},}\ }\href {\doibase 10.48550/arXiv.2506.16529} {\bibfield  {journal} {\bibinfo  {journal} {arXiv e-prints}\ ,\ \bibinfo {eid} {arXiv:2506.16529}} (\bibinfo {year} {2025})},\ \Eprint {http://arxiv.org/abs/2506.16529} {arXiv:2506.16529 [astro-ph.SR]} \BibitemShut {NoStop}%
\bibitem [{\citenamefont {Raffelt}(1996)}]{raffelt1996stars}%
  \BibitemOpen
  \bibfield  {author} {\bibinfo {author} {\bibfnamefont {Georg~G}\ \bibnamefont {Raffelt}},\ }\href@noop {} {\emph {\bibinfo {title} {Stars as laboratories for fundamental physics: The astrophysics of neutrinos, axions, and other weakly interacting particles}}}\ (\bibinfo  {publisher} {University of Chicago press},\ \bibinfo {year} {1996})\BibitemShut {NoStop}%
\bibitem [{\citenamefont {Isern}\ \emph {et~al.}(2009)\citenamefont {Isern}, \citenamefont {Catalan}, \citenamefont {Garcia-Berro},\ and\ \citenamefont {Torres}}]{Isern:2008fs}%
  \BibitemOpen
  \bibfield  {author} {\bibinfo {author} {\bibfnamefont {J.}~\bibnamefont {Isern}}, \bibinfo {author} {\bibfnamefont {S.}~\bibnamefont {Catalan}}, \bibinfo {author} {\bibfnamefont {E.}~\bibnamefont {Garcia-Berro}}, \ and\ \bibinfo {author} {\bibfnamefont {S.}~\bibnamefont {Torres}},\ }\bibfield  {title} {\enquote {\bibinfo {title} {{Axions and the white dwarf luminosity function}},}\ }\href {\doibase 10.1088/1742-6596/172/1/012005} {\bibfield  {journal} {\bibinfo  {journal} {J. Phys. Conf. Ser.}\ }\textbf {\bibinfo {volume} {172}},\ \bibinfo {pages} {012005} (\bibinfo {year} {2009})},\ \Eprint {http://arxiv.org/abs/0812.3043} {arXiv:0812.3043 [astro-ph]} \BibitemShut {NoStop}%
\bibitem [{\citenamefont {Hansen}\ \emph {et~al.}(2015)\citenamefont {Hansen}, \citenamefont {Richer}, \citenamefont {Kalirai}, \citenamefont {Goldsbury}, \citenamefont {Frewen},\ and\ \citenamefont {Heyl}}]{hansen2015constraining}%
  \BibitemOpen
  \bibfield  {author} {\bibinfo {author} {\bibfnamefont {Brad~MS}\ \bibnamefont {Hansen}}, \bibinfo {author} {\bibfnamefont {Harvey}\ \bibnamefont {Richer}}, \bibinfo {author} {\bibfnamefont {Jason}\ \bibnamefont {Kalirai}}, \bibinfo {author} {\bibfnamefont {Ryan}\ \bibnamefont {Goldsbury}}, \bibinfo {author} {\bibfnamefont {Shane}\ \bibnamefont {Frewen}}, \ and\ \bibinfo {author} {\bibfnamefont {Jeremy}\ \bibnamefont {Heyl}},\ }\bibfield  {title} {\enquote {\bibinfo {title} {Constraining neutrino cooling using the hot white dwarf luminosity function in the globular cluster 47 tucanae},}\ }\href@noop {} {\bibfield  {journal} {\bibinfo  {journal} {The Astrophysical Journal}\ }\textbf {\bibinfo {volume} {809}},\ \bibinfo {pages} {141} (\bibinfo {year} {2015})}\BibitemShut {NoStop}%
\bibitem [{\citenamefont {Corsico}\ \emph {et~al.}(2012{\natexlab{b}})\citenamefont {Corsico}, \citenamefont {Althaus}, \citenamefont {Miller~Bertolami}, \citenamefont {Romero}, \citenamefont {Garc{\'\i}a-Berro}, \citenamefont {Isern},\ and\ \citenamefont {Kepler}}]{corsico2012rate}%
  \BibitemOpen
  \bibfield  {author} {\bibinfo {author} {\bibfnamefont {Alejandro~H}\ \bibnamefont {Corsico}}, \bibinfo {author} {\bibfnamefont {Leandro~G}\ \bibnamefont {Althaus}}, \bibinfo {author} {\bibfnamefont {Marcelo~Miguel}\ \bibnamefont {Miller~Bertolami}}, \bibinfo {author} {\bibfnamefont {Alejandra~D}\ \bibnamefont {Romero}}, \bibinfo {author} {\bibfnamefont {Enrique}\ \bibnamefont {Garc{\'\i}a-Berro}}, \bibinfo {author} {\bibfnamefont {Jordi}\ \bibnamefont {Isern}}, \ and\ \bibinfo {author} {\bibfnamefont {Souza~Oliveira}\ \bibnamefont {Kepler}},\ }\bibfield  {title} {\enquote {\bibinfo {title} {The rate of cooling of the pulsating white dwarf star g117- b15a: a new asteroseismological inference of the axion mass},}\ }\href@noop {} {\bibfield  {journal} {\bibinfo  {journal} {Monthly Notices of the Royal Astronomical Society}\ }\textbf {\bibinfo {volume} {424}},\ \bibinfo {pages} {2792--2799} (\bibinfo {year} {2012}{\natexlab{b}})}\BibitemShut {NoStop}%
\bibitem [{\citenamefont {Dolan}\ \emph {et~al.}(2021)\citenamefont {Dolan}, \citenamefont {Hiskens},\ and\ \citenamefont {Volkas}}]{Dolan:2021rya}%
  \BibitemOpen
  \bibfield  {author} {\bibinfo {author} {\bibfnamefont {Matthew~J.}\ \bibnamefont {Dolan}}, \bibinfo {author} {\bibfnamefont {Frederick~J.}\ \bibnamefont {Hiskens}}, \ and\ \bibinfo {author} {\bibfnamefont {Raymond~R.}\ \bibnamefont {Volkas}},\ }\bibfield  {title} {\enquote {\bibinfo {title} {{Constraining axion-like particles using the white dwarf initial-final mass relation}},}\ }\href {\doibase 10.1088/1475-7516/2021/09/010} {\bibfield  {journal} {\bibinfo  {journal} {JCAP}\ }\textbf {\bibinfo {volume} {09}},\ \bibinfo {pages} {010} (\bibinfo {year} {2021})},\ \Eprint {http://arxiv.org/abs/2102.00379} {arXiv:2102.00379 [hep-ph]} \BibitemShut {NoStop}%
\bibitem [{\citenamefont {{Paxton}}\ \emph {et~al.}(2011)\citenamefont {{Paxton}}, \citenamefont {{Bildsten}}, \citenamefont {{Dotter}}, \citenamefont {{Herwig}}, \citenamefont {{Lesaffre}},\ and\ \citenamefont {{Timmes}}}]{Paxton2011}%
  \BibitemOpen
  \bibfield  {author} {\bibinfo {author} {\bibfnamefont {B.}~\bibnamefont {{Paxton}}}, \bibinfo {author} {\bibfnamefont {L.}~\bibnamefont {{Bildsten}}}, \bibinfo {author} {\bibfnamefont {A.}~\bibnamefont {{Dotter}}}, \bibinfo {author} {\bibfnamefont {F.}~\bibnamefont {{Herwig}}}, \bibinfo {author} {\bibfnamefont {P.}~\bibnamefont {{Lesaffre}}}, \ and\ \bibinfo {author} {\bibfnamefont {F.}~\bibnamefont {{Timmes}}},\ }\bibfield  {title} {\enquote {\bibinfo {title} {{Modules for Experiments in Stellar Astrophysics (MESA)}},}\ }\href {\doibase 10.1088/0067-0049/192/1/3} {\bibfield  {journal} {\bibinfo  {journal} {\apjs}\ }\textbf {\bibinfo {volume} {192}},\ \bibinfo {eid} {3} (\bibinfo {year} {2011})},\ \Eprint {http://arxiv.org/abs/1009.1622} {arXiv:1009.1622 [astro-ph.SR]} \BibitemShut {NoStop}%
\bibitem [{\citenamefont {{Paxton}}\ \emph {et~al.}(2013)\citenamefont {{Paxton}}, \citenamefont {{Cantiello}}, \citenamefont {{Arras}}, \citenamefont {{Bildsten}}, \citenamefont {{Brown}}, \citenamefont {{Dotter}}, \citenamefont {{Mankovich}}, \citenamefont {{Montgomery}}, \citenamefont {{Stello}}, \citenamefont {{Timmes}},\ and\ \citenamefont {{Townsend}}}]{Paxton2013}%
  \BibitemOpen
  \bibfield  {author} {\bibinfo {author} {\bibfnamefont {B.}~\bibnamefont {{Paxton}}}, \bibinfo {author} {\bibfnamefont {M.}~\bibnamefont {{Cantiello}}}, \bibinfo {author} {\bibfnamefont {P.}~\bibnamefont {{Arras}}}, \bibinfo {author} {\bibfnamefont {L.}~\bibnamefont {{Bildsten}}}, \bibinfo {author} {\bibfnamefont {E.~F.}\ \bibnamefont {{Brown}}}, \bibinfo {author} {\bibfnamefont {A.}~\bibnamefont {{Dotter}}}, \bibinfo {author} {\bibfnamefont {C.}~\bibnamefont {{Mankovich}}}, \bibinfo {author} {\bibfnamefont {M.~H.}\ \bibnamefont {{Montgomery}}}, \bibinfo {author} {\bibfnamefont {D.}~\bibnamefont {{Stello}}}, \bibinfo {author} {\bibfnamefont {F.~X.}\ \bibnamefont {{Timmes}}}, \ and\ \bibinfo {author} {\bibfnamefont {R.}~\bibnamefont {{Townsend}}},\ }\bibfield  {title} {\enquote {\bibinfo {title} {{Modules for Experiments in Stellar Astrophysics (MESA): Planets, Oscillations, Rotation, and Massive Stars}},}\ }\href {\doibase 10.1088/0067-0049/208/1/4} {\bibfield  {journal} {\bibinfo  {journal} {\apjs}\
  }\textbf {\bibinfo {volume} {208}},\ \bibinfo {eid} {4} (\bibinfo {year} {2013})},\ \Eprint {http://arxiv.org/abs/1301.0319} {arXiv:1301.0319 [astro-ph.SR]} \BibitemShut {NoStop}%
\bibitem [{\citenamefont {{Paxton}}\ \emph {et~al.}(2015)\citenamefont {{Paxton}}, \citenamefont {{Marchant}}, \citenamefont {{Schwab}}, \citenamefont {{Bauer}}, \citenamefont {{Bildsten}}, \citenamefont {{Cantiello}}, \citenamefont {{Dessart}}, \citenamefont {{Farmer}}, \citenamefont {{Hu}}, \citenamefont {{Langer}}, \citenamefont {{Townsend}}, \citenamefont {{Townsley}},\ and\ \citenamefont {{Timmes}}}]{Paxton2015}%
  \BibitemOpen
  \bibfield  {author} {\bibinfo {author} {\bibfnamefont {B.}~\bibnamefont {{Paxton}}}, \bibinfo {author} {\bibfnamefont {P.}~\bibnamefont {{Marchant}}}, \bibinfo {author} {\bibfnamefont {J.}~\bibnamefont {{Schwab}}}, \bibinfo {author} {\bibfnamefont {E.~B.}\ \bibnamefont {{Bauer}}}, \bibinfo {author} {\bibfnamefont {L.}~\bibnamefont {{Bildsten}}}, \bibinfo {author} {\bibfnamefont {M.}~\bibnamefont {{Cantiello}}}, \bibinfo {author} {\bibfnamefont {L.}~\bibnamefont {{Dessart}}}, \bibinfo {author} {\bibfnamefont {R.}~\bibnamefont {{Farmer}}}, \bibinfo {author} {\bibfnamefont {H.}~\bibnamefont {{Hu}}}, \bibinfo {author} {\bibfnamefont {N.}~\bibnamefont {{Langer}}}, \bibinfo {author} {\bibfnamefont {R.~H.~D.}\ \bibnamefont {{Townsend}}}, \bibinfo {author} {\bibfnamefont {D.~M.}\ \bibnamefont {{Townsley}}}, \ and\ \bibinfo {author} {\bibfnamefont {F.~X.}\ \bibnamefont {{Timmes}}},\ }\bibfield  {title} {\enquote {\bibinfo {title} {{Modules for Experiments in Stellar Astrophysics (MESA): Binaries, Pulsations, and
  Explosions}},}\ }\href {\doibase 10.1088/0067-0049/220/1/15} {\bibfield  {journal} {\bibinfo  {journal} {\apjs}\ }\textbf {\bibinfo {volume} {220}},\ \bibinfo {eid} {15} (\bibinfo {year} {2015})},\ \Eprint {http://arxiv.org/abs/1506.03146} {arXiv:1506.03146 [astro-ph.SR]} \BibitemShut {NoStop}%
\bibitem [{\citenamefont {{Paxton}}\ \emph {et~al.}(2018)\citenamefont {{Paxton}}, \citenamefont {{Schwab}}, \citenamefont {{Bauer}}, \citenamefont {{Bildsten}}, \citenamefont {{Blinnikov}}, \citenamefont {{Duffell}}, \citenamefont {{Farmer}}, \citenamefont {{Goldberg}}, \citenamefont {{Marchant}}, \citenamefont {{Sorokina}}, \citenamefont {{Thoul}}, \citenamefont {{Townsend}},\ and\ \citenamefont {{Timmes}}}]{Paxton2018}%
  \BibitemOpen
  \bibfield  {author} {\bibinfo {author} {\bibfnamefont {B.}~\bibnamefont {{Paxton}}}, \bibinfo {author} {\bibfnamefont {J.}~\bibnamefont {{Schwab}}}, \bibinfo {author} {\bibfnamefont {E.~B.}\ \bibnamefont {{Bauer}}}, \bibinfo {author} {\bibfnamefont {L.}~\bibnamefont {{Bildsten}}}, \bibinfo {author} {\bibfnamefont {S.}~\bibnamefont {{Blinnikov}}}, \bibinfo {author} {\bibfnamefont {P.}~\bibnamefont {{Duffell}}}, \bibinfo {author} {\bibfnamefont {R.}~\bibnamefont {{Farmer}}}, \bibinfo {author} {\bibfnamefont {J.~A.}\ \bibnamefont {{Goldberg}}}, \bibinfo {author} {\bibfnamefont {P.}~\bibnamefont {{Marchant}}}, \bibinfo {author} {\bibfnamefont {E.}~\bibnamefont {{Sorokina}}}, \bibinfo {author} {\bibfnamefont {A.}~\bibnamefont {{Thoul}}}, \bibinfo {author} {\bibfnamefont {R.~H.~D.}\ \bibnamefont {{Townsend}}}, \ and\ \bibinfo {author} {\bibfnamefont {F.~X.}\ \bibnamefont {{Timmes}}},\ }\bibfield  {title} {\enquote {\bibinfo {title} {{Modules for Experiments in Stellar Astrophysics (MESA): Convective Boundaries,
  Element Diffusion, and Massive Star Explosions}},}\ }\href {\doibase 10.3847/1538-4365/aaa5a8} {\bibfield  {journal} {\bibinfo  {journal} {\apjs}\ }\textbf {\bibinfo {volume} {234}},\ \bibinfo {eid} {34} (\bibinfo {year} {2018})},\ \Eprint {http://arxiv.org/abs/1710.08424} {arXiv:1710.08424 [astro-ph.SR]} \BibitemShut {NoStop}%
\bibitem [{\citenamefont {{Paxton}}\ \emph {et~al.}(2019)\citenamefont {{Paxton}}, \citenamefont {{Smolec}}, \citenamefont {{Schwab}}, \citenamefont {{Gautschy}}, \citenamefont {{Bildsten}}, \citenamefont {{Cantiello}}, \citenamefont {{Dotter}}, \citenamefont {{Farmer}}, \citenamefont {{Goldberg}}, \citenamefont {{Jermyn}}, \citenamefont {{Kanbur}}, \citenamefont {{Marchant}}, \citenamefont {{Thoul}}, \citenamefont {{Townsend}}, \citenamefont {{Wolf}}, \citenamefont {{Zhang}},\ and\ \citenamefont {{Timmes}}}]{Paxton2019}%
  \BibitemOpen
  \bibfield  {author} {\bibinfo {author} {\bibfnamefont {Bill}\ \bibnamefont {{Paxton}}}, \bibinfo {author} {\bibfnamefont {R.}~\bibnamefont {{Smolec}}}, \bibinfo {author} {\bibfnamefont {Josiah}\ \bibnamefont {{Schwab}}}, \bibinfo {author} {\bibfnamefont {A.}~\bibnamefont {{Gautschy}}}, \bibinfo {author} {\bibfnamefont {Lars}\ \bibnamefont {{Bildsten}}}, \bibinfo {author} {\bibfnamefont {Matteo}\ \bibnamefont {{Cantiello}}}, \bibinfo {author} {\bibfnamefont {Aaron}\ \bibnamefont {{Dotter}}}, \bibinfo {author} {\bibfnamefont {R.}~\bibnamefont {{Farmer}}}, \bibinfo {author} {\bibfnamefont {Jared~A.}\ \bibnamefont {{Goldberg}}}, \bibinfo {author} {\bibfnamefont {Adam~S.}\ \bibnamefont {{Jermyn}}}, \bibinfo {author} {\bibfnamefont {S.~M.}\ \bibnamefont {{Kanbur}}}, \bibinfo {author} {\bibfnamefont {Pablo}\ \bibnamefont {{Marchant}}}, \bibinfo {author} {\bibfnamefont {Anne}\ \bibnamefont {{Thoul}}}, \bibinfo {author} {\bibfnamefont {Richard H.~D.}\ \bibnamefont {{Townsend}}}, \bibinfo {author} {\bibfnamefont
  {William~M.}\ \bibnamefont {{Wolf}}}, \bibinfo {author} {\bibfnamefont {Michael}\ \bibnamefont {{Zhang}}}, \ and\ \bibinfo {author} {\bibfnamefont {F.~X.}\ \bibnamefont {{Timmes}}},\ }\bibfield  {title} {\enquote {\bibinfo {title} {{Modules for Experiments in Stellar Astrophysics (MESA): Pulsating Variable Stars, Rotation, Convective Boundaries, and Energy Conservation}},}\ }\href {\doibase 10.3847/1538-4365/ab2241} {\bibfield  {journal} {\bibinfo  {journal} {\apjs}\ }\textbf {\bibinfo {volume} {243}},\ \bibinfo {eid} {10} (\bibinfo {year} {2019})},\ \Eprint {http://arxiv.org/abs/1903.01426} {arXiv:1903.01426 [astro-ph.SR]} \BibitemShut {NoStop}%
\bibitem [{\citenamefont {{Jermyn}}\ \emph {et~al.}(2023)\citenamefont {{Jermyn}}, \citenamefont {{Bauer}}, \citenamefont {{Schwab}}, \citenamefont {{Farmer}}, \citenamefont {{Ball}}, \citenamefont {{Bellinger}}, \citenamefont {{Dotter}}, \citenamefont {{Joyce}}, \citenamefont {{Marchant}}, \citenamefont {{Mombarg}}, \citenamefont {{Wolf}}, \citenamefont {{Sunny Wong}}, \citenamefont {{Cinquegrana}}, \citenamefont {{Farrell}}, \citenamefont {{Smolec}}, \citenamefont {{Thoul}}, \citenamefont {{Cantiello}}, \citenamefont {{Herwig}}, \citenamefont {{Toloza}}, \citenamefont {{Bildsten}}, \citenamefont {{Townsend}},\ and\ \citenamefont {{Timmes}}}]{Jermyn2023}%
  \BibitemOpen
  \bibfield  {author} {\bibinfo {author} {\bibfnamefont {Adam~S.}\ \bibnamefont {{Jermyn}}}, \bibinfo {author} {\bibfnamefont {Evan~B.}\ \bibnamefont {{Bauer}}}, \bibinfo {author} {\bibfnamefont {Josiah}\ \bibnamefont {{Schwab}}}, \bibinfo {author} {\bibfnamefont {R.}~\bibnamefont {{Farmer}}}, \bibinfo {author} {\bibfnamefont {Warrick~H.}\ \bibnamefont {{Ball}}}, \bibinfo {author} {\bibfnamefont {Earl~P.}\ \bibnamefont {{Bellinger}}}, \bibinfo {author} {\bibfnamefont {Aaron}\ \bibnamefont {{Dotter}}}, \bibinfo {author} {\bibfnamefont {Meridith}\ \bibnamefont {{Joyce}}}, \bibinfo {author} {\bibfnamefont {Pablo}\ \bibnamefont {{Marchant}}}, \bibinfo {author} {\bibfnamefont {Joey S.~G.}\ \bibnamefont {{Mombarg}}}, \bibinfo {author} {\bibfnamefont {William~M.}\ \bibnamefont {{Wolf}}}, \bibinfo {author} {\bibfnamefont {Tin~Long}\ \bibnamefont {{Sunny Wong}}}, \bibinfo {author} {\bibfnamefont {Giulia~C.}\ \bibnamefont {{Cinquegrana}}}, \bibinfo {author} {\bibfnamefont {Eoin}\ \bibnamefont {{Farrell}}}, \bibinfo
  {author} {\bibfnamefont {R.}~\bibnamefont {{Smolec}}}, \bibinfo {author} {\bibfnamefont {Anne}\ \bibnamefont {{Thoul}}}, \bibinfo {author} {\bibfnamefont {Matteo}\ \bibnamefont {{Cantiello}}}, \bibinfo {author} {\bibfnamefont {Falk}\ \bibnamefont {{Herwig}}}, \bibinfo {author} {\bibfnamefont {Odette}\ \bibnamefont {{Toloza}}}, \bibinfo {author} {\bibfnamefont {Lars}\ \bibnamefont {{Bildsten}}}, \bibinfo {author} {\bibfnamefont {Richard H.~D.}\ \bibnamefont {{Townsend}}}, \ and\ \bibinfo {author} {\bibfnamefont {F.~X.}\ \bibnamefont {{Timmes}}},\ }\bibfield  {title} {\enquote {\bibinfo {title} {{Modules for Experiments in Stellar Astrophysics (MESA): Time-dependent Convection, Energy Conservation, Automatic Differentiation, and Infrastructure}},}\ }\href {\doibase 10.3847/1538-4365/acae8d} {\bibfield  {journal} {\bibinfo  {journal} {\apjs}\ }\textbf {\bibinfo {volume} {265}},\ \bibinfo {eid} {15} (\bibinfo {year} {2023})},\ \Eprint {http://arxiv.org/abs/2208.03651} {arXiv:2208.03651 [astro-ph.SR]}
  \BibitemShut {NoStop}%
\bibitem [{\citenamefont {Bauer}(2023)}]{Bauer_2023}%
  \BibitemOpen
  \bibfield  {author} {\bibinfo {author} {\bibfnamefont {Evan~B.}\ \bibnamefont {Bauer}},\ }\bibfield  {title} {\enquote {\bibinfo {title} {Carbon–oxygen phase separation in modules for experiments in stellar astrophysics (mesa) white dwarf models},}\ }\href {\doibase 10.3847/1538-4357/acd057} {\bibfield  {journal} {\bibinfo  {journal} {The Astrophysical Journal}\ }\textbf {\bibinfo {volume} {950}},\ \bibinfo {pages} {115} (\bibinfo {year} {2023})}\BibitemShut {NoStop}%
\bibitem [{\citenamefont {Euchner}\ \emph {et~al.}(2002)\citenamefont {Euchner}, \citenamefont {Jordan}, \citenamefont {Beuermann}, \citenamefont {Gänsicke},\ and\ \citenamefont {Hessman}}]{Euchner_2002}%
  \BibitemOpen
  \bibfield  {author} {\bibinfo {author} {\bibfnamefont {F.}~\bibnamefont {Euchner}}, \bibinfo {author} {\bibfnamefont {S.}~\bibnamefont {Jordan}}, \bibinfo {author} {\bibfnamefont {K.}~\bibnamefont {Beuermann}}, \bibinfo {author} {\bibfnamefont {B.~T.}\ \bibnamefont {Gänsicke}}, \ and\ \bibinfo {author} {\bibfnamefont {F.~V.}\ \bibnamefont {Hessman}},\ }\bibfield  {title} {\enquote {\bibinfo {title} {Zeeman tomography of magnetic white dwarfs: I. reconstruction of the field geometry from synthetic spectra},}\ }\href {\doibase 10.1051/0004-6361:20020726} {\bibfield  {journal} {\bibinfo  {journal} {Astronomy \& Astrophysics}\ }\textbf {\bibinfo {volume} {390}},\ \bibinfo {pages} {633–647} (\bibinfo {year} {2002})}\BibitemShut {NoStop}%
\bibitem [{\citenamefont {Braithwaite}\ and\ \citenamefont {Spruit}(2004)}]{Braithwaite_2004}%
  \BibitemOpen
  \bibfield  {author} {\bibinfo {author} {\bibfnamefont {Jonathan}\ \bibnamefont {Braithwaite}}\ and\ \bibinfo {author} {\bibfnamefont {Hendrik~C.}\ \bibnamefont {Spruit}},\ }\bibfield  {title} {\enquote {\bibinfo {title} {A fossil origin for the magnetic field in a stars and white dwarfs},}\ }\href {\doibase 10.1038/nature02934} {\bibfield  {journal} {\bibinfo  {journal} {Nature}\ }\textbf {\bibinfo {volume} {431}},\ \bibinfo {pages} {819–821} (\bibinfo {year} {2004})}\BibitemShut {NoStop}%
\bibitem [{\citenamefont {Braithwaite}(2009)}]{Braithwaite_2009}%
  \BibitemOpen
  \bibfield  {author} {\bibinfo {author} {\bibfnamefont {Jonathan}\ \bibnamefont {Braithwaite}},\ }\bibfield  {title} {\enquote {\bibinfo {title} {Axisymmetric magnetic fields in stars: relative strengths of poloidal and toroidal components},}\ }\href {\doibase 10.1111/j.1365-2966.2008.14034.x} {\bibfield  {journal} {\bibinfo  {journal} {Monthly Notices of the Royal Astronomical Society}\ }\textbf {\bibinfo {volume} {397}},\ \bibinfo {pages} {763–774} (\bibinfo {year} {2009})}\BibitemShut {NoStop}%
\bibitem [{\citenamefont {Fujisawa}\ \emph {et~al.}(2012)\citenamefont {Fujisawa}, \citenamefont {Yoshida},\ and\ \citenamefont {Eriguchi}}]{Fujisawa_2012}%
  \BibitemOpen
  \bibfield  {author} {\bibinfo {author} {\bibfnamefont {Kotaro}\ \bibnamefont {Fujisawa}}, \bibinfo {author} {\bibfnamefont {Shin’ichirou}\ \bibnamefont {Yoshida}}, \ and\ \bibinfo {author} {\bibfnamefont {Yoshiharu}\ \bibnamefont {Eriguchi}},\ }\bibfield  {title} {\enquote {\bibinfo {title} {Axisymmetric and stationary structures of magnetized barotropic stars with extremely strong magnetic fields deep inside: Magnetized barotropic star structures},}\ }\href {\doibase 10.1111/j.1365-2966.2012.20614.x} {\bibfield  {journal} {\bibinfo  {journal} {Monthly Notices of the Royal Astronomical Society}\ }\textbf {\bibinfo {volume} {422}},\ \bibinfo {pages} {434–448} (\bibinfo {year} {2012})}\BibitemShut {NoStop}%
\bibitem [{\citenamefont {Bera}\ and\ \citenamefont {Bhattacharya}(2014)}]{Bera_2014}%
  \BibitemOpen
  \bibfield  {author} {\bibinfo {author} {\bibfnamefont {Prasanta}\ \bibnamefont {Bera}}\ and\ \bibinfo {author} {\bibfnamefont {Dipankar}\ \bibnamefont {Bhattacharya}},\ }\bibfield  {title} {\enquote {\bibinfo {title} {Mass–radius relation of strongly magnetized white dwarfs: nearly independent of landau quantization},}\ }\href {\doibase 10.1093/mnras/stu2014} {\bibfield  {journal} {\bibinfo  {journal} {Monthly Notices of the Royal Astronomical Society}\ }\textbf {\bibinfo {volume} {445}},\ \bibinfo {pages} {3951–3958} (\bibinfo {year} {2014})}\BibitemShut {NoStop}%
\bibitem [{\citenamefont {Peterson}\ \emph {et~al.}(2021)\citenamefont {Peterson}, \citenamefont {Dexheimer}, \citenamefont {Negreiros},\ and\ \citenamefont {Castanheira}}]{Peterson_2021}%
  \BibitemOpen
  \bibfield  {author} {\bibinfo {author} {\bibfnamefont {J.}~\bibnamefont {Peterson}}, \bibinfo {author} {\bibfnamefont {V.}~\bibnamefont {Dexheimer}}, \bibinfo {author} {\bibfnamefont {R.}~\bibnamefont {Negreiros}}, \ and\ \bibinfo {author} {\bibfnamefont {B.~G.}\ \bibnamefont {Castanheira}},\ }\bibfield  {title} {\enquote {\bibinfo {title} {Effects of magnetic fields in hot white dwarfs},}\ }\href {\doibase 10.3847/1538-4357/ac1ba7} {\bibfield  {journal} {\bibinfo  {journal} {The Astrophysical Journal}\ }\textbf {\bibinfo {volume} {921}},\ \bibinfo {pages} {1} (\bibinfo {year} {2021})}\BibitemShut {NoStop}%
\bibitem [{\citenamefont {Drewes}\ \emph {et~al.}(2022)\citenamefont {Drewes}, \citenamefont {McDonald}, \citenamefont {Sablon},\ and\ \citenamefont {Vitagliano}}]{Drewes:2021fjx}%
  \BibitemOpen
  \bibfield  {author} {\bibinfo {author} {\bibfnamefont {Marco}\ \bibnamefont {Drewes}}, \bibinfo {author} {\bibfnamefont {Jamie}\ \bibnamefont {McDonald}}, \bibinfo {author} {\bibfnamefont {Lo\"\i{}c}\ \bibnamefont {Sablon}}, \ and\ \bibinfo {author} {\bibfnamefont {Edoardo}\ \bibnamefont {Vitagliano}},\ }\bibfield  {title} {\enquote {\bibinfo {title} {{Neutrino Emissivities as a Probe of the Internal Magnetic Fields of White Dwarfs}},}\ }\href {\doibase 10.3847/1538-4357/ac7874} {\bibfield  {journal} {\bibinfo  {journal} {Astrophys. J.}\ }\textbf {\bibinfo {volume} {934}},\ \bibinfo {pages} {99} (\bibinfo {year} {2022})},\ \Eprint {http://arxiv.org/abs/2109.06158} {arXiv:2109.06158 [astro-ph.SR]} \BibitemShut {NoStop}%
\bibitem [{\citenamefont {Franzon}\ and\ \citenamefont {Schramm}(2015)}]{Franzon:2015gda}%
  \BibitemOpen
  \bibfield  {author} {\bibinfo {author} {\bibfnamefont {Bruno}\ \bibnamefont {Franzon}}\ and\ \bibinfo {author} {\bibfnamefont {Stefan}\ \bibnamefont {Schramm}},\ }\bibfield  {title} {\enquote {\bibinfo {title} {{Effects of strong magnetic fields and rotation on white dwarf structure}},}\ }\href {\doibase 10.1103/PhysRevD.92.083006} {\bibfield  {journal} {\bibinfo  {journal} {Phys. Rev. D}\ }\textbf {\bibinfo {volume} {92}},\ \bibinfo {pages} {083006} (\bibinfo {year} {2015})},\ \Eprint {http://arxiv.org/abs/1507.05557} {arXiv:1507.05557 [astro-ph.SR]} \BibitemShut {NoStop}%
\bibitem [{\citenamefont {Chamel}\ \emph {et~al.}(2014)\citenamefont {Chamel}, \citenamefont {Molter}, \citenamefont {Fantina},\ and\ \citenamefont {Arteaga}}]{PhysRevD.90.043002}%
  \BibitemOpen
  \bibfield  {author} {\bibinfo {author} {\bibfnamefont {N.}~\bibnamefont {Chamel}}, \bibinfo {author} {\bibfnamefont {E.}~\bibnamefont {Molter}}, \bibinfo {author} {\bibfnamefont {A.~F.}\ \bibnamefont {Fantina}}, \ and\ \bibinfo {author} {\bibfnamefont {D.~Pe\~na}\ \bibnamefont {Arteaga}},\ }\bibfield  {title} {\enquote {\bibinfo {title} {Maximum strength of the magnetic field in the core of the most massive white dwarfs},}\ }\href {\doibase 10.1103/PhysRevD.90.043002} {\bibfield  {journal} {\bibinfo  {journal} {Phys. Rev. D}\ }\textbf {\bibinfo {volume} {90}},\ \bibinfo {pages} {043002} (\bibinfo {year} {2014})}\BibitemShut {NoStop}%
\bibitem [{\citenamefont {Das}\ and\ \citenamefont {Mukhopadhyay}(2012)}]{PhysRevD.86.042001}%
  \BibitemOpen
  \bibfield  {author} {\bibinfo {author} {\bibfnamefont {Upasana}\ \bibnamefont {Das}}\ and\ \bibinfo {author} {\bibfnamefont {Banibrata}\ \bibnamefont {Mukhopadhyay}},\ }\bibfield  {title} {\enquote {\bibinfo {title} {Strongly magnetized cold degenerate electron gas: Mass-radius relation of the magnetized white dwarf},}\ }\href {\doibase 10.1103/PhysRevD.86.042001} {\bibfield  {journal} {\bibinfo  {journal} {Phys. Rev. D}\ }\textbf {\bibinfo {volume} {86}},\ \bibinfo {pages} {042001} (\bibinfo {year} {2012})}\BibitemShut {NoStop}%
\bibitem [{\citenamefont {Li}\ and\ \citenamefont {Xu}(2023)}]{Li:2023vpv}%
  \BibitemOpen
  \bibfield  {author} {\bibinfo {author} {\bibfnamefont {Shao-Ping}\ \bibnamefont {Li}}\ and\ \bibinfo {author} {\bibfnamefont {Xun-Jie}\ \bibnamefont {Xu}},\ }\bibfield  {title} {\enquote {\bibinfo {title} {{Production rates of dark photons and Z' in the Sun and stellar cooling bounds}},}\ }\href {\doibase 10.1088/1475-7516/2023/09/009} {\bibfield  {journal} {\bibinfo  {journal} {JCAP}\ }\textbf {\bibinfo {volume} {09}},\ \bibinfo {pages} {009} (\bibinfo {year} {2023})},\ \Eprint {http://arxiv.org/abs/2304.12907} {arXiv:2304.12907 [hep-ph]} \BibitemShut {NoStop}%
\bibitem [{\citenamefont {Di~Lella}\ \emph {et~al.}(2000)\citenamefont {Di~Lella}, \citenamefont {Pilaftsis}, \citenamefont {Raffelt},\ and\ \citenamefont {Zioutas}}]{DiLella:2000dn}%
  \BibitemOpen
  \bibfield  {author} {\bibinfo {author} {\bibfnamefont {L.}~\bibnamefont {Di~Lella}}, \bibinfo {author} {\bibfnamefont {A.}~\bibnamefont {Pilaftsis}}, \bibinfo {author} {\bibfnamefont {G.}~\bibnamefont {Raffelt}}, \ and\ \bibinfo {author} {\bibfnamefont {K.}~\bibnamefont {Zioutas}},\ }\bibfield  {title} {\enquote {\bibinfo {title} {{Search for solar Kaluza-Klein axions in theories of low scale quantum gravity}},}\ }\href {\doibase 10.1103/PhysRevD.62.125011} {\bibfield  {journal} {\bibinfo  {journal} {Phys. Rev. D}\ }\textbf {\bibinfo {volume} {62}},\ \bibinfo {pages} {125011} (\bibinfo {year} {2000})},\ \Eprint {http://arxiv.org/abs/hep-ph/0006327} {arXiv:hep-ph/0006327} \BibitemShut {NoStop}%
\bibitem [{\citenamefont {Carenza}\ \emph {et~al.}(2020)\citenamefont {Carenza}, \citenamefont {Straniero}, \citenamefont {D\"obrich}, \citenamefont {Giannotti}, \citenamefont {Lucente},\ and\ \citenamefont {Mirizzi}}]{Carenza:2020zil}%
  \BibitemOpen
  \bibfield  {author} {\bibinfo {author} {\bibfnamefont {Pierluca}\ \bibnamefont {Carenza}}, \bibinfo {author} {\bibfnamefont {Oscar}\ \bibnamefont {Straniero}}, \bibinfo {author} {\bibfnamefont {Babette}\ \bibnamefont {D\"obrich}}, \bibinfo {author} {\bibfnamefont {Maurizio}\ \bibnamefont {Giannotti}}, \bibinfo {author} {\bibfnamefont {Giuseppe}\ \bibnamefont {Lucente}}, \ and\ \bibinfo {author} {\bibfnamefont {Alessandro}\ \bibnamefont {Mirizzi}},\ }\bibfield  {title} {\enquote {\bibinfo {title} {{Constraints on the coupling with photons of heavy axion-like-particles from Globular Clusters}},}\ }\href {\doibase 10.1016/j.physletb.2020.135709} {\bibfield  {journal} {\bibinfo  {journal} {Phys. Lett. B}\ }\textbf {\bibinfo {volume} {809}},\ \bibinfo {pages} {135709} (\bibinfo {year} {2020})},\ \Eprint {http://arxiv.org/abs/2004.08399} {arXiv:2004.08399 [hep-ph]} \BibitemShut {NoStop}%
\bibitem [{\citenamefont {Raffelt}(1986{\natexlab{a}})}]{Raffelt:1985nk}%
  \BibitemOpen
  \bibfield  {author} {\bibinfo {author} {\bibfnamefont {Georg~G.}\ \bibnamefont {Raffelt}},\ }\bibfield  {title} {\enquote {\bibinfo {title} {{ASTROPHYSICAL AXION BOUNDS DIMINISHED BY SCREENING EFFECTS}},}\ }\href {\doibase 10.1103/PhysRevD.33.897} {\bibfield  {journal} {\bibinfo  {journal} {Phys. Rev. D}\ }\textbf {\bibinfo {volume} {33}},\ \bibinfo {pages} {897} (\bibinfo {year} {1986}{\natexlab{a}})}\BibitemShut {NoStop}%
\bibitem [{\citenamefont {Bottaro}\ \emph {et~al.}(2023)\citenamefont {Bottaro}, \citenamefont {Caputo}, \citenamefont {Raffelt},\ and\ \citenamefont {Vitagliano}}]{Bottaro:2023gep}%
  \BibitemOpen
  \bibfield  {author} {\bibinfo {author} {\bibfnamefont {Salvatore}\ \bibnamefont {Bottaro}}, \bibinfo {author} {\bibfnamefont {Andrea}\ \bibnamefont {Caputo}}, \bibinfo {author} {\bibfnamefont {Georg}\ \bibnamefont {Raffelt}}, \ and\ \bibinfo {author} {\bibfnamefont {Edoardo}\ \bibnamefont {Vitagliano}},\ }\bibfield  {title} {\enquote {\bibinfo {title} {{Stellar limits on scalars from electron-nucleus bremsstrahlung}},}\ }\href {\doibase 10.1088/1475-7516/2023/07/071} {\bibfield  {journal} {\bibinfo  {journal} {JCAP}\ }\textbf {\bibinfo {volume} {07}},\ \bibinfo {pages} {071} (\bibinfo {year} {2023})},\ \Eprint {http://arxiv.org/abs/2303.00778} {arXiv:2303.00778 [hep-ph]} \BibitemShut {NoStop}%
\bibitem [{\citenamefont {Raffelt}(1986{\natexlab{b}})}]{Raffelt:1985nj}%
  \BibitemOpen
  \bibfield  {author} {\bibinfo {author} {\bibfnamefont {Georg~G.}\ \bibnamefont {Raffelt}},\ }\bibfield  {title} {\enquote {\bibinfo {title} {{Axion Constraints From White Dwarf Cooling Times}},}\ }\href {\doibase 10.1016/0370-2693(86)91588-1} {\bibfield  {journal} {\bibinfo  {journal} {Phys. Lett. B}\ }\textbf {\bibinfo {volume} {166}},\ \bibinfo {pages} {402--406} (\bibinfo {year} {1986}{\natexlab{b}})}\BibitemShut {NoStop}%
\bibitem [{\citenamefont {Nakagawa}\ \emph {et~al.}(1987)\citenamefont {Nakagawa}, \citenamefont {Kohyama},\ and\ \citenamefont {Itoh}}]{Nakagawa:1987pga}%
  \BibitemOpen
  \bibfield  {author} {\bibinfo {author} {\bibfnamefont {Masayuki}\ \bibnamefont {Nakagawa}}, \bibinfo {author} {\bibfnamefont {Yasuharu}\ \bibnamefont {Kohyama}}, \ and\ \bibinfo {author} {\bibfnamefont {Naoki}\ \bibnamefont {Itoh}},\ }\bibfield  {title} {\enquote {\bibinfo {title} {{Axion Bremsstrahlung in Dense Stars}},}\ }\href {\doibase 10.1086/165724} {\bibfield  {journal} {\bibinfo  {journal} {Astrophys. J.}\ }\textbf {\bibinfo {volume} {322}},\ \bibinfo {pages} {291} (\bibinfo {year} {1987})}\BibitemShut {NoStop}%
\bibitem [{\citenamefont {Nakagawa}\ \emph {et~al.}(1988)\citenamefont {Nakagawa}, \citenamefont {Adachi}, \citenamefont {Kohyama},\ and\ \citenamefont {Itoh}}]{Nakagawa:1988rhp}%
  \BibitemOpen
  \bibfield  {author} {\bibinfo {author} {\bibfnamefont {Masayuki}\ \bibnamefont {Nakagawa}}, \bibinfo {author} {\bibfnamefont {Tomoo}\ \bibnamefont {Adachi}}, \bibinfo {author} {\bibfnamefont {Yasuharu}\ \bibnamefont {Kohyama}}, \ and\ \bibinfo {author} {\bibfnamefont {Naoki}\ \bibnamefont {Itoh}},\ }\bibfield  {title} {\enquote {\bibinfo {title} {{Axion bremsstrahlung in dense stars. II - Phonon contributions}},}\ }\href {\doibase 10.1086/166085} {\bibfield  {journal} {\bibinfo  {journal} {Astrophys. J.}\ }\textbf {\bibinfo {volume} {326}},\ \bibinfo {pages} {241} (\bibinfo {year} {1988})}\BibitemShut {NoStop}%
\bibitem [{\citenamefont {Navas}\ \emph {et~al.}(2024)\citenamefont {Navas} \emph {et~al.}}]{ParticleDataGroup:2024cfk}%
  \BibitemOpen
  \bibfield  {author} {\bibinfo {author} {\bibfnamefont {S.}~\bibnamefont {Navas}} \emph {et~al.} (\bibinfo {collaboration} {Particle Data Group}),\ }\bibfield  {title} {\enquote {\bibinfo {title} {{Review of particle physics}},}\ }\href {\doibase 10.1103/PhysRevD.110.030001} {\bibfield  {journal} {\bibinfo  {journal} {Phys. Rev. D}\ }\textbf {\bibinfo {volume} {110}},\ \bibinfo {pages} {030001} (\bibinfo {year} {2024})}\BibitemShut {NoStop}%
\bibitem [{\citenamefont {Dine}\ and\ \citenamefont {Fischler}(1983{\natexlab{b}})}]{Dine:1982ah}%
  \BibitemOpen
  \bibfield  {author} {\bibinfo {author} {\bibfnamefont {Michael}\ \bibnamefont {Dine}}\ and\ \bibinfo {author} {\bibfnamefont {Willy}\ \bibnamefont {Fischler}},\ }\bibfield  {title} {\enquote {\bibinfo {title} {{The Not So Harmless Axion}},}\ }\href {\doibase 10.1016/0370-2693(83)90639-1} {\bibfield  {journal} {\bibinfo  {journal} {Phys. Lett. B}\ }\textbf {\bibinfo {volume} {120}},\ \bibinfo {pages} {137--141} (\bibinfo {year} {1983}{\natexlab{b}})}\BibitemShut {NoStop}%
\bibitem [{\citenamefont {Shifman}\ \emph {et~al.}(1980)\citenamefont {Shifman}, \citenamefont {Vainshtein},\ and\ \citenamefont {Zakharov}}]{Shifman:1979if}%
  \BibitemOpen
  \bibfield  {author} {\bibinfo {author} {\bibfnamefont {Mikhail~A.}\ \bibnamefont {Shifman}}, \bibinfo {author} {\bibfnamefont {A.~I.}\ \bibnamefont {Vainshtein}}, \ and\ \bibinfo {author} {\bibfnamefont {Valentin~I.}\ \bibnamefont {Zakharov}},\ }\bibfield  {title} {\enquote {\bibinfo {title} {{Can Confinement Ensure Natural CP Invariance of Strong Interactions?}}}\ }\href {\doibase 10.1016/0550-3213(80)90209-6} {\bibfield  {journal} {\bibinfo  {journal} {Nucl. Phys. B}\ }\textbf {\bibinfo {volume} {166}},\ \bibinfo {pages} {493--506} (\bibinfo {year} {1980})}\BibitemShut {NoStop}%
\bibitem [{\citenamefont {Srednicki}(1985)}]{Srednicki:1985xd}%
  \BibitemOpen
  \bibfield  {author} {\bibinfo {author} {\bibfnamefont {Mark}\ \bibnamefont {Srednicki}},\ }\bibfield  {title} {\enquote {\bibinfo {title} {{Axion Couplings to Matter. 1. CP Conserving Parts}},}\ }\href {\doibase 10.1016/0550-3213(85)90054-9} {\bibfield  {journal} {\bibinfo  {journal} {Nucl. Phys. B}\ }\textbf {\bibinfo {volume} {260}},\ \bibinfo {pages} {689--700} (\bibinfo {year} {1985})}\BibitemShut {NoStop}%
\bibitem [{\citenamefont {Berlin}\ and\ \citenamefont {Kahn}(2024)}]{Berlin:2024pzi}%
  \BibitemOpen
  \bibfield  {author} {\bibinfo {author} {\bibfnamefont {Asher}\ \bibnamefont {Berlin}}\ and\ \bibinfo {author} {\bibfnamefont {Yonatan}\ \bibnamefont {Kahn}},\ }\bibfield  {title} {\enquote {\bibinfo {title} {{New Technologies for Axion and Dark Photon Searches}},}\ }\href {\doibase 10.1146/annurev-nucl-121423-101015} {\  (\bibinfo {year} {2024}),\ 10.1146/annurev-nucl-121423-101015},\ \Eprint {http://arxiv.org/abs/2412.08704} {arXiv:2412.08704 [hep-ph]} \BibitemShut {NoStop}%
\bibitem [{\citenamefont {Berlin}\ and\ \citenamefont {Trickle}(2024)}]{Berlin:2023ppd}%
  \BibitemOpen
  \bibfield  {author} {\bibinfo {author} {\bibfnamefont {Asher}\ \bibnamefont {Berlin}}\ and\ \bibinfo {author} {\bibfnamefont {Tanner}\ \bibnamefont {Trickle}},\ }\bibfield  {title} {\enquote {\bibinfo {title} {{Absorption of Axion Dark Matter in a Magnetized Medium}},}\ }\href {\doibase 10.1103/PhysRevLett.132.181801} {\bibfield  {journal} {\bibinfo  {journal} {Phys. Rev. Lett.}\ }\textbf {\bibinfo {volume} {132}},\ \bibinfo {pages} {181801} (\bibinfo {year} {2024})},\ \Eprint {http://arxiv.org/abs/2305.05681} {arXiv:2305.05681 [hep-ph]} \BibitemShut {NoStop}%
\bibitem [{\citenamefont {Alonso-{\'A}lvarez}\ and\ \citenamefont {Curtin}(2025)}]{Alonso-Alvarez:2024ypq}%
  \BibitemOpen
  \bibfield  {author} {\bibinfo {author} {\bibfnamefont {Gonzalo}\ \bibnamefont {Alonso-{\'A}lvarez}}\ and\ \bibinfo {author} {\bibfnamefont {David}\ \bibnamefont {Curtin}},\ }\bibfield  {title} {\enquote {\bibinfo {title} {{Dark astronomy with dark matter detectors}},}\ }\href {\doibase 10.1088/1475-7516/2025/05/082} {\bibfield  {journal} {\bibinfo  {journal} {JCAP}\ }\textbf {\bibinfo {volume} {05}},\ \bibinfo {pages} {082} (\bibinfo {year} {2025})},\ \Eprint {http://arxiv.org/abs/2412.06883} {arXiv:2412.06883 [astro-ph.CO]} \BibitemShut {NoStop}%
\end{thebibliography}%


%
\end{document}